\documentstyle[preprint,pre,aps,eqsecnum,epsf,graphicx]{revtex}
\begin{document}
\tightenlines

\title{Liquid-gas Phase Transition in Nuclear Multifragmentation}

\author{S. Das Gupta$^1$, A. Z. Mekjian$^2$ and M. B. Tsang$^3$}

\address{
$^1$Physics Dept., McGill University, 3600 University Street, Montreal,
Canada H3A 2T8\\
$^2$Physics Dept., Rutgers University, Piscataway, New Jersey 08854, USA\\
$^3$National Superconducting Cyclotron Laboratory,
Michigan State University, East Lansing, Michigan 48824, USA}

\date{ \today }

\maketitle

\begin{abstract}

The equation of state of nuclear matter suggests that at suitable beam 
energies the disassembling hot system formed in heavy ion collisions
will pass through a liquid-gas coexistence region.  Searching for
the signatures of the phase transition has been a very
important focal point of experimental endeavours in heavy-ion collisions,
in the last fifteen years. Simultaneously theoretical models have been 
developed to provide information about the equation of state 
and reaction mechanisms consistent with the experimental observables.  
This article is a review of this endeavour.
\end{abstract}


\section{Introduction}
Heavy-ion collisions allow one to pump energy into a nuclear system.
In central collisions of equal size nuclei one can also create a significant
amount of compression using high energy nuclear beams.  
The possibility of studying nuclei far from 
normal conditions raises the question: can we study phase transitions
in nuclei similar, for instance, 
to the way, one can study phase transition in water?
This is the subject of the present article.

Two important phase transitions are being studied using heavy-ion collisions
from medium to very high energies.  One phase transition occurs at
densities that are subnormal and at temperatures of a 
few MeV ( 1 MeV$\approx 10^{10}K$).
Nuclei at normal density and zero temperature behave
like Fermi liquids so that this transition is a liquid to gas phase
transition.  The second phase transition of current interest is expected
at a much higher temperature ($\approx$150 MeV) and at a much higher density
(several times normal density) and will be the  subject of intense
experimental investigation at the Relativistic Heavy Ion Collider (RHIC)
and at CERN in the coming decade.
There one expects to see transition from hadronic matter to a quark-gluon
plasma.  In very high energy collisions many new particles are created.  
This is a domain very much beyond the limits of non-relativistic
quantum mechanics with conservation of particles.  Thus, we will not treat
this phenomenon at all.  Instead at an energy scale of tens of MeV, we 
should be able to stretch
or compress pieces of nuclear matter and we expect to see Van der Waals type
of behavior.  As a Van der Waals gas is considered to be a classic example of
a liquid-gas phase transition, we have a situation similar to that in
condensed matter physics.

Unfortunately, the experimental conditions in the nuclear physics case
are quite severe.  The collisions which produce
different phases of nuclear matter are over in $10^{-22}$ seconds.  
Thus, we can not keep matter in an ``abnormal'' state long enough to
study the properties.
Furthermore, the detectors 
measure only the products of these collisions where all the final products
are in normal states.  We have to extrapolate from the end products to
what happened during disassembly.  This is a difficult task which
complicates confirmation of theoretical predictions.

This article is written so that it is suitable for nuclear physicists 
not specialised
in the area of heavy-ion collisions.  We hope it is also accessible to
non nuclear physicists since the ideas are quite general and
well-known from statistical physics.  We hope practising heavy-ion
collision physicists will also find this a useful reference.  The plan
of the article is as follows.  Section II deals with early theoretical
discussions which showed that well established models predict that during
disassembly after heavy-ion collisions bulk matter will enter liquid-gas
coexistence region provided the beam energies are suitably chosen.  In
Section III an experimental overview is provided.  Sections IV to IX
bring us in contact with some experimental results.  
Here we show, for example,
how estimates of temperature or freeze-out density are extracted from 
experiments.  Sections X to XXII are primarily theoretical.  We introduce
and develop some models; some we simply sketch without providing all the
details, as that would make the article extremely long.

\section{Liquid Gas Phase Transition In Nuclear Mean-Field Theory}
Nuclear matter is an idealised system of equal number of
neutrons N and protons Z.  The system is vary large and
the Coulomb interaction between protons
is switched off.  For light nuclei the Coulomb interaction 
has a very small effect and N=Z nuclei have the highest binding energy.  As
nuclei get bigger the Coulomb energy shifts the highest binding energy
towards nuclei with N$>$Z.  This brings into play the symmetry energy which
is repulsive and is proportional to (N-Z)$^2$/(N+Z).  Stable systems are
scarce after mass number A=N+Z$>$260.  Thus no known nuclei approach
the limit of nuclear matter.  However,  
extrapolation from known nuclei leads one to deduce that nuclear matter has
density $\rho_0\approx 0.16 fm^{-3}$ and binding energy $\approx$16 MeV/A.  
We will choose an Equation of State (EOS) of this idealised nuclear matter
to examine if a liquid-gas phase transition can be expected and at
what temperature and density.

The following
parametrisation, called the Skyrme parametrisation for the interaction 
potential energy, has been demonstrated 
\cite {Ring} to be a good approximation for Hartree-Fock calculations. 
We take the potential energy density arising from nuclear forces to be
\begin{eqnarray}
v(\rho)=\frac{a}{2}\frac{\rho ^2}{\rho_0}+\frac{b}{\sigma +1}\frac{\rho^{
\sigma+1}}{\rho_0^{\sigma}}
\end{eqnarray}
Our unit of length is $fm$ and unit of energy is MeV.  In the above
$\rho_0$=0.16$fm^{-3}$, $a,b$ are in MeV, $a$ is attractive, $b$
repulsive and $\sigma$ is a parameter.  The constants should be chosen
such that in nuclear matter the minimum energy is obtained
at $\rho=\rho_0$ with  energy E/A=-16 MeV.
This fixes two of the three parameters and the third can be
obtained by the compressibility coefficient (nuclear physics has its own
unique definition of compressibility coefficient 
$k\partial p/\partial \rho$ at $\rho_0$).  The Skyrme parametrisation 
is simple enough that we will write down all the relevant formulae.  
From Eq.(2.1) the energy per particle as a function of 
density $\rho$ at zero temperature is given by
\begin{eqnarray}
\frac {E}{A}(\rho)=\frac{a}{2}\frac{\rho}{\rho_0}+\frac{b}{\sigma +1}
(\frac {\rho}{\rho_0})^{\sigma}+22.135(\frac{\rho}{\rho_0})^{2/3}
\end{eqnarray}
In Eq.(2.2) the last term on the right hand side 
is the zero-temperature
Fermi-gas value for kinetic energy.  The pressure due to the interaction
at zero or any temperature is 
\begin{eqnarray}
p=[\frac{a}{2}\frac{\rho}{\rho_0}+\frac{\sigma b}{\sigma+1}
(\frac{\rho}{\rho_0})^{\sigma}]\times \rho
\end{eqnarray}
The condition that $E/A$ minimise at $\rho/\rho_0=1$ gives
\begin{eqnarray}
\frac{a}{2}+\frac{b\sigma}{\sigma+1}+(2/3)\times 22.135=0
\end{eqnarray}
The condition that $E/A$ is -16 MeV at $\rho_0$ gives
\begin{eqnarray}
-16=\frac{a}{2}+\frac{b}{\sigma +1}+22.135
\end{eqnarray}
Lastly, compressibility is given by
\begin{eqnarray}
k=9\times (a+\sigma b+\frac{p_F^2}{3m})
\end{eqnarray}
The EOS for the Skyrme parametrisation with $a$=-356.8 MeV, $b$=303.9 MeV
and $\sigma$=7/6 (this gives $k=201$MeV) is shown in Fig. 1.  In the
figure isotherms are drawn for various temperatures (10, 12, 14, 15,
15.64 and 17 MeV).  The pressure contributed by kinetic energy was
calculated in the finite temperature Fermi-gas model.  The similarity with
Van der Waals EOS is obvious; for a more quantitative comparison we refer
to Jaqaman et. al \cite {Jaquaman}.  With the parameters chosen here the
critical temperature is 15.64 MeV.  The spinodal region ($\partial p/
\partial \rho<0$) can be seen clearly.
The coexistence curve which is shown in the figure is obtained using a
Maxwell construction \cite {Reif}.

\vskip 0.2in

\epsfxsize=3.5in
\epsfysize=3.5in
\centerline{\epsffile{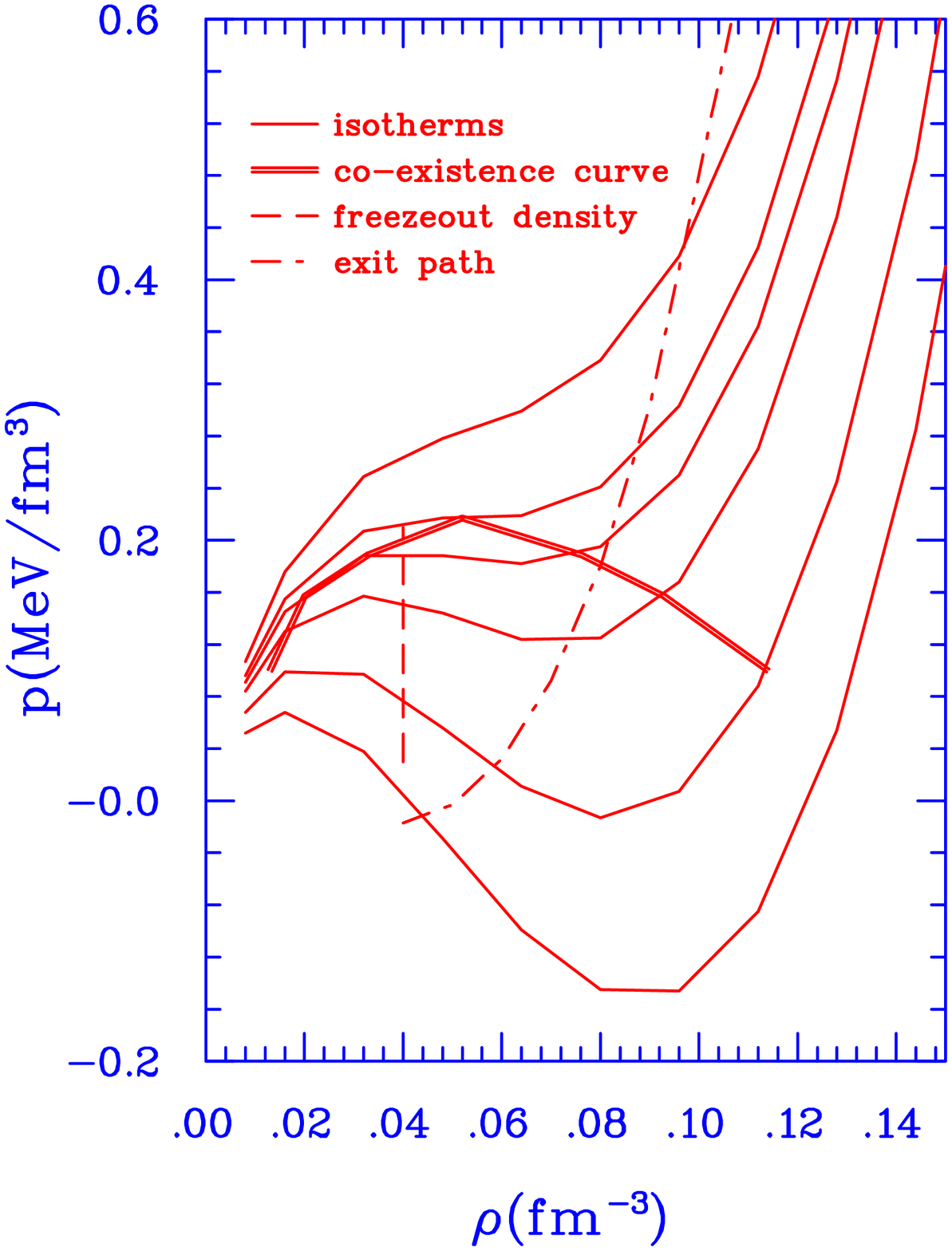}}
\vskip 0.2in

Fig. 1. Nuclear matter equation of state with Skyrme interaction with
compressibility 201 MeV.  In ascending order the isothermals are at
temperatures 10, 12, 14, 15, 15.64 (critical isotherm) and 17 MeV.
The coexistence curve obtained from a Maxwell construction is shown.
The vertical line is drawn at assumed freeze-out density 0.04 fm$^{-3}$.
The dot-dash line is obtained by assuming that the excited system 
expands isentropically (see Fig.2).  This is an idealisation.

\vskip 0.2in
We now describe how in heavy-ion collisions one will sweep across the
$p-\rho$ plane.  In heavy-ion collisions one distinguishes between
spectators and participants.  Imagine two equal ions  colliding at
zero impact parameter.  Some highly excited nucleons are emitted first.
The other nucleons are called participants because
each nucleon will collide with at least one nucleon
in its path if all nucleons are assumed
to move in straight line paths.  
In peripheral collisions where the impact parameters are non-zero,
nucleons outside the overlapping zone would not have collided with nucleons
from the other nucleus.  These ``non-interacting'' nucleons are
defined as spectators.  For the
same beam energy much more energy is pumped into the participating zone.  There
will even be compression in this zone if the excitation is very
high.  Spectators are only mildly excited.
They are excited for many reasons: highly non-spherical shapes,
unfavourable N/Z ratios,  migration 
from participants etc..  
In general, spectators should have little compression.  
Both central and peripheral collisions have been studied
to find signals of phase transitions.

Imagine then, as a result of heavy-ion collisions, nuclear matter has been
excited to a high temperature, with or without compression.  Looking at 
Fig. 1 we see that the pressure will be positive and matter will begin
to expand \cite {Siemens83,Curtin,Bertsch83}.  The exit path is hard
to guess but the simplest expectation supported by transport models is 
that it is
approximately isentropic  in the beginning part of the expansion. 
However, if the system reaches the spinodal region,  
the mean-field description is inappropriate
and the system is expected to break up into chunks.  In Fig. 1
we have nonetheless followed the isentropic trajectory.  Expansion continues
till it reaches a freeze-out volume, a theoretical  
idealisation.  Once the freeze-out volume is reached there is no 
exchange of matter between different fragments.  Since the fragments are still
hot, they will get rid of their excitation by evaporation (sequential
two body decays \cite {Friedman83,Weisskopf}) 
before they reach the detector.  The freeze-out density
is significantly lower than the normal density.  It is probably
not as low as one-tenth the normal density because interactions between
fragments (except for Coulomb forces) will cease well before that.  The
freeze-out density is often a parameter in the theory adjusted to get
the best fit and is model dependent.
The `best' choice seems to be always less than half the normal density.
In Fig. 1 we have shown this arbitrarily to be $\rho_{fr}=0.04 fm^{-3}$, 
that is, one-quarter of normal nuclear density.

Normally, the EOS is drawn with isotherms but some additional insight can be
gained by looking at isentrops \cite {Bertsch83}.  For this we refer to
Fig. 2 where we have drawn, for the same Skyrme interaction, $\frac{E}{A}(\rho)
$ and $p(\rho)$ but now for constant entropy instead of constant temperature.

\epsfxsize=4.5in
\epsfysize=5.5in
\centerline{\epsffile{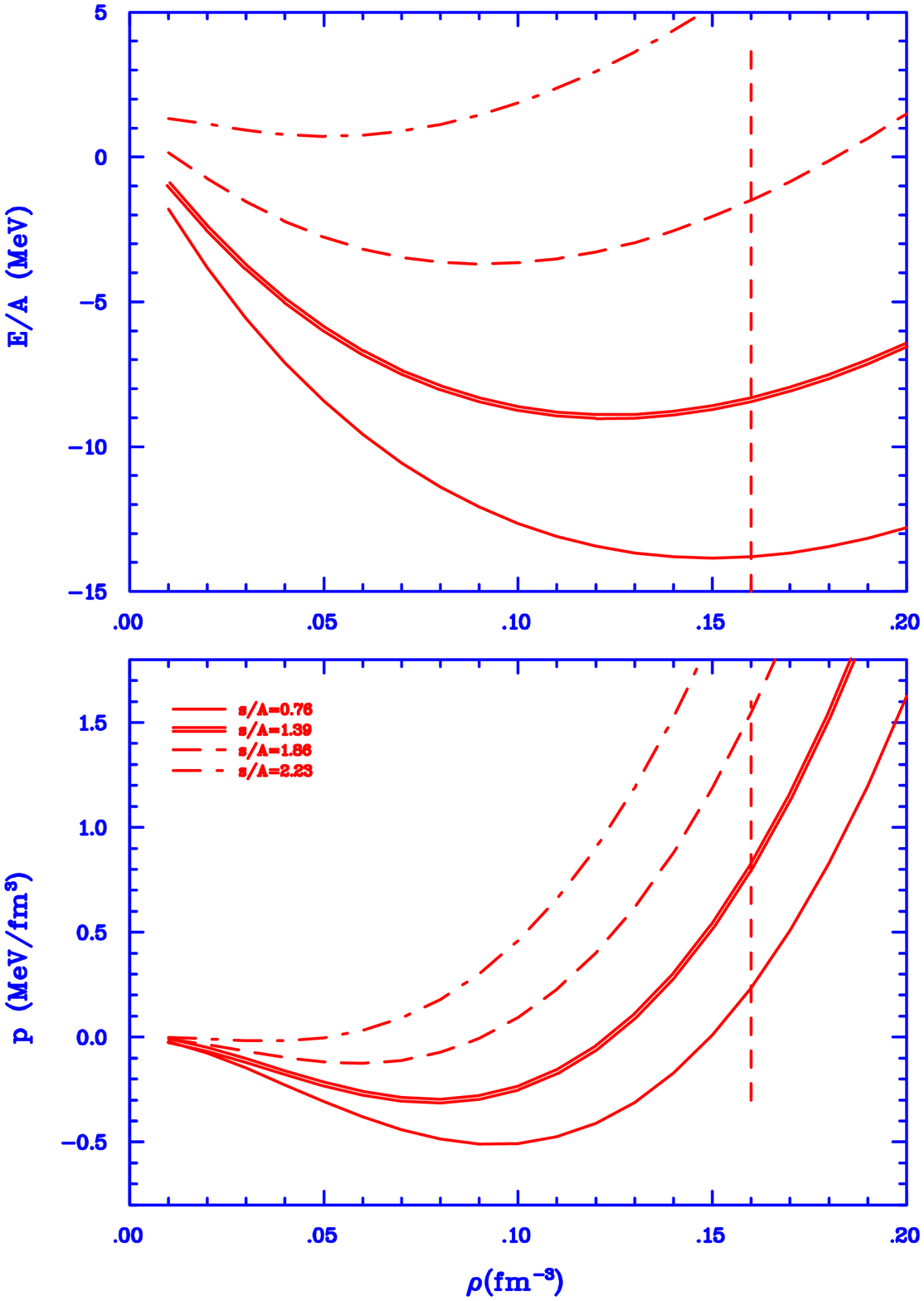}}
\vskip 0.2in

Fig.2. For the same Skyrme interaction as used in Fig.1, we draw
$\frac{E}{A}(\rho)$ and $p(\rho)$ but now for fixed entropy rather
than fixed temperature.  The vertical dashed line is along normal nuclear
density.  We consider disassembly of an excited spectator.  If the system
starts from $E/A\approx -2$MeV (entropy 1.86 per particle), the initial
pressure is positive and it begins to expand (in this idealisation)
along the isentrop.  The thermal $E/A$ will fall, and to compensate,
 collective velocity will develop.  This collective velocity will take
the system beyond the minimum of $E/A$ and drive it to the region of
spinodal instability.  For small excitation energy and entropy(0.76)
the system starts with positive pressure but does not attain enough
collective velocity to reach the spinodal region.  It will de-excite
by sequential decay.

\vskip 0.2in
Imagine then an excited spectator is formed at normal density
indicated by the vertical dashed line.  Let us
focus on two isentrops, $S/A$=1.86 and $S/A$=0.76.  In the first case
the system starts with $E/A\approx -2$ MeV and positive pressure.  It
will begin to expand; the value of ``thermal'' $E/A$ as it expands
along the isentrop, drops.  For conservation of energy it must then
develop a collective flow.  This collective flow will take it beyond
the minimum of $E/A$ and drives it to the spinodal region.
For the isentrop with value 0.76, even though
it starts with positive pressure, it does not gain enough collective
energy to drive it to the spinodal region.  It will therefore oscillate
around zero pressure and has to de-excite by other means
( two-body decay \cite {Friedman83}).  The intermediate case with $S/A$=
1.39 just makes it to the spinodal region.

It is also clear both from Fig. 1 and Fig. 2 that if the starting point
is too high (i.e., too much excitation energy) the system will entirely
miss the liquid region and probe only the gas region at the time
of dissociation.  This was the case at Bevalac \cite {Daswest}
(incident energy in GeV)
where the goal of studying heavy-ion reactions was quite different.

While mean-field theory, as described above, easily leads to a liquid-gas
phase transition picture, one clearly needs to go much beyond. 
There is
hardly any observable that one can calculate using mean-field theory alone.
Most common experimental observables are the clusters, their compositions, 
their excitations, velocities etc.
Mean-field theory does not give these values although it suggests
that the system must break up because of spinodal instability.   
However, with the Maxwell construction obtained from the mean-field EOS
one can draw a coexistence curve (Fig. 1)
and this has experimental relevance.  As we will describe in much greater
detail later, one may measure the caloric curve \cite{Poch} defined as
$T$ vs. $E^*/A$ in heavy ion experiments.
The experiment gives a measure of the specific heat, $dE^*/dT$ in the
vicinity of temperature 5 MeV.
Indeed many models ( to be described in later sections)
produce a peak in the specific heat at about this temperature.  The peak
is reminiscent of the crossing of the coexistence curve \cite {Heller}.  
These models are also able to calculate many other observables with 
reasonable success.  Let us refer to Fig.1 to see where we would expect
to see the peak in mean-field theory.  If we consider the freeze-out
density to be 0.04$fm^{-3}$, the intersection of the 0.04$fm^{-3}$ line 
and the coexistence curve suggests a temperature of about 15 MeV.
This ``boiling'' temperature will come down if a lower freeze-out
density is used but even at one tenth the normal density the boiling
temperature is still 12 MeV.  Since we have used nuclear matter
theory, the temperature is expected to be lower due
to finite particle number and Coulomb interaction.  
In ref \cite {Jaquaman} and later
in \cite{jaquaman2}, the effect of
finite particle number was estimated to be significant.  In addition
we must remember that mean-field theories normally overestimate the
critical temperature.  For example, in the Ising model, this 
overestimation is about 50 per cent \cite {Huang}.  
In mean-field
Thomas-Fermi theory that includes the Coulomb interaction De {\it et. al.}
\cite {De97}
find the peak in specific heat at 10 MeV for $^{150}$Sm.  
Without the Coulomb 
interaction, in bulk matter with the same isospin asymmetry as $^{150}$Sm,
the peak is located at 13 MeV.  As will be described in greater detail
later, both experimental data and more realistic models point to
much lower temperature.  Thus in mean-field theory interesting things
seem to happen at too high a temperature.

\section{Experimental Overview}

Experimentally the following features are well known.  At excitation 
energy $\epsilon \approx$ 1 MeV/nucleon, successive emissions 
of particles by
evaporation of the compound nucleus or its fission are the basic
deexcitation mechanisms.  The picture can be justified by saying 
that there is enough time between successive emissions so that the
nucleus can relax ($\tau_{re}\approx 2R/c_s$) to a new equilibrium
state where $R$ is the radius of the compound nucleus and $c_s$ is
the velocity of sound.  
At $\epsilon \approx$ 3 MeV/nucleon, the time interval between successive
emissions is comparable with $\tau_{re}$.  At excitation energy
comparable to binding energy $\epsilon \approx 8$ MeV/nucleon
the very existence
of a long-lived compound nucleus is unlikely which leads to the scenario of
an explosion-like process involving the whole nucleus.
This will lead to multiple emission of nuclear
fragments of different masses.  This is what is called ``multifragmentation''
where `multi' is more than two.  Associated with multifragmentation
is a term Intermediate Mass Fragment (IMF) that we will use often.
This refers to particles with charge $Z$ between 3 and 20 to 30.  The lower
charge limit is set to 3 because of exceptional binding of the alpha
particle.  The upper limit is set not to include 
fission like fragments; if the nucleus broke up into several chunks
in the spinodal region, we could expect some of them to be IMF's.
In the mean-field scenario described earlier,  
multifragmentation is associated with the co-existence region.
Thus, it is considered to be the most promising experimental observable
to study the liquid-gas phase transition in nuclear matter.  However, while
phase transition signals will always be weakened by finite particle
number effects, multifragmentation is usually found at the appropriate
energy and occurs in nuclear collisions even when the thermodynamic
limit is not reached.  Thus multifragmentation is a more general
process than phase transition.

In the co-existence region, light particles such as neutrons,
hydrogen isotopes (p, d, t) and helium isotopes ($^3$He, $^4$He, $^6$He)
are considered as gas while the IMF's are treated as droplet forms
of the liquid.  In collisions where larger residues remain, they are
the liquid remnants from the original colliding nuclei.  Since
nuclei are two-component systems consisting of neutrons and protons,
the isotopic contents of the gas and the liquid phase will be different.
This is specially so when bound nuclei of smaller sizes are usually
found along the valley of stability and have nearly equal number of
protons and neutrons.  Thus if the initial collisions consist of heavy
nuclei which have more neutrons than protons, one would expect the excess
neutrons to diffuse out to the gas region resulting in a neutron
enriched nucleon gas.  This has already been seen in experiments and will
be discussed later.  Thus a preliminary glimpse of the phase transition
in nuclei suggests a much richer structure than what has been implied by
nuclear matter alone.  

Multifragmentation was seen in high-energy proton-nucleus collisions
\cite {Poskanzer71,Westfall78,Finn82,Hirsch84} before systematic
studies were undertaken in nucleus-nucleus collisions.  For a proton
incident on a nucleus the picture is as follows.  Shortly after the
collision between the proton and the target nucleus, several prompt
nucleons leave the system and carry off much of the energy of the
collision.  At low incident proton energies only remnants near the
mass of the target are produced.  For incident proton energies around
0.5 GeV, the system undergoes fission leaving two large fragments.
When the incident proton energies are between 1.0 GeV and 10 GeV, the
cross-section for multifragmentation rises by an order of magnitude.
At energies above 20 GeV the cross-section becomes independent of
energy, reaching the limiting fragmentation region.

Systematic studies of multifragmentation have been undertaken using
heavy-ion beams since the mid eighties, when these beams became
routinely available and large detection arrays were built.  The
production of fragments from central collisions reaches a maximum
around 100A MeV.  In the following section we will examine 
various aspects of the
multifragmentation process, which may be employed to signal the
liquid-gas phase transition.

\section{Event Selection}

Most early multifragmentation experiments are
inclusive measurements \cite
{Poskanzer71,Westfall78,Finn82,Hirsch84,Fields84,sobotka83},
 i.e. particles are identified with no
requirements that other particles from the same event
should be detected in coincidence. These types of
experiments do not provide information about the
collision dynamics or properties of the emission
sources from the nuclear reaction. Since
multifragmentation of spectators, produced in
peripheral collisions has different characteristics from
fragments emitted from the participant zone formed in
central collisions, it is important that the emission
sources be identified. There are both advantages and
disadvantages of using central or peripheral collisions
to find the signals of phase transition. In this section,
the methods used to select central and peripheral
collisions will be discussed.

\subsection{Central Collisions}

In central collisions, the excitation energy
pumped into the participant zone is higher and the
source characteristics, i.e. selection of a single source,
can be accomplished easily with large detector arrays
which provide nearly 4$\pi$ angular coverage.
Intuitively, one expects more particles to be produced
in violent or central collisions than peripheral
collisions. Thus the number of emitted particles can be
related to the collision geometry and the simplest
observable is the number of charged particles detected,
$N_c$. There are variations of the observable $N_c$, such as
the hydrogen multiplicity, $N_1$ or light charge particle
multiplicity, $N_{LCP}$. All these observables work
reasonably well in distinguishing central collisions
with small impact parameters from peripheral collisions
\cite {Phair92}. However, the neutron
multiplicity, $N_n$ \cite {Kunde90} and the IMF multiplicity,
$N_{IMF}$ \cite {Phair99} do not work as well. Aside from
multiplicities, there are other observables such as the
mid-rapidity charge, $Z_y$, \cite {Llope95} and the total
transverse kinetic energy, $E_t$, of the identified particles
\cite {Lukasik97}, which can be used as an impact parameter
filter. $Z_y$ is the summed charge of particles with
rapidity between that of the target and projectile. This
quantity reflects properties of the participant zone. The
total transverse energy is defined as
$E_t =\sum_i E_isin^2\theta_i =\sum_i(p_isin\theta_i)^2/2m_i$
where $E_i, p_i$, and $\theta_i$ denote the kinetic energy,
momentum and emission angle of particle $i$ with
respect to the beam axis.

The most common way to relate an
experimental observable to the impact parameter is to
assume a monotonic relationship between the
observable and the impact parameter \cite {Cavata90,Phair92}.
In general, a reduced impact-parameter scale,
$\hat{b}$ which ranges between 0 (head-on collisions) to 1
(glancing collisions), is defined as
\begin{equation}
\hat{b}=b(X)/b_{max}=(\int_X^{\infty}[dP(X')/dX']dX')^{1/2}
\end{equation}
where $X=N_C, N_1, N_n, N_{IMF}, N_{LCP}, E_t, Z_y, dP(X)/dX$ is
the normalized probability distribution for the
measured quantity $X$, and $b_{max}$ is the maximum impact
parameter for which particles were detected in the near
$4\pi$ detection array.

For illustration of the impact parameter
determination, the top panel of Fig.3 shows the
charged particle multiplicity distribution of the $^{84}$Kr
induced reaction on $^{197}$Au at 35 MeV per nucleon
incident energy \cite {Williams98}. The bottom panel shows
the relationship between the reduced impact
parameter, $\hat{b}$, and $N_c$ with a lower cut of $N_c>2$
applied in the analysis.

\vskip 0.2in

\epsfxsize=4.5in
\epsfysize=5.5in
\centerline{\rotatebox{180}{\epsffile{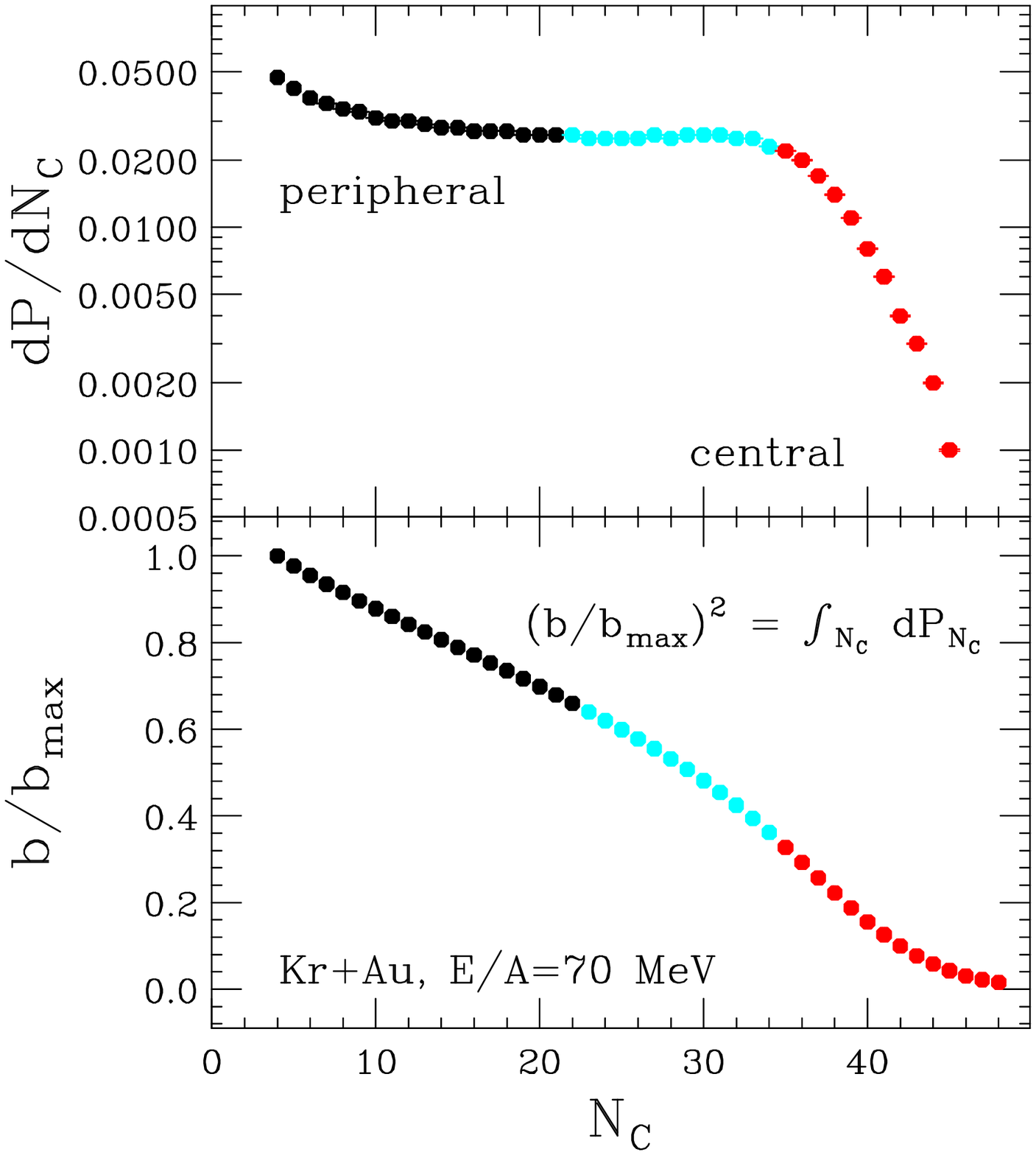}}}
\vskip 0.2in

Fig. 3: (Top) Charged particle multiplicity distribution of the $^{84}$Kr 
induced
reaction on $^{197}$Au at 35 MeV per nucleon incident energy. 
(Bottom) $\hat{b}$ as a
function of $N_c$ \cite {Williams98}.

\vskip 0.2in

While $N_c$ is the most simple observable to
measure the impact parameters, it is not very precise
due to fluctuations and geometric efficiencies of the
detection device . In cases where the single source
from the central collision needs to be better defined or
determined, additional constraints are applied. In
head-on collisions, the angular momentum transfer is
zero and all the emitted particles are emitted
isotropically in the azimuthal angle \cite{Phair93}. Thus
additional constraints on central collisions can be
placed by requiring the detected particles to have
isotropic emission pattern. Other constraints include
requiring the total charge detected to be a substantial
fraction of that of the initial system \cite {D'Agostino95}, the ratio
of total transverse momentum to longitudinal kinetic
energy or by requiring the velocity of the emitted
particles to be about half of the center of mass velocity
\cite {Llope95,Xu00}. Obviously, each additional constraint
reduces the number of events available for analysis.
Too many constraints may reduce the data to the
extreme tails of the distributions where large
fluctuations of the observable become a problem.

During the compression stage, pressure in the
central region causes the participant zone to expand
\cite {Danielewicz95}. Thus part of the available energy in
central collision is converted to collective energy such
as radial flow which expands outward, or transverse
flow caused by the spectators being pushed by the
participant region to the side, and the squeeze-out of
nucleons from the participant region perpendicular to
the reaction plane due to blocking by the spectators
\cite {Reisdorf97}. Clearly, all these collective motions
strongly affect the signals of the phase-transition
observed in central collisions and reduce the amount
of excitation energy available for heating up the
system. They must be understood and taken into
account in the study of phase transition.

\subsection{Peripheral Collisions}

Theoretically, spectators should be less affected
by the effects of collective motion than the participants.
The collision kinematics focus the emitted fragments
from the projectile to the forward direction in the
laboratory. These fragments are generally detected
with spectrometers or detectors placed at forward
angles and the charges, velocities etc. are identified.  The
decay of a projectile spectator is easier to study
experimentally than the target spectator which is
emitted backward with very low energy in the
laboratory frame.

Unlike central collisions, the impact parameter
is strongly correlated with the size of the source in
peripheral collisions. Thus the size of the projectile-like
residue such as the charge, $Z_{PLF}$, provides some
indication of the impact parameter \cite {Dempsey96}. In
the case where most of the projectiles fragment into
many small pieces, the quantity $Z_{bound}$ defined
as the sum of atomic numbers $Z_i$ of all fragments
with $Z_i\ge 2$ has been found to be a good measure
for impact parameter \cite{Hubele91}.  It represents the
charge of the spectator system reduced by the number
of hydrogen isotopes emitted during its decay and
thus, it is the complement of the hydrogen multiplicity,
$N_1$. In an experiment where both $Z_{bound}$ was
measured by the forward spectrometer and $N_c$ was
measured by a $4\pi$ array in the reaction of Au+Au at
E/A=400 MeV, the two observables are anti-correlated
\cite {Hsi95}. Thus, like $N_c$ and other observables discussed
in previous section, $Z_{bound}$ can be used in eq.(4.1) 
to provide a quantitative measure of the impact
parameter.

\section{Evidence for Nuclear expansion}

Around incident energy of 50A MeV, fragment
multiplicities increase with the size of the emission
source and excitation energy. In examining reactions of
Xe on various targets, $^{12}$C, $^{27}$Al, 
$^{51}$V, $^{nat}$Cu, $^{89}$Y and $^{197}$Au,
even though the targets span a range of N/Z from 1.0
to 1.5, a near-universal correlation has been observed
between the average number of emitted IMFs, $<N_{IMF}>$,
and the charge-particle multiplicity, $N_c$, for non-
central collisions \cite{Tso95}. 
Fig.4 shows the mean number of IMF
detected in the collision of $^{129}$Xe+$^{197}$Au at 50A MeV
as a function of the detected charge particle
multiplicity, $N_c$ \cite {Bowman91}. In the most central
collision, $N_c>$33, the mean number of $<N_{IMF}>$ is 7 but
up to 14 IMF fragments have been observed. The large
fragment multiplicities cannot be reproduced by the
break-up of the hot system at normal nuclear matter
density with either the dynamical or statistical models.
(Predictions of various statistical models are lower
than the data as shown by the dashed lines, open
circles and crosses.) Calculations requiring expansion
to less than 1/3 of the normal nuclear matter density is
needed to explain the large increase in $<N_{imf}>$ as shown
by the solid lines.

\epsfxsize=4in
\epsfysize=4in
\centerline{\rotatebox{-90}{\epsffile{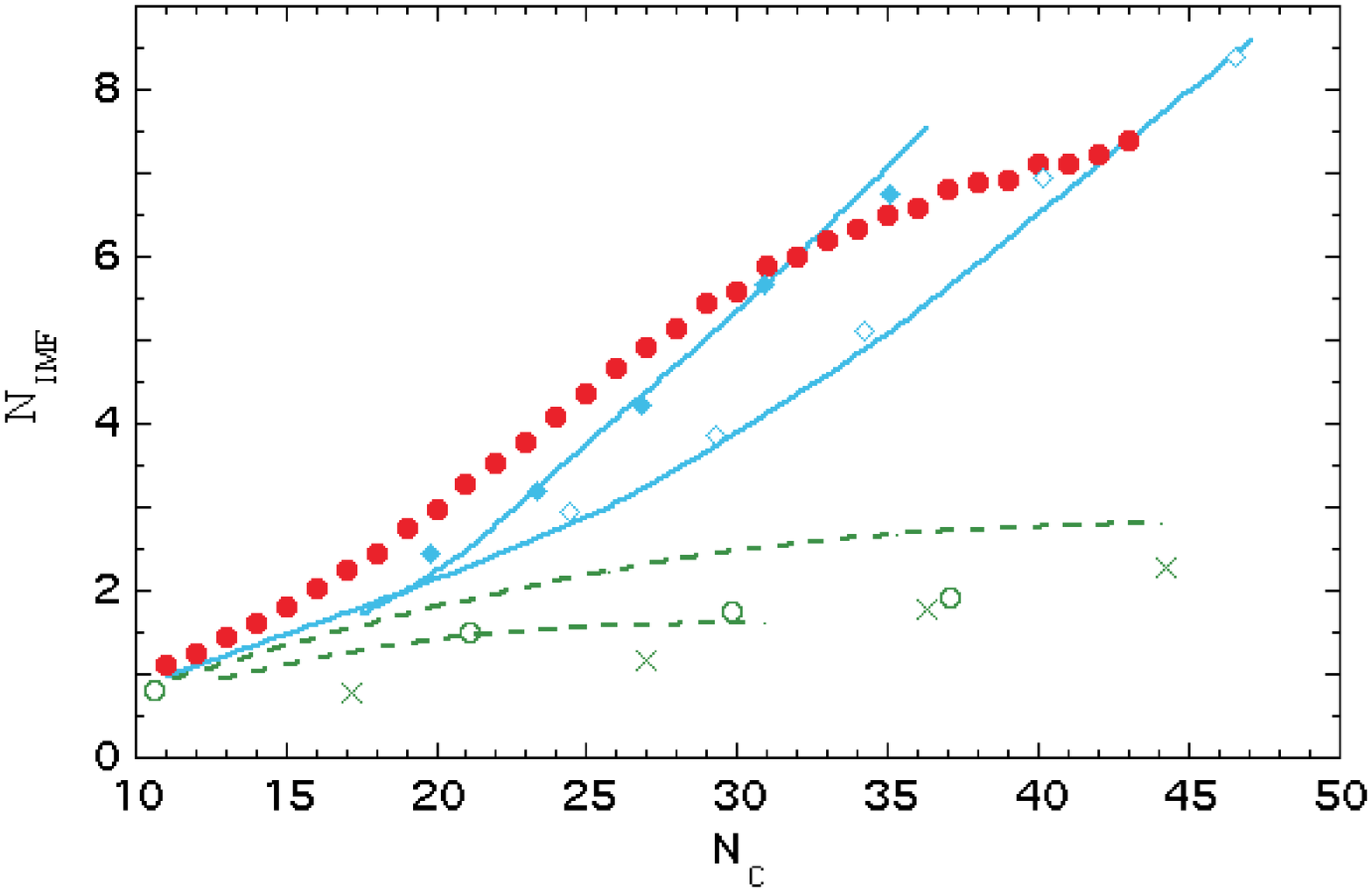}}}
\vskip 0.2in
Fig. 4: The mean number of IMF detected $N_{IMF}$ in the collision of
$^{129}$Xe+$^{197}$Au at 50A MeV as a function of the 
detected charge particle multiplicity, $N_c$.
Data are represented by the solid points. The dashed
lines, open circles and crosses are data from various statistical models
assuming normal nuclear matter density. They all under-predict the number of
IMF emitted. The solid lines are calculations from an expanding nuclear
system. See \cite{Bowman91} for more details about the calculations.

\vskip 0.2in
If the hot nuclear system expands, the ``radial"
component of the velocity should be evidenced in the
particle energy spectra. Without the influence of radial
expansion, the energy spectra resulting from the collision
of a target and projectile at intermediate energy are
composed of three isotropically emitting thermal
sources corresponding to the projectile-like and target-
like spectators in addition to the participant region
formed by the overlap of the projectile and target.
Instead, the IMF and light particle energy spectra from
the central collisions of Au+Au reaction show a
shoulder like shape \cite{Hsi94,Lisa95}. To fit the energy
spectra, large radial expansion velocities are required
in addition to the three sources \cite{Hsi94,Lisa95}.
Similarly, the mean kinetic and transverse energy of
emitted fragments also provide measure of the radial
collective velocities when compared to the predictions
of thermal models \cite{Williams97,Lauret98,D'Agostino96}. 
Fig.5 shows a nearly linear
relationship between the radial velocities with the
incident energy \cite{Williams97}. The plot suggests that
30\% to 60\% of the available energy is used in the radial
expansion. This energy is thus not available for
thermal heating of the nuclear matter. Evidence for the
``nuclear expansion" of the hot nuclear systems is a
necessary but not sufficient condition for the
occurrence of a liquid-gas phase transition.

\vskip 0.2in
\epsfxsize=4in
\epsfysize=4in
\centerline{\rotatebox{90}{\epsffile{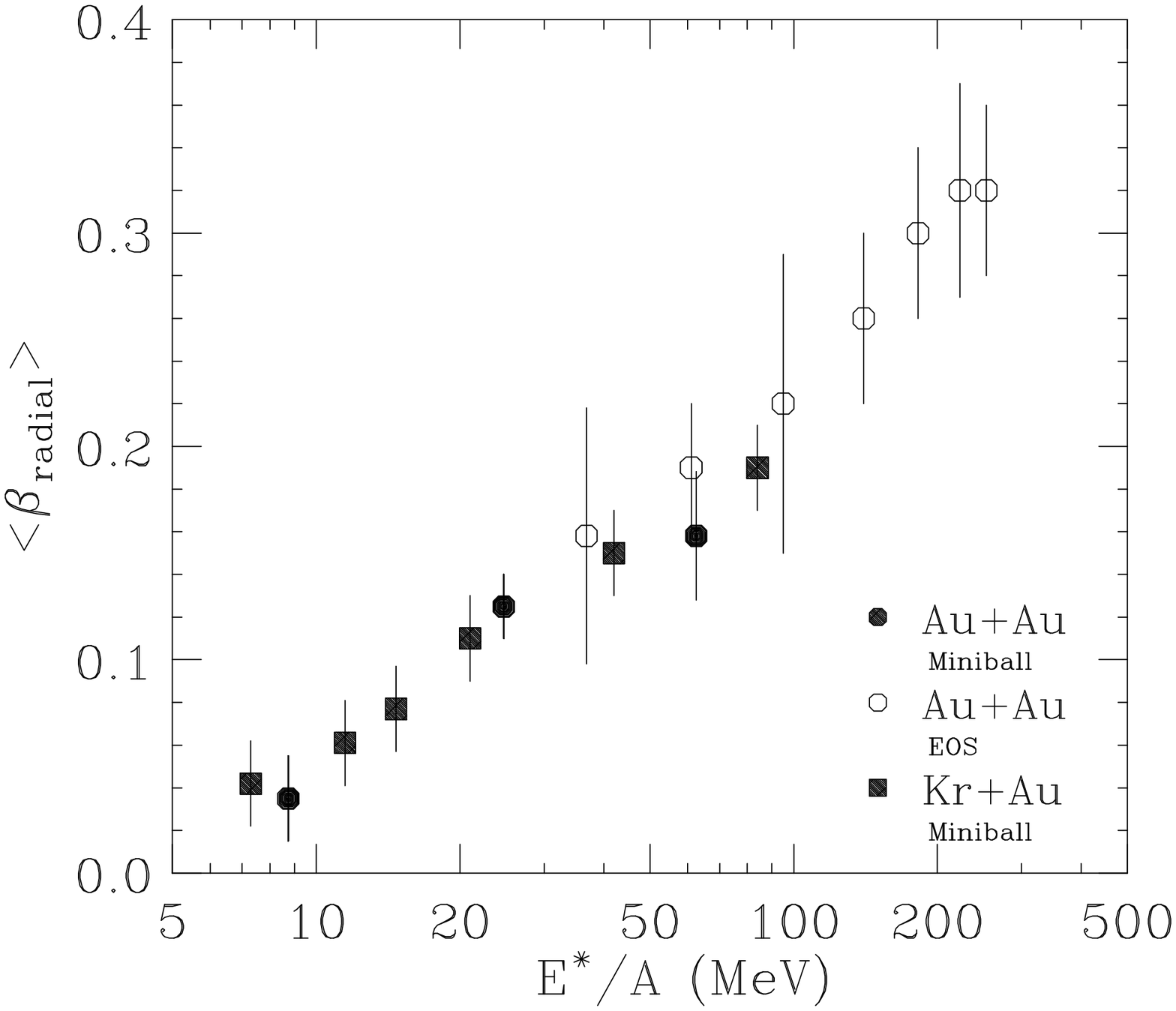}}}
\vskip 0.2in

Fig. 5: Systematics of average radial velocity as a function of incident 
energy for various systems \cite{Williams97}.

\vskip 0.2in
\section{Space-Time Determination}

The average radial velocity plotted in Fig.5 indicates
that the nuclear expansion occurs in a rather short time
($10^{-22}$ second). As a result of the fast expansion, the
density of the reaction zone is below normal nuclear
matter density. Information about the space-time
evolution of the reaction zone can be obtained via
intensity interferometry. The principle behind such
experiments is similar to the intensity
interferometry \cite{Hanbury-Brown56} employed to
determine the radius of stars, where both singles $(Y_i)$
and coincident $(Y_{12})$ yields of photons from the same
source (star) are measured. Intuitively, one expects the
correlation to be small if the source size is large and
a large correlation from a small source.
 In nuclear physics, particles are detected
instead of photons. A correlation function constructed
from these yields is defined as,
$1+R(p_1, p_2)= Y_{12}(p_{12})/(Y_1(p_1)Y_2(p_2))$
where $p_i$ is the laboratory momentum of particle $i$.
At large relative momenta where the final interaction
is negligible, $R(p_1, p_2)$ should be zero. Unlike
astronomy where the space-time evolution of stars is
slow, the time scale involved in nuclear physics is very
short. Thus there are ambiguities in determining the
size and time-scale of nuclear reactions using intensity
interferometry, because a small source emitting over a
long period of time behaves like a large source emitting
over a short period of time \cite{Fox93}.

The space-time information of the emitting
source can be obtained by measuring the correlation
function. An example of the fragment-fragment
correlation from the Ar+Au reaction at E/A=50 MeV is
shown in Fig.6 \cite{Glasmacher94} as a function of the
reduced velocity, $v_{red}=v_{rel}/\sqrt{(Z_1+ Z_2)}$ where 
$v_{rel}$ is the
relative velocity between fragment 1 and 2. The use of
$v_{red}$ allows summing over different combinations of
fragment-fragment correlations. Basic features of the
correlation functions for different particle pairs depend
on details of the final state interaction between the two
particles. For intermediate mass-fragments, the most
important interaction is the Coulomb interaction
between the particles as shown by the suppression of
the correlation functions at small $v_{red}$. However, if the
fragments are emitted in the vicinity of a heavy
reaction residue, the Coulomb interaction with the
residue may not be neglected \cite{Durand95}.
This space-time ambiguity is illustrated by the
calculations shown as lines in Fig.6 \cite{Glasmacher94}.

\epsfxsize=4.5in
\epsfysize=4.5in
\centerline{\rotatebox{90}{\epsffile{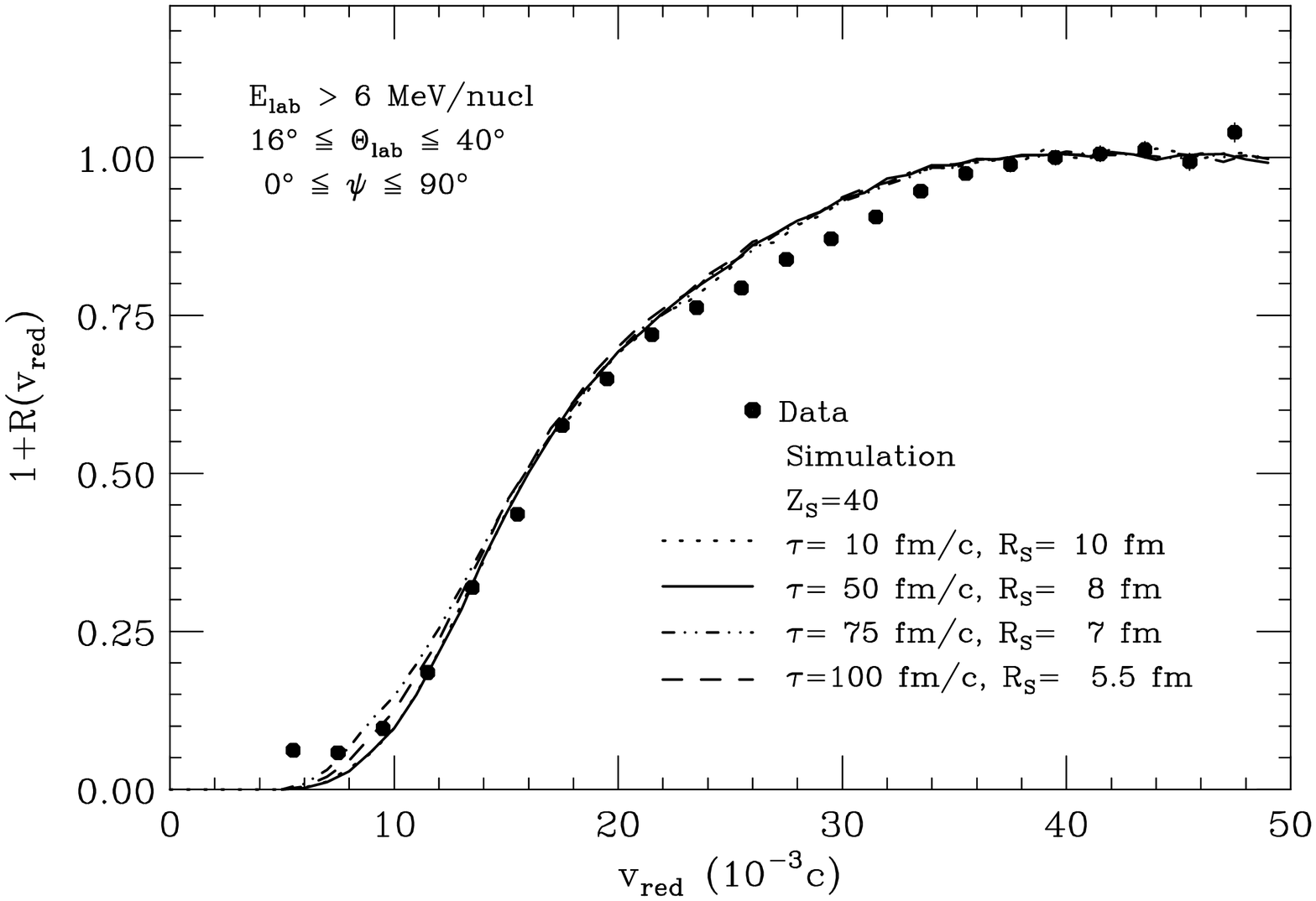}}}
\vskip 0.5in

Fig. 6: The fragment-fragment correlation from the Ar+Au reaction at E/A=50
MeV is shown as a function of the reduced velocity, 
$v_{red}= v_{rel}/\sqrt{(Z1+ Z2)}$
where $v_{rel}$ is the relative velocity between 
fragment 1 and 2 \cite{Glasmacher94}.
The lines are Monte Carlo simulations of many body Coulomb trajectory
calculations of fragments emitted from a spherical source of radius $R_s$ and
lifetime $\tau$.

\vskip 0.2in

The calculations are Monte Carlo simulations of many
body Coulomb trajectory calculations of fragments
emitted from a spherical source of radius $R_s$ and
lifetime $\tau$. The data are equally well described by
calculations using four different combinations of $R_s$ and
$\tau$ as shown in the figure. Even with this ambiguity, the
``valley" exhibited at low $v_{red}$ provides some measure
of the space-time extent of the source. As the energy
increases, the width of this ``valley" increases
suggesting emission from a smaller and may be a faster
source.
In order to get more definite results about the
emission time, information about the source size must
be obtained independently. Such information is most
commonly extracted by comparing predictions with
data and thus is model dependent. For example,
assuming that the source sizes can be obtained from
the linear momentum transferred to the system, one
can obtain the emission time from fragment-fragment
correlation functions. The left panel of fig.7 shows the dependence
of mean emission time as a function of incident energy
for the system Kr+Nb \cite{Bauge93}. Above 55 MeV per
nucleon, multifragmentation seems to occur in a time
scale that saturates at $\approx$125 fm/c. The result is
consistent with breakup of a fragmenting source at low
density including those driven by Coulomb
instabilities as in the Au+Au reaction at E/A=35 MeV
\cite{D'Agostino96}. In the latter experiment, the source
size was obtained by comparing various experimental
observables to the prediction of the statistical
multifragmentation model. Recent analysis of the IMF
correlation functions from high energy hadron induced
multifragmentation suggests the saturation time occurs
at much shorter time scale ($<$100 fm/c) as shown
in the right panel of fig. 7\cite{Beaulieu00}.
Considering the space-time ambiguity and model
dependence in extracting the time information, the
correlation analysis is probably not reliable in
extracting a time scale less than 50 fm/c.

\vskip 0.2in
\epsfxsize=4.5in
\epsfysize=4.5in
\centerline{\rotatebox{90}{\epsffile{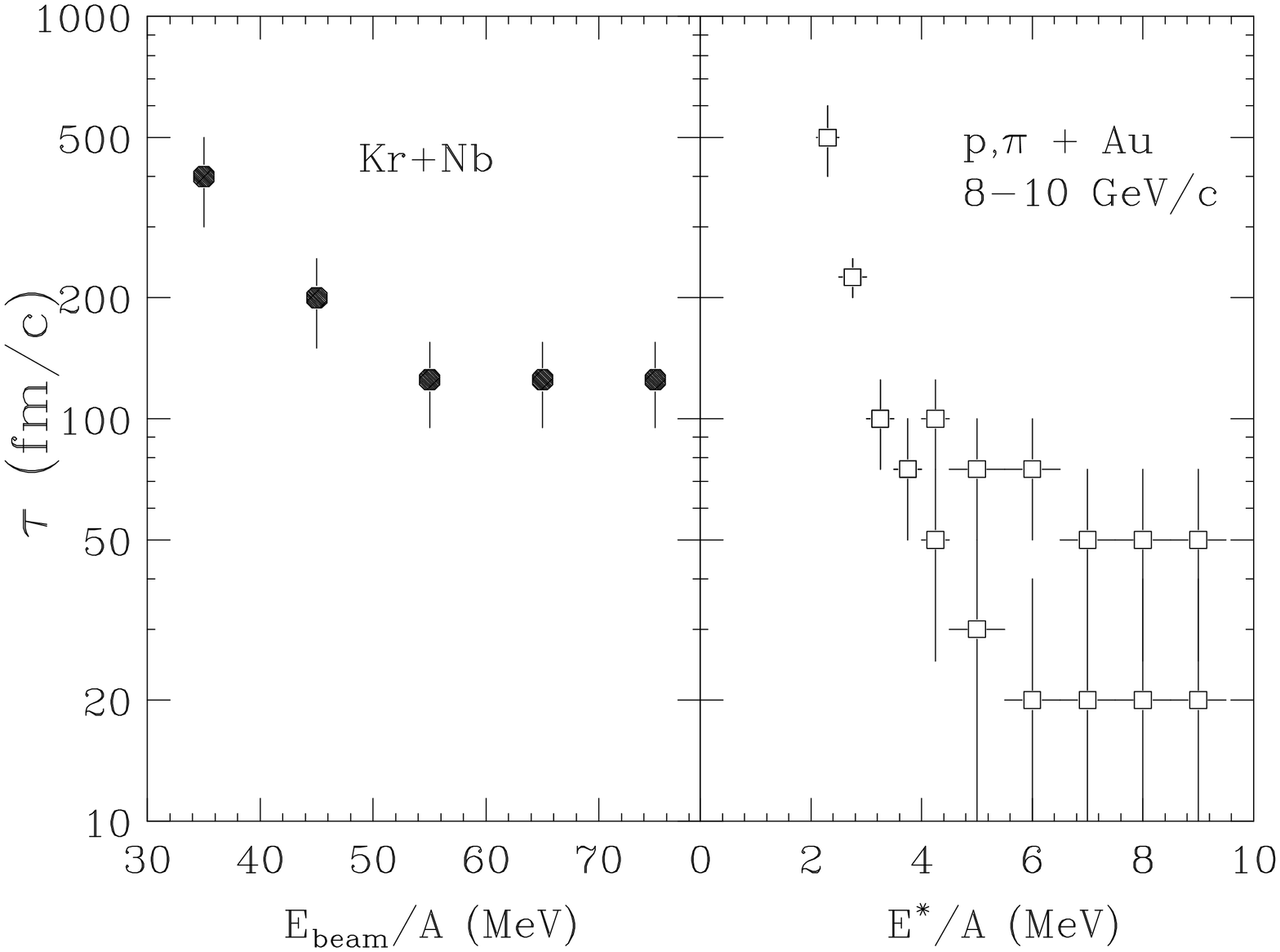}}}
\vskip 0.2in

Fig. 7: Dependence of mean emission time as a function of incident energy
for the system Kr+Nb \cite{Bauge93} (Left panel) 
and as a function of excitation
energy for hadron induced multifragmentation 
(right panel) \cite{Beaulieu00}.

\vskip 0.2in

Without precise time information, 
quantitative measurements of freeze-out densities have
been difficult to obtain, since the density is quite
sensitive to the emission time and volume of the
source. Assuming zero lifetime, the density or source
sizes can be obtained from light charged particle
correlation measurements \cite{Pratt98,Zhu91,Fritz99}.
The left panel of Fig.8 shows the radii extracted for
different reactions using the p-p correlation as a
function of the proton velocity normalized by the beam
velocity. The middle data set with lots of data points
are experimental results from the $^{16}$O and $^{14}$N
induced reaction on Au. The solid diamonds and solid
circles are radii extracted from the $^{40}$Ar induced
reaction on Au and $^3$He induced reaction on Ag,
respectively. The dot-dashed and the dash lines are
scaled from the solid lines by the radii of the projectile.
At high velocity where the protons originate from the
projectile, the scaled predictions agree with the data
very well, suggesting that the method of using p-p
correlations to extract source size information is
consistent within the same method.

\vskip 0.2in
\epsfxsize=5.5in
\epsfysize=4.5in
\centerline{\rotatebox{90}{\epsffile{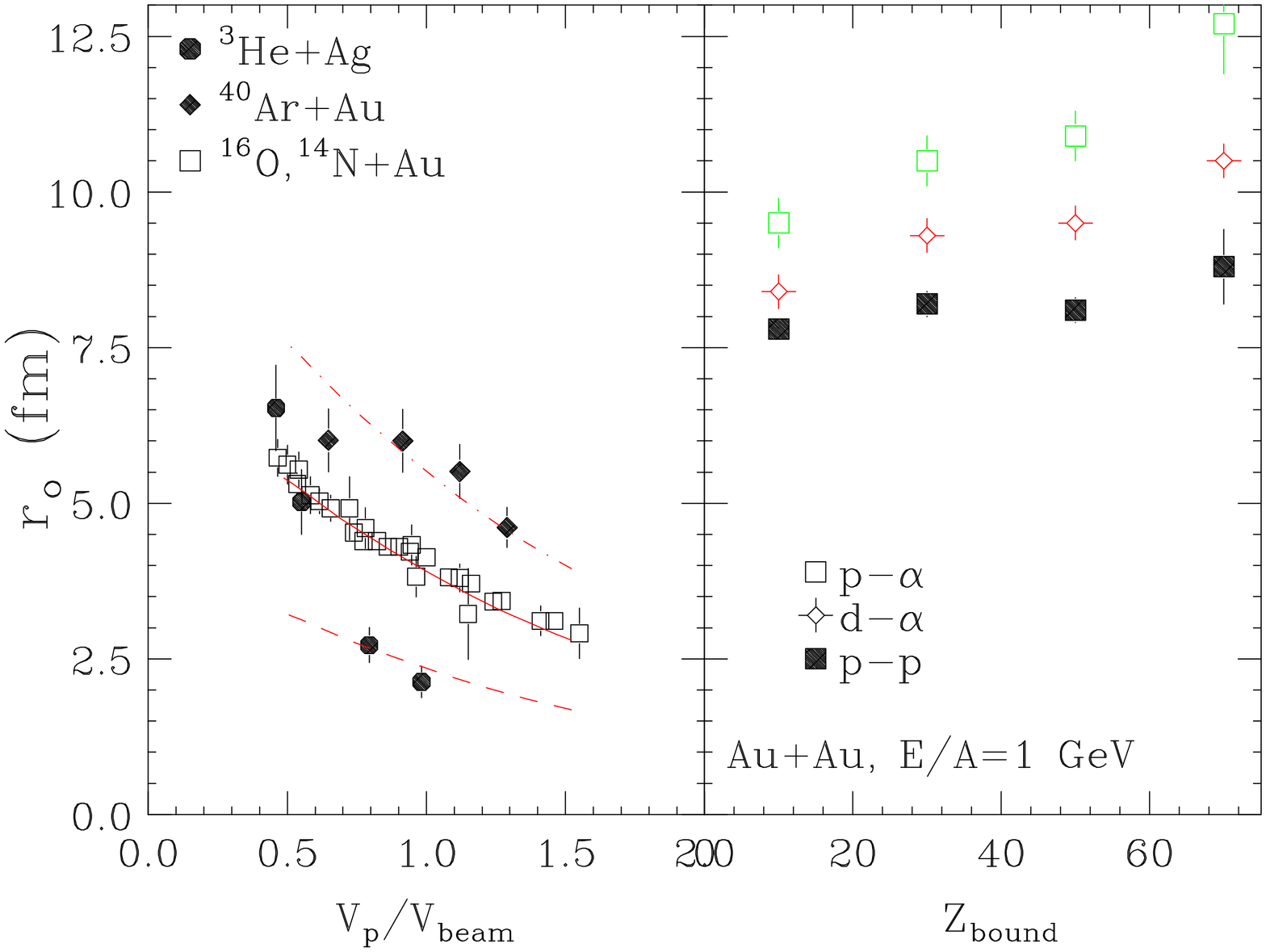}}}
\vskip 0.2in

Fig. 8: This shows the radii extracted for different reactions using the
p-p correlation as a function of the proton velocity normalized by the beam
velocity. The open symbols are radii extracted from 
$^{16}$O and $^{14}$N induced
reactions on Au. The solid line is the interpolation of this set of data.
The solid diamonds and solid circles are radii extracted from the $^{40}$Ar
induced reaction on Au and $^3$He induced reaction on Ag, respectively. The
dot-dashed and the dash lines are scaled from the solid lines with the radii
of the projectile \cite{Zhu91}. 
The right panel shows the radii deduced
from the p-p, p-$\alpha$, d-$\alpha$ correlations as a function of 
$Z_{bound}$ for the
projectile fragmentation of Au+Au at E/A=1000 MeV \cite{Fritz99}.

\vskip 0.2in
The right panel of Fig.8 shows the radii
deduced from the p-p [solid squares], p-$\alpha$ [open stars],
d-$\alpha$ [open squares] correlations as a
function of $Z_{bound}$ for the projectile decay of Au+Au collisions
at E/A=1 GeV.  The radii obtained from
the p-p correlation are smaller than the radii obtained
from p-$\alpha$ correlations which are in turn smaller than the
radii obtained from d-$\alpha$ correlations. There is no
logical explanation for such an observation. The
inconsistencies of the measurements, regarding the
different radius values obtained from different particle
correlations, illustrate the present experimental
difficulties in extracting the precise values of the
freeze-out densities since the analysis is highly model
dependent. Other methods, such as comparing IMF
multiplicities or mean kinetic energies with statistical
models have been employed to determine the source
sizes \cite{D'Agostino96}. Independent of different analysis
methods, densities lower than 1/3 of the normal
nuclear matter density provide the best agreement
with the data. Consistent with such a conjecture, models
that assume normal nuclear matter density
underpredict the fragment multiplicities \cite
{Bowman91,Hubele92,Hagel94,Botvina95}.

Considering the complexities of the issue, the
available experimental data suggest a time scale for the
multifragmentation process of the order of 100 fm/c
and density values of less than a third of the normal
nuclear matter density. Obviously, it is desirable to
have more exact values.  Further experimental work
and better understanding of the reaction mechanisms
in the coming years will allow more precise
measurements in this area.

\section{Temperature Measurements}

\subsection{Kinetic Temperatures}

How valid is the concept of temperature in
heavy ion reactions?  Long before intermediate energy
collisions which are best described by
multifragmenation mechanisms, the concept of
temperature was being used routinely to describe
heavy ion collisions at Bevalac (see for example
\cite{Westfall78}).  In cascade \cite {Cugnon81} or
transport calculations \cite {Bertsch88} one can follow in microscopic
models how the original ordered motion of the beam
gets dispersed into a Maxwell-Boltzmann distribution
through two-body collisions.  In the Purdue
experiment of proton on Xe \cite{Hirsch84}, the high
energy tails of the kinetic energy spectra provide
evidence that the fragments originate from a common
remnant system somewhat lighter than the target
which disassembles simultaneously into a multibody
final system.  Theoretically, the slopes of the particle
kinetic energy spectra assuming a Maxwellian
distribution, should be sensitive to the initial
temperature.
\begin{eqnarray}
\frac{d^2\sigma}{dEd\Omega}=N_0(E-V_c)^{1/2}.\exp(-E_s/T_{kin})
\end{eqnarray}
where
$E_s =E-V_c+E_0-2[E_0(E-V_c)]^{1/2}$cos$(\theta)$
Here $N_0$ is a normalization constant; $E$ and $m$
are the energy and mass of the emitted particle; $\theta$ is the
detection angle relative to the incident beam; 
$T_{kin}$ is the slope or kinetic temperature;
$E_0=mv_0^2/2$; and $V_c$ corrects for the Coulomb repulsion
from the target residue. As discussed earlier, collective
motions complicate the energy spectra. Furthermore,
fluctuations in Coulomb barriers, sequential feedings
from higher-lying states \cite{Nayak92}, Fermi motion
\cite{Bauer95} and pre-equilibrium emissions all
contribute to the complications associated with
extracting emission temperatures from the energy
spectra.

\subsection{Excited state temperature}

To circumvent some of these problems, other
thermometers, which are less sensitive to the collective
motion, have been sought. Thermometers based upon
the relative populations of excited states of emitted
light particles have been used quite extensively in
extracting the temperatures of the hot nuclear systems.
\begin{eqnarray}
T=\frac{E_1-E_2}{ln(a'Y_1/Y_2)}
\end{eqnarray}
Here $a' =(2J_2 + 1)/(2J_1 + 1), E_i$ is the excitation
energy, $Y_i$  is the measured yield and $J_i$ is the spin 
of the state $i$. To minimize the influence of sequential
decays, nuclei with levels that are widely separated are
often chosen \cite{Schwarz93}. One of the most commonly
used nuclei is the unstable $^5$Li isotope. The ground
state $3/2^+$ decays to p and $\alpha$ particles while the
16.66 MeV $3/2^+$ excited state decays into d and
$^3$He particles. Statistical models incorporating the
effect of sequential decays suggest that temperatures up
to 6 MeV should be obtainable with this nucleus based
on the excited states population \cite{Xi99}. Other nuclei
include $\alpha$ particles, $^{10}$B and $^8$Be which all have
relatively widely separated states. Even though the
ground state and first excited state of the alpha
particle are separated by 20.1 MeV, a substantial part of
the measured ground state alpha yields can be
attributed to sequential decays from heavy nuclei due
to the unusual binding energy of the alpha particles.
The consistency of the method is normally checked by
measuring the temperatures of several nuclei.

\subsection{Isotope temperature}

Another thermometer, $T_{iso}$, which utilizes the
yield ratios of two pairs of isotopes have been under
intense study in the past few years \cite
{Albergo85,Pochodzalla85}. If chemical and thermal equilibrium
are achieved, in the limit of the Grand Canonical
Ensemble, one can obtain the isotope temperature
information from a double isotope ratio defined by
\begin{eqnarray}
T=\frac{B}{ln(a(Y_1/Y_2)/(Y_3/Y_4))}
\end{eqnarray}
where $Y_1, Y_2$ are the yields of one isotope pair
and $Y_3, Y_4$  is another isotope pair. To cancel the nucleon
chemical potential terms, the mass number
differences of isotope pair (1,2) must be the same as
the mass number differences of isotope pair (3,4); $B$ is
the binding energy difference, $B=BE_1+BE_2-(BE_3+BE_4); a$
contains the statistical weighting factor. This equation
assumes that the sequential decay corrections to the
yields are negligible. This assumption is rather
problematic as the experimental measured yields are
``cold" fragments containing contributions from the
decays of many excited nuclei. 
In experiments where a large number of isotope
yields are measured such as the proton induced
reaction on Xe \cite{Hirsch84}, thousands of $T_{iso}$ values can
be extracted \cite{Tsang97} using Eq.(7.3). If Eq.(7.3) is correct,
all the values of $T_{iso}$ thus obtained should be the same.
However, the experimental values fluctuate over a
large range of $T_{iso}$ values, including negative numbers.
These fluctuations arise from sequential decays and
can be minimized by selecting double ratios with large
binding energy differences $(B>10$ MeV). However,
such requirements select mainly isotope pairs that
involve proton rich isotopes such as $^3$He or $^{11}$C which
are not well bound. Thus instead of having many
independent thermometers, there are in general two
classes of isotopes thermometers, those involving the
($^3$He, $^4$He) pair and those using ($^{11}$C, $^{12}$C) pair. The fact
that $^3$He and to a lesser extent $^{11}$C have been found to
exhibit anomalous energy spectra may invalidate the
simple relationship of Eq.(7.3).

The difference in shape of the energy spectra
between $^3$He and $^4$He means that the isotope
temperature depends on experimental energy thresholds. It has been
shown that the temperature depends strongly on the
energy gates used \cite {Tsang96,Xi98c,Viola99,Hauger00}. This dependence
has been exploited to examine the evaporative cooling
of the Xe on Cu collisions at E/A=30 MeV \cite{Xi98c}. Since
energy thresholds are often employed to minimize the
contributions of the pre-equilibrium emissions \cite {Viola99,Hauger00},
this directly affects the temperature values
measured.

\subsection{Effect of Sequential decays}

In recent years, many models have been
developed to describe the emission of particles from the
multifragmentation process successfully. However, to
compare with experimental data, these models must take into
account the effects of sequential decays. Inclusion of nuclear
spectral information into the calculations to simulate the
effects of secondary decay has not been fully successful
because the task is not only computationally difficult, but it is
hampered by the lack of complete information about the
resonances in many nuclei.
Fig.9 shows the effect of sequential feedings
on the temperature determination using two
assumptions about the unstable states \cite{Xi98}. The horizontal axis is
the emission temperature used in the statistical
calculations \cite{Xi99} while the vertical axis is the
apparent temperatures obtained using different classes
of thermometers. In the left panel, sequential decay
calculations including only known bound states and
resonances are shown for T$_{iso}$(He-Li) and T$_{\Delta E}$($^4$He),
denoted by the solid and dashed lines respectively.
The apparent isotope temperatures increase
monotonically with the input temperature. With
inclusion of continuum states (right panel), both
temperatures flatten out at an asymptotic value of
about 6 MeV. Thus, inclusion of sequential decay
contributions from the continuum enhances decays to
low-lying states and renders temperatures involving
alpha particles insensitive to the emission temperature
at high excitation energy. For comparison, the
calculated excited state temperature of $^5$Li (dot-dashed
line) is plotted in the right panel: T$_{\Delta E}(^5$Li) continues to
increase monotonically beyond 6 MeV emission
temperature. The rate of increase becomes much less
only after 9 MeV temperature.

\vskip 0.2in
\epsfxsize=4in
\epsfysize=4in
\centerline{\rotatebox{90}{\epsffile{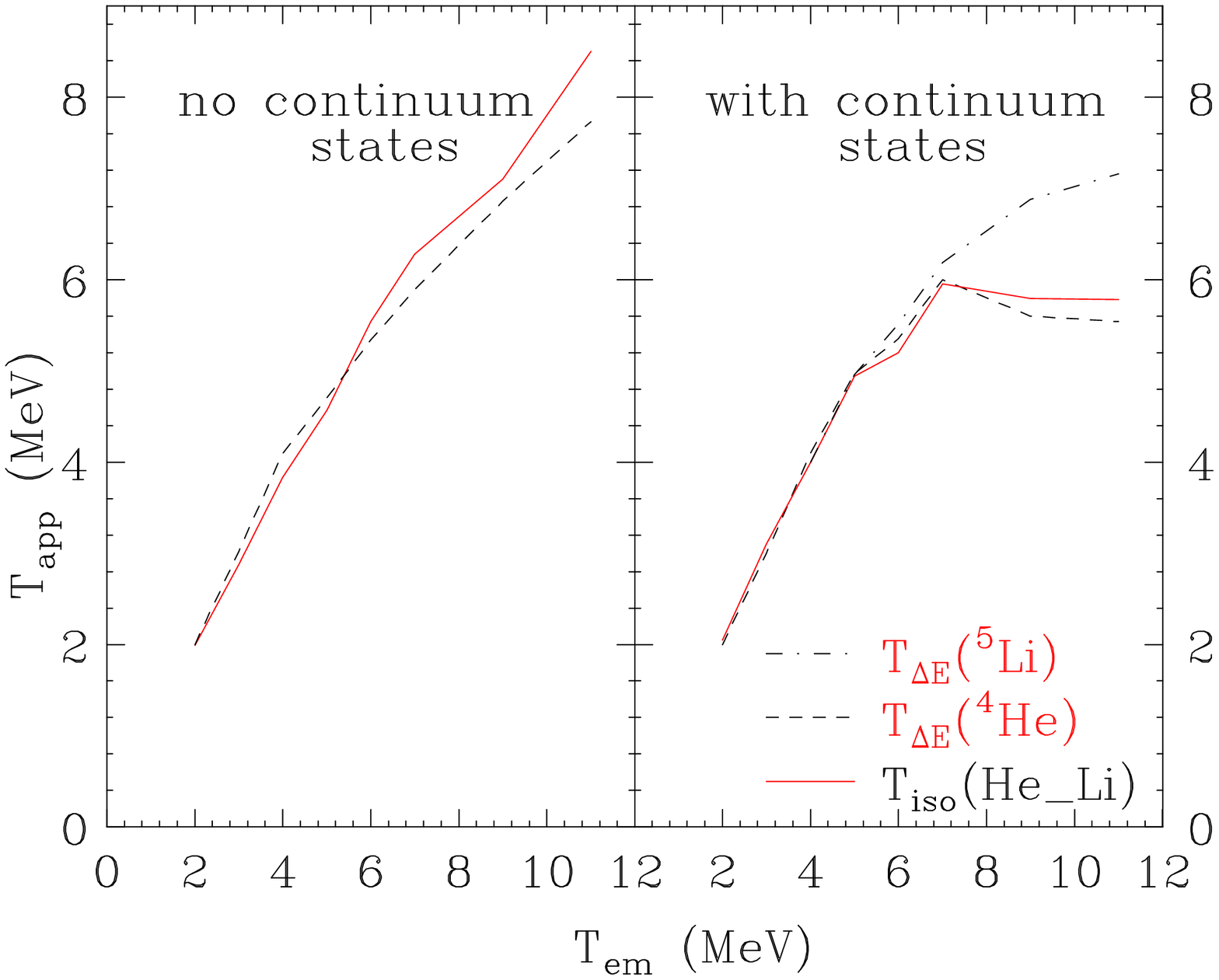}}}
\vskip 0.2in

Fig. 9: Effect of sequential feedings on the temperatures determination
using two assumptions on the unstable states \cite{Xi98}. 
In the left panel,
sequential decay calculations including only known bound states and
resonances are shown for $T_{iso}$(He-Li) and T$_{\Delta E}(^4$He), 
denoted by the solid and
dashed lines repectively. With inclusion of continuum states (right panel),
both temperatures flatten out at an asymptotic value of about 6 MeV while
the calculated excited state temperature of $^5$Li (dot-dashed line) 
continues
to increase monotonically until 9 MeV before flattening out \cite{Xi98}.

\vskip 0.2in

Of course, the dependence of the apparent
temperature on the emission temperature is model
dependent. When the sequential decay effect is large
such as at high temperature, reaction models with
accurate description of the sequential decay processes
are needed to relate the measured temperature to the
emission temperature. Efforts have been made to
include structural information to describe secondary
decays. However, even with the best code available,
the disagreement between measured and predicted
isotope yields could be a factor of 10 or more especially
for the neutron rich or proton-rich isotopes \cite{Souza00}.

\subsection{Cross-comparisons between thermometers}

Cross-comparisons between different
thermometers exist. In the case of kinetic temperature
measurement, the slope of the energy spectra
measured at backward angles and at low incident
energy give reliable temperature information for
systems with low excitation energy. Under these
conditions, the collective flow effect and
pre-equilibrium contributions are minimized.
At higher energy, the kinetic temperatures are
not reliable but one can cross-compare the isotope ratio
and excited state temperatures \cite
{Tsang96,Huang97,Serfling98,Durand98,Xi98b}. Careful measurements of
these temperatures in various systems suggest that
below E/A=35 MeV, there are good agreements
between T$_{\Delta E}$ and T$_{iso}$ \cite {Huang97}. However, the
disagreement increases with incident energy \cite
{Serfling98,Xi98b} as shown in Fig.10. For the system of
Kr+Nb, temperatures obtained from excited state
populations and isotope yields have been measured as
a function of the incident energy \cite{Xi98b}.
The open symbols represent the temperatures
extracted from the excited state populations of $^5$Li, $^4$He
and $^8$Be respectively. Within experimental
uncertainties, they are the same. The consistencies of
the experimental results from different nuclei render
credence to this thermometer. The closed symbols
represent temperatures extracted from isotope yield
ratios; T$_{iso}$(HeLi), (closed squares) rely on the double
ratio [Y($^6$Li)/Y($^7$Li)] /[Y($^3$He)/Y($^4$He)] while T$_{iso}$(C-Li),
(closed circles) use the double ratio [Y($^6$Li)/Y($^7$Li)]
/[Y($^{11}$C)/Y($^{12}$C)]. Values for T$_{iso}$(C-Li) vary little with
incident energy, similar to the trends exhibited by the excited states
temperatures of $^5$Li,  $^4$He,  and  $^8$Be. In
contrast, values of T$_{iso}$(He-Li) (closed squares) increase
monotonically with incident or excitation energy.
Similar discrepancies between T$_{\Delta E}$ and T$_{iso}$ have been
observed in Au + Au central collision from E/A=50 to
200 MeV \cite{Serfling98}, Ar+Sc reaction from E/A=50 to
150 MeV \cite{Xi00} and in Ar+Ni system \cite{Durand98} at
E/A=95 MeV.

\vskip 0.2in
\epsfxsize=4in
\epsfysize=4in
\centerline{\rotatebox{90}{\epsffile{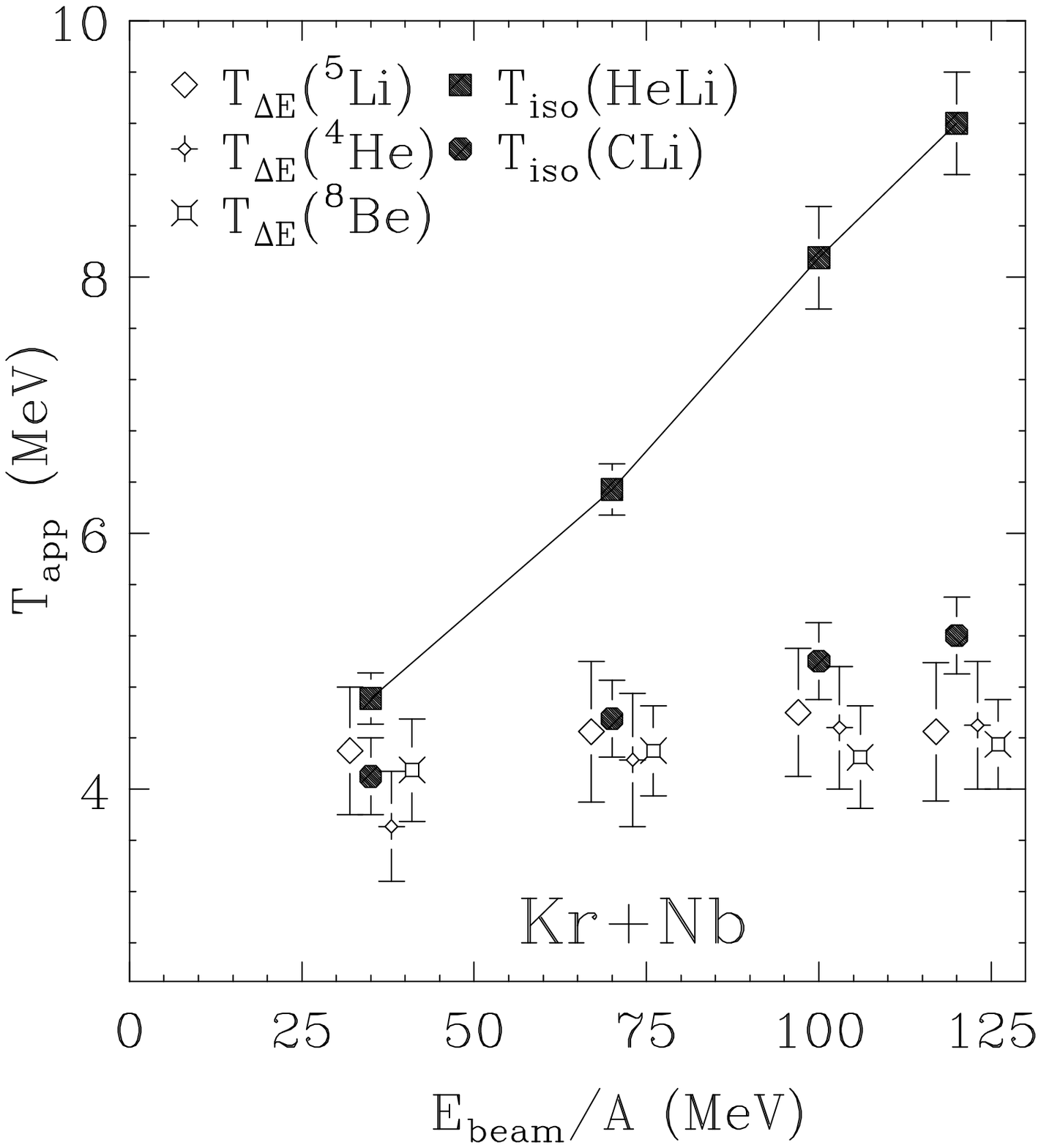}}}
\vskip 0.2in

Fig. 10: Excitation function of different thermometers for the system Kr+Nb.
The solid symbols are isotope ratio temperatures while the open symbols are
temperatures extracted from excited states populations \cite{Xi98b}.

\vskip 0.2in

Independent of models describing sequential
decays, thermometers using alpha particles (e.g. 
$T_{\Delta E}(^4$He),$T_{iso}$(He-Li)) should be
affected by sequential decays in the same manner and
should give the same experimental temperature.
However, current data \cite{Xi99,Serfling98,Durand98}
show that there are substantial differences between
these two thermometers when the excitation energy or
incident energy increases. This may indicate that
different reaction mechanisms may be involved in the
production of primary $^3$He and $^4$He
particles. In such case, isotope yield temperatures
constructed from Eq.(7.3) are problematic.

\subsection{Summary of temperature measurements}

Fig.11 shows an overall picture provided by
the present data using different thermometers. The
kinetic temperatures extracted from fitting the charged
particle energy spectra with an intermediate rapidity
source exhibit a smooth trend over a wide range of
incident energies from a few MeV to nearly 1GeV per
nucleons \cite{Chitwood86}. The open diamond points
shown in Fig.11 are the proton kinetic temperatures
extracted from \cite {Chitwood86} over a narrow range
of incident energies for comparison purposes. The
dashed lines are drawn to guide the eye. The
temperature values depend slightly on the particle
types.  However, the other light charged particles, d and
t, exhibit similar trends, namely, the kinetic energy
temperature increases rapidly with the incident
energy.  A collection of the T$_{\Delta E}(^5$Li) over a range of
incident energies from 30 to 200 AMeV are plotted as
solid points in Fig.11 \cite{Schwarz93,Serfling98,Xi98b}.
Contrary to the kinetic temperature, there is only a
slight increase from 3 to 5.5 MeV, in the excited state
temperature as a function of the incident energy
spanning over one order of magnitude.
The open circles in Fig.11 represent the most
commonly used isotope ratio temperature, T$_{iso}$(He-Li),
\cite{Serfling98,Xi98b,Xi00} as a function of incident
energy from 35 to 200 AMeV. The increase from
4 to 10 MeV as a function of incident energy is much
less than T$_{kin}$ but the increase is larger than the nearly
constant value of T$_{\Delta E}(^5$Li). T$_{iso}$(C-Li) are plotted as
open diamonds in Fig.11. This latter isotope thermometer
does not agree with T$_{iso}$(He-Li). Instead, T$_{iso}$(C-Li)
remain relatively constant over the incident energy
studied. They behave more like the excited state
temperature.

\vskip 0.2in
\epsfxsize=5in
\epsfysize=4in
\centerline{\rotatebox{90}{\epsffile{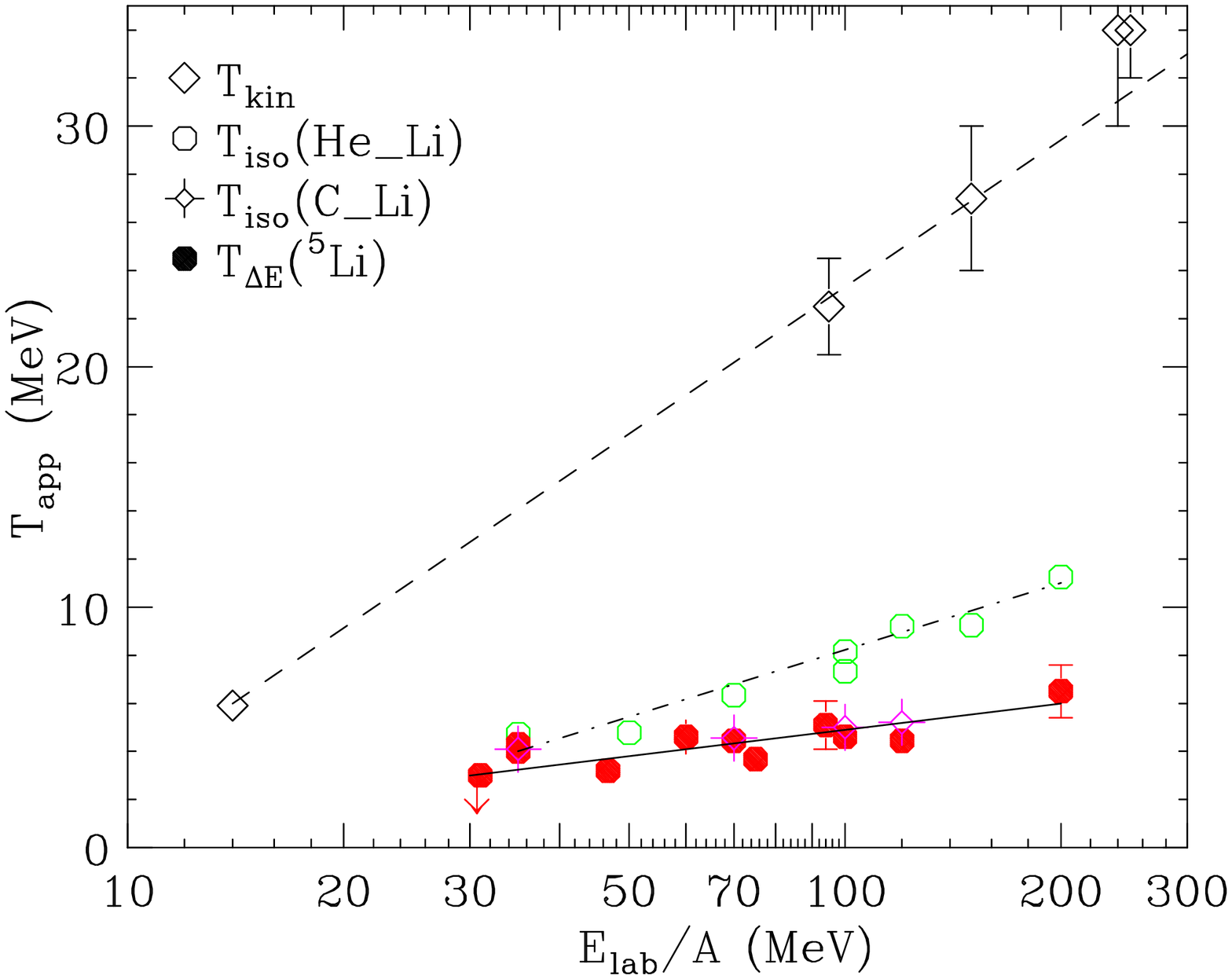}}}
\vskip 0.2in

Fig. 11: Excitation function of proton kinetic temperature, (open diamonds
with dashed lines drawn through the data), excited state 
temperature of $^5$Li
(solid circles and solid line), isotope ratio temperature of 
$^{3,4}$He and $^{6,7}$Li
(open circles with dot-dashed lines) \cite{Gelbke00}

\vskip 0.2in

Experimentally, temperatures extracted from
excited states or yield ratios involving carbon isotopes
are around 4-5 MeV\cite
{Chen87,Schwarz93,Xi98b,Souza00,Tsang97}. Around 4 MeV emission
temperatures, the secondary decay effects are small
and can be corrected with current models
incorporating sequential decays. The near constant
temperature may signal enhanced specific heat.
However, if the low temperature of 4 MeV is caused by
the limiting temperature due to sequential decays, it
becomes difficult to deduce the freeze-out temperature
from the measured quantities.

\section{Excitation Energy Determination}

In nuclear physics experiments, the collision
conditions are reconstructed from the particles
detected. Even if all the emitted particles can be
measured experimentally, it is still difficult to
disentangle the contributions from various emitting
sources arising from the spectator and participant
zones. Before multifragmentation occurs, the hot
systems first de-excite by emitting neutrons and light
charged particles (including very light IMF's). These
particles are normally emitted in the forward
directions and in a very short time scale ($<$30 fm/c)
before equilibrium is established. It is important that
these pre-equilibrium particles are not included in the
determination of excitation energy characterizing the
fragmenting source.

Experimentally the contamination of the
observables used in characterizing the emitting source
is very difficult to assess without the use of model
assumptions. Cascade and transport calculations 
can be used to estimate the number of particles emitted
and the energy lost in the ``prompt" or early stage of
the reaction. Such calculations may suggest some
optimum ways of estimating the number
such as imposing energy thresholds
on the data to minimize the ``pre-equilibrium"
contributions \cite{Beaulieu00,Hauger98}. However, the
prompt contributions cannot be completely eliminated
from the data using the energy threshold gates.  This
increases the uncertainties and fluctuations in the
excitation energy determined. The model calculations
may also indicate the size, mass, N/Z ratio and energy
of the residues which undergo ``multifragmentation".
Assume we can detect and identify all the particles
emitted from this excited source, conservation of
energy suggests that
\begin{eqnarray}
E^*=\sum E_i + \sum E_n + \sum E_{\gamma} + Q
\end{eqnarray}
Where $E^*$ is the excitation energy, $E_i$ is the
kinetic energy of the charged particles, $E_n$ is the energy
of the neutrons, $E_{\gamma}$ is the energy of the gamma
rays emitted during the de-excitation and $Q$ is the
mass difference between the parent nucleus and all the
emitted particles. Gamma energies are relatively small
and contribute little to the total excitation energy
compared to the other terms. 

In reality, most
experimental apparatus does not have a complete 4$\pi$
coverage. Furthermore, thresholds in energy and
geometry exist in the detection arrays. Thus there are
uncertainties in determining the terms $\sum E_i$ and
$Q$. More importantly, neutrons are often not measured.
Neutrons do not interact with matter as much as
the charged particles so they are more difficult to
detect. Very often, $\sum E_n$ is estimated using the
average number of neutrons emitted and the mean
neutron energy, $\sum E_n =N_n<E_n>$ \cite
{Beaulieu00,Poch,Hauger98}. Conservation of particles
imposes some constraints on the value of neutron multiplicity $N_n$. 
As the neutron data is difficult to obtain, 
the mean neutron energy $<E_n>$ values are
usually adopted from other experiments or assumed to
be the same as the proton mean energy. Therefore in general,
$\sum E_n$ poses the largest uncertainty to Eq.(8.1) and
determination of excitation energy of the fragmenting
system becomes quite a difficult task.

Recently, intense effort has been placed in
extracting the excitation energy of heavy nuclei
induced by high-energy hadron beams such as
protons, pions and anti-protons at high energy
\cite{Beaulieu99}. In these reactions, the collective
excitation and existence of multiple sources are
minimal. Even with such "simplified" systems, it is
difficult to extract precise excitation energy without
the use of model assumptions. For heavy ion reactions,
the task is much more daunting. In addition, all
collective motions strongly affect the signals of the
phase-transition and must be understood before exact
values of the excitation energy can be assigned.
With so many uncertainties associated with
extracting the excitation energy, any experimental
observables that utilize the fluctuations of excitation
energy such as measuring the heat capacities are
subjected to the same problems \cite{D'Agostino00}. The
results must be viewed cautiously. While more work is
needed to determine the excitation energy accurately,
one consensus is that increasing incident energy
corresponds to increasing excitation energy.

It is quite common to extract the excitation
enegy using projectile fragmentation. Assuming that
the projectile is only modestly excited, only one source,
the projectile-like residue, exists. Then the excitation
energy of the projectile-like residue should be
inversely proportional to impact parameter. Thus it
has been argued that $Z_{bound}$, rather directly,
reflects the energy transfer to the excited spectator
system \cite{Poch}.  Larger energy transfers, then,
correspond to smaller value of $Z_{bound}$ and vice
versa. However, geometrical arguments suggest that
the source size also varies with impact parameter
\cite{Natowitz95}.  Furthermore, the energy spectra of light
charged particles are inconsistent with one source but
require multiple sources to fit the spectra \cite{Xi98b}. 
It is thus incorrect to assume that there are no
contributions from pre-equilibrium emissions.  As
discussed previously, collective flow also plays a role
in projectile fragmentation.  Besides, all the uncertainties
associated with Eq.(8.1) as described above apply to
these reactions.

Currently determination of the excitation
energy presents the biggest challenge to the
experimenters. With care and the increasing
availability of large neutron detection devices, the
excitation energy measurements will be improved in
the coming years.  A firm grip of this parameter is very
important in our understanding of the liquid-gas phase-transition.

\section{Signals for Liquid-Gas Phase Transitions}

Over the years, many experimental observables
involving IMF have been used to study the nuclear
liquid-gas phase transition. Discussions of all the
proposed experimental signatures will require too
much time and space. Instead, we will focus our
discussions on four experimental signatures, which
have attracted most attention in the past years.

\subsection{Rise and Fall of IMF}

Copious emission of intermediate mass
fragments is one predicted consequence of the 
liquid-gas phase transition of nuclear matter, both by
statistical models and transport models. At low
excitation energy, few fragments are ``evaporated"
from the liquid while at very high excitation energy,
the liquid ``vaporizes" to produce a nucleon gas. The
``rise and fall" of IMF multiplicities has been observed
in both central and peripheral collisions. For central
collisions, maximum fragment productions occur
around incident energy of 100A MeV as shown in the
left panel of Fig. 12 for the Kr+Au reaction
\cite{Peaslee94}.
At incident energy above 400A MeV, production of
IMF shifts from central to more peripheral collisions
\cite{Ogilvie91,Tsang93}. The right panel of Fig.12
shows the impact parameter dependence (obtained by
measuring $Z_{bound}$ for the fragmentation of Au
projectiles at incident energy from 400 MeV to 1 GeV
\cite{Ogilvie91}. In both panels of Fig.12, fragment
multiplicities increase to a maximum with increasing
excitation energy. The fragment production then
declines and ``vaporizes" completely into nucleons and
light particles.

\vskip 0.2in
\epsfxsize=6in
\epsfysize=4.5in
\centerline{\epsffile{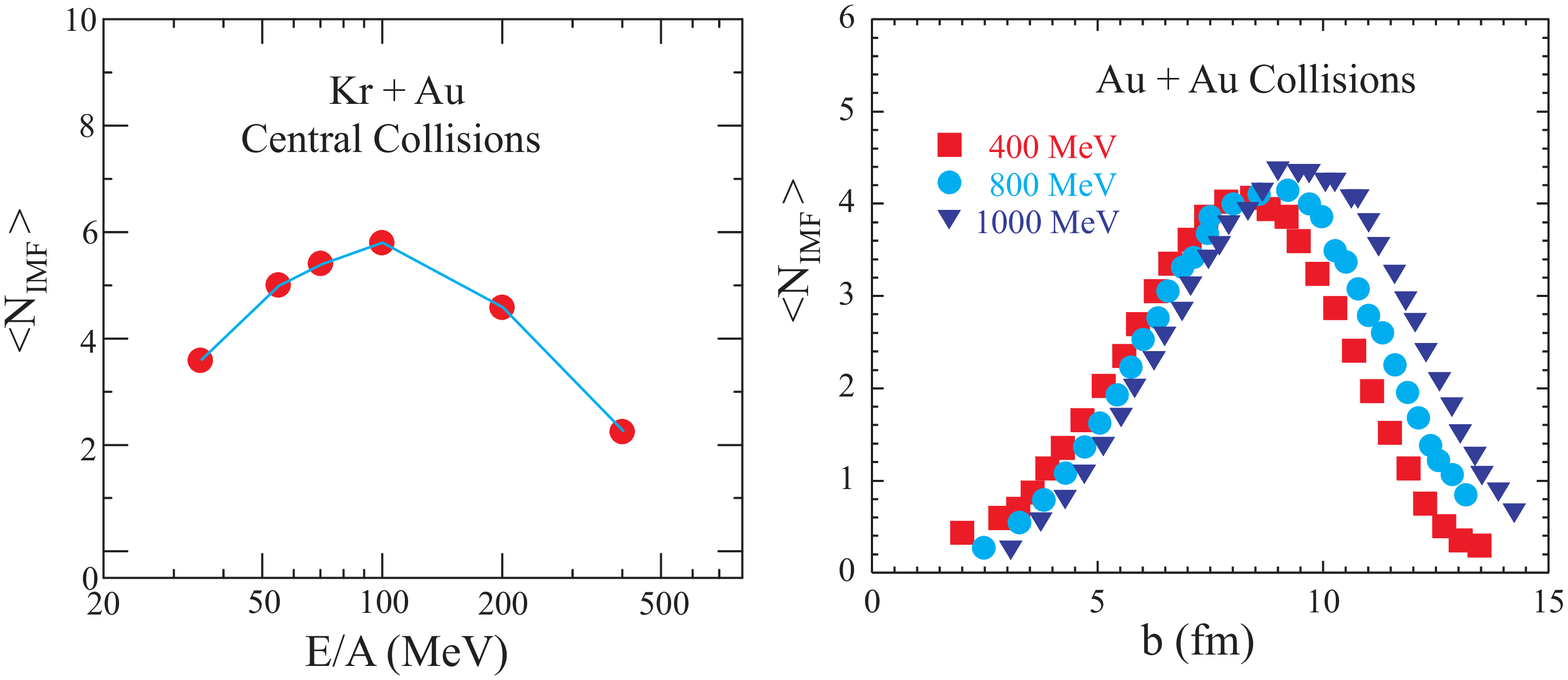}}
\vskip 0.2in

Fig. 12: Rise and fall of IMF multiplicity as a function of incident energy
for the central collisions of Kr + Au (left panel) 
\cite{Peaslee94} and as a
function of impact parameters for the projectile multifragmentation of Au+Au
reactions (right panel) \cite{Ogilvie91}.

\vskip 0.2in

\subsection{Critical Exponents}

The observation by the Purdue group \cite
{Finn82} that the yields of the fragments produced in
p+Xe and p+Kr obeyed a power law $Y(A_f)\propto
A_f^{-\tau}$ led to a  conjecture that the fragmenting
target was near the critical point of liquid-gas phase
transition.  The origin of this conjecture is the Fisher
model \cite {Fisher} which predicts that at the critical
point the yields of droplets will be given by a power
law.  The power law has since then  been established
very firmly in collisions between heavy ions \cite
{Ogilvie91} with the value of the exponent $\tau$ being close to 2.
But the power law is no longer taken as the `proof' of
criticality. There are many systems that exhibit this sort
of power law: mass distributions of asteroids in the
solar system, debris from the crushing of basalt pellets
\cite {Hufner86} and the fragmentation of frozen
potatoes \cite {Oddershede}.  In fact, the lattice gas
model which has been used a great deal for
calculations of phase transitions and
multifragmentation in nuclei \cite {Pan195,Pan295}
gives a power law at the critical point, on the
coexistence curve (that is a first order phase transition
provided the freeze-out density is less than the critical
density) and also along a line in the $T-\rho$ plane
away from the coexistence curve.  Nonetheless, the
occurrence of a power law is an experimental fact and
it is therefore desirable that models which aim to
describe multifragmentation produce a power law,
phase transition or not.

Even if we expect to see a phase transition in
nuclear collisions it is unlikely that the system
dissociates at the critical point.  Much of the literature
in intermediate energy heavy-ion collisions assumes
that when one is seeing a phase transition one is actually
seeing critical phenomena \cite{Gilkes94,Bauer95b}.  To
reach the critical point one has to hit the right
temperature and the right density.  While one may be
able to hit the right temperature by varying the beam
energy, one has no control over the freeze-out density.
Thus it is unlikely that the dissociation takes place at
the critical point.  We think that this strong emphasis
on critical phenomenon rather than first order phase
transition in nuclear multifragmentation came about
for several reasons.  Firstly, the experimentally
observed power law was interpreted in terms of
critical phenomena.  Secondly, a bond percolation
model \cite
{Campi84,Bauer84,Campi88,Bauer88,Stauffer92} was
among the first to be applied to
multifragmentation in nuclear collisions.  This model
has only a continuous phase transition.  The bond
percolation model can be demonstrated to be a special
case of a lattice gas model \cite{Dasgupta97} which is
more versatile and has both first order phase transition
and critical phenomena.

An excellent review of the early history of this
topic exists \cite {Csernai}. This covered the period
to the end of 1984. More recently, the study of the
liquid-gas phase transition in nuclear matter focuses
more on measuring the thermodynamical properties,
such as the temperature and densities, of the
disassembling system.

\subsection{Nuclear Caloric Curve}

Experimentally, production of particles in
multifragmentation appears to be dominated by their
phase space \cite{D'Agostino96,Botvina95,Moretto95}.
Thus, one should be able to measure temperature and
densities, basic quantities in statistical physics. If the
liquid-gas phase transition is of first order, one would
expect to see enhanced specific heat $dE^*/dT$
corresponding to a plateau region in the caloric curve
defined as temperature, $T$ vs. heat or excitation
energy $E^*/A$. Aside from the experimental
difficulties associated with measuring both quantities
as discussed in sections 7 and 8,
the simple caloric curve of temperature vs.
excitation energy with a plateau in the temperature
assumes that the pressure is constant \cite{Moretto96}.
There is no experimental evidence that such a condition
is met in nuclear collisions.  Thus even without
introducing the isospin degree of freedom, the caloric
curves depend on three variables, pressure, volume and
temperature. Such complicated, three-dimensional
nuclear caloric curves have been recently calculated 
\cite {Choprivate}. Different shapes of the caloric curves
have been obtained depending on the conditions of the
experiments and analysis. Therefore, one-dimensional
caloric curves are useful only if the exact conditions can
be determined or modeled. By themselves, these
curves can be misleading and definitely do not
constitute a signature for the liquid-gas phase
transition even though the idea is very attractive.

One of the purposes of this review article is to
examine the experimental efforts in extracting the
liquid-gas phase transition signals. Since many 
caloric curves have been measured since 1995, we 
will discuss the experimentally obtained curves
keeping the above ``warnings" in mind.

If the incident energy is assumed to be related to
the excitation energy, (this is particularly true for
central collisions), then Fig.11, which is a plot of
temperature versus incident energy from E/A=30 to
200 MeV, is one form of caloric curve. It shows that the
trend depends highly on the specific thermometers
chosen to measure the temperature; the kinetic
temperature increases rapidly (from 12 to 30 MeV); the
excited states temperatures are nearly constant (from 3
to 7 MeV); the isotope temperatures involving He isotopes 
increase moderately (from 4 to 10 MeV) but isotope temperatures
involving $^{11}$C,$^{12}$C
stay nearly constant at 4 MeV. Fig.11 sums up the
most serious experimental problems we are faced with
i.e. the discrepancies in the temperature
measurements. However, this figure also shows that
the excitation functions of the temperature
measurement exhibit a smooth behavior within each
class of thermometer. Moreover, the trends are
consistent from experiment to experiment since the
data shown in Fig.11 come from many different sources.

\vskip 0.2in
\epsfxsize=5in
\epsfysize=4.5in
\centerline{\rotatebox{90}{\epsffile{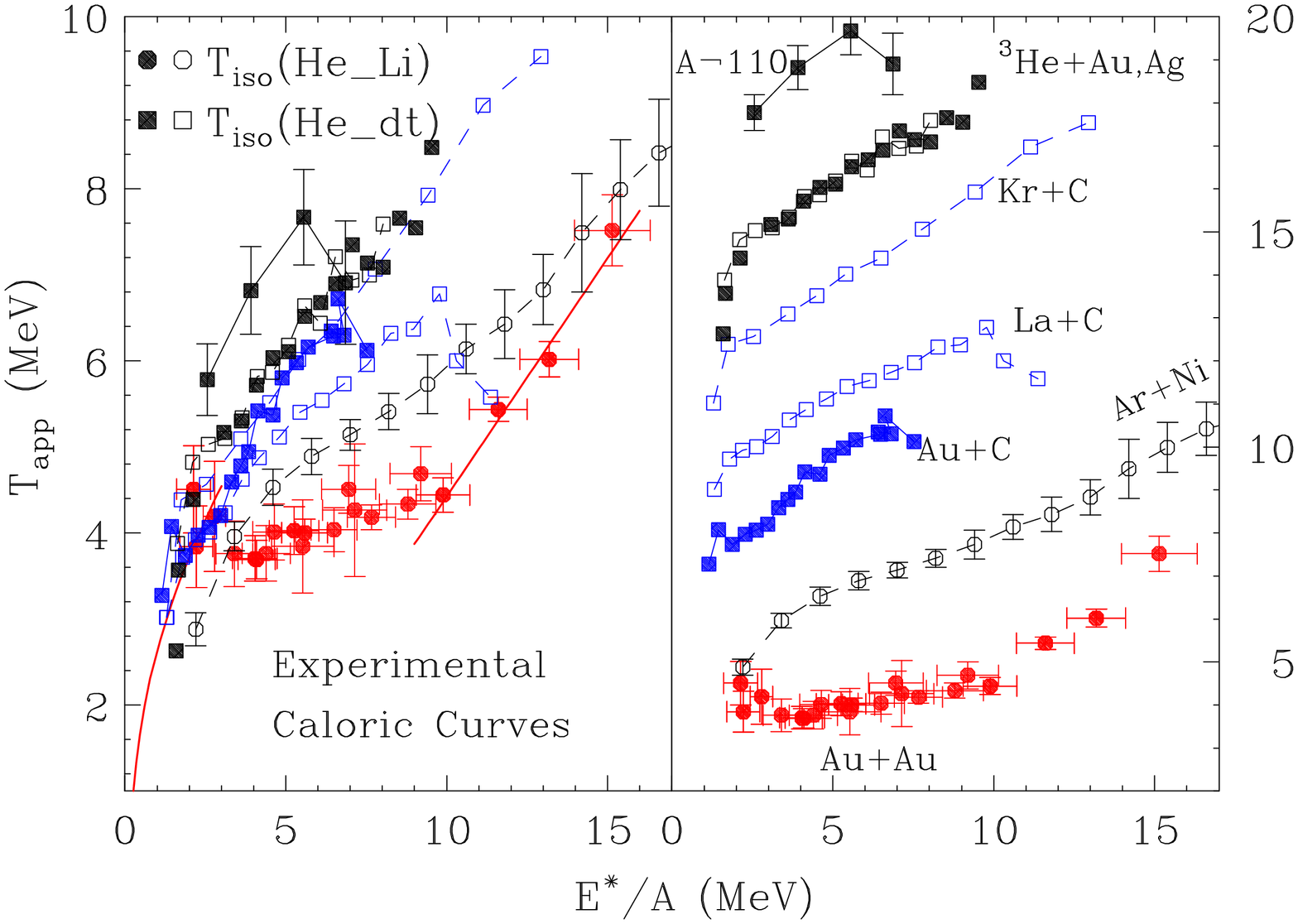}}}
\vskip 0.2in

Fig. 13: Summary of caloric curves measured. The curves in the right panel
have been offset by 2 MeV for each successive curve, starting with
Ar+Ni.

\vskip 0.2in

The more traditional caloric curves which plott
temperatures versus the extracted excitation energy are
shown in Fig.13. In the left hand panel, all the
curves are plotted on the same scale. The temperatures
obtained have been extracted using the isotope yield
ratios; T$_{iso}$(He-Li) are denoted by circles and T$_{iso}$(He-dt)
are represented by the squares. To avoid confusions,
all the temperatures plotted are the experimental
apparent temperatures since sequential decay
corrections are highly model dependent. Sequential
decays account for part of the differences between
T$_{iso}$(He-Li) and T$_{iso}$(He-dt) and there are empirical
correction factors to reduce such differences.
However, in this plot, the differences between the
curves constructed with T$_{iso}$(He-Li) or T$_{iso}$(He-dt) are
much larger than the correction factors. Thus, to keep
the discussion simple, only the reported raw data are
shown.

In order to view each successive curve better, they
are re-plotted in the right panel with a scale
compressed by a factor of 2. Each curve is offset from
from its predecessor by 2 MeV and the corresponding reactions
are labelled close to the curves. The most interesting 
curve is the one labeled ``Au+Au" plotted at the
bottom of both the right and left panels. It was
obtained from the spectator decays of the Au+Au
reaction at E/A=600 MeV \cite{Poch}. T$_{iso}$(He-Li),
remains relatively constant as a function of deduced
excitation energy, E*/A, between 3 and 10 MeV but
increases rapidly at E*/A greater than 10 MeV.  The
resemblance to the first order phase transition of liquid
raises a lot of excitement in the field.  It also resembles
the prediction from the statistical model 
\cite{Bondorf95}.

Unlike the predicted caloric curves from realistic models 
which will be discussed in more details in Section XII or 
the temperature excitation functions of Figure 11, the 
experimental caloric curves depend strongly on the reaction systems
and analysis. (Due to the effect of sequential decay, caloric
curves determined from the experimentally extracted isotope 
yields may not resemble the curves of the deduced primary
fragments \cite{Hauger00}.) On the other hand,
if one ignores the highest excitation data point
in the Au+Au system, all caloric curves exhibit a
smooth increase of temperature with excitation energy.
This trend is very similar to the increase of T$_{iso}$(He-Li)
as a function of incident energy as shown in fig.11.
However in that case, the excitation function of the
T$_{iso}$(He-Li) is nearly independent of reaction systems,
Au+Au, Kr+Nb and Ar+Sc reactions, measured by
different experimental groups. The differences in the 
curves shown in Figure 13 again point to the uncertainties
associated with the experimental procedures
in extracting temperature and excitation energy. 

All the caloric curves measured so far
suffer from the same uncertainties in
determining the excitation energy. Some data may
have better handle on the excitation energy because of
better detector coverage or simpler reaction
mechanisms. For example, the caloric curves obtained
from the projectile fragmentation of Au, La and Kr on
C, \cite{Hauger00} have been extracted with a 
time-projection chamber
(EOS-TPC) where a complete reconstruction of the
projectile charge can be accomplished. The curves 
obtained from Au+C and Kr+C overlap very nicely with 
each other even though the La+C system shows lower 
temperatures measured. (The experimenters of 
Ref. \cite{Hauger00} claim that the discrepancies 
observed in the temperatures (~15\%)
are within the experimental uncertainties.)
It is also encouraging to see that these two curves
overlap with the $^3$He+Au and $^3$He+Ag data which
used similar procedures in determining the temperatures
and excitation energy \cite {Kwi98}. 

Many of the extracted caloric curves
do not agree with each other. Part of the differences can
be attributed to the energy thresholds applied to
extract the isotope yields. The high energy thresholds used in
the A$\approx$100 systems to isolate the prompt component 
\cite {Cibor00} probably account for the highest temperatures
obtained in all the curves. For the Au+C, La+C, Kr+C, 
\cite{Hauger00} $^3$He+Au, and $^3$He+Ag \cite {Kwi98} systems, 
assumptions have been made regarding 
the pre-equilibrium contributions and 
the missing neutrons. Energy thresholds are used to eliminate
pre-equilibrium emissions. This might account for the relatively
high temperatures measured as compared to the temperatures
extracted from Au+Au reaction. In the latter case, the
pre-equilibrium contributions were minimized using other
methods and attempts were made to extrapolate yields to zero
energy thresholds.

In the past year, some of the caloric curves have been 
revised \cite {Hauger00} and others including the original ``caloric curve" 
data are being reanalyzed \cite {Poch,Ma97}. With more attention
paid to the experimental problems associated with 
determining these curves, some of the discrepancies 
might be resolved in the near future. Future studies
might extract other underlining physics from these data.
Without further understanding of the reaction dynamics and
experimental limitations, one should be extremely cautious
in interpreting these curves as experimental signatures for
the liquid-gas phase transitions of nuclear matter.

\subsection{Isospin Fractionation}

Since nuclei are composed of neutrons and protons,
isospin effects may be very important for the nuclear
liquid-gas phase transition \cite{Li97}. As the asymmetry
between neutron and proton densities becomes a local
property in the system, calculations predict neutrons
and protons to be inhomogenously distributed within
the system resulting in a relatively neutron-rich gas
and relatively neutron-poor liquid \cite
{Muller95,Ch99,Sam}. The critical temperature may
also be reduced with increasing neutron excess
reflecting the fact that a pure neutron liquid probably
does not exist \cite{Muller95}. While, recent calculations
suggest that the rather narrow range of isospin values
available in the laboratory might not allow us to
observe the decrease in critical temperature \cite{Ma99},
efforts are underway to study the fractionation of the
isospin in the co-existence region.
As the isospin effects are not large, the influence
of sequential decays becomes important and may
obscure the isospin fractionation effect one wishes to
study. To minimize such problems, isobar pairs, such
as (t,$^3$He), which have the same number of internal
excited states have been used. Some indications for
isospin fractionation are provided by the sensitivity of
Y(t)/Y($^3$He) distributions to the overall N/Z ratio of
the system \cite{Dempsey96}. The ratios of Y(t)/Y($^3$He)
also have been observed to decrease with incident
energies, in qualitative agreement with the predictions
from the isospin dependent lattice gas model
\cite{Sam,Yennello00,Kunde00}.
Light isobars such as (t,$^3$He) pair may suffer
from contamination of pre-equilibrium processes.
Attempts have been made to use additional mirror
isobar pairs such as ($^7$Li, $^7$Be) and ($^{11}$B, $^{11}$C) 
\cite{Xu00}.

\vskip 0.2in
\epsfxsize=4in
\epsfysize=4in
\centerline{\rotatebox{90}{\epsffile{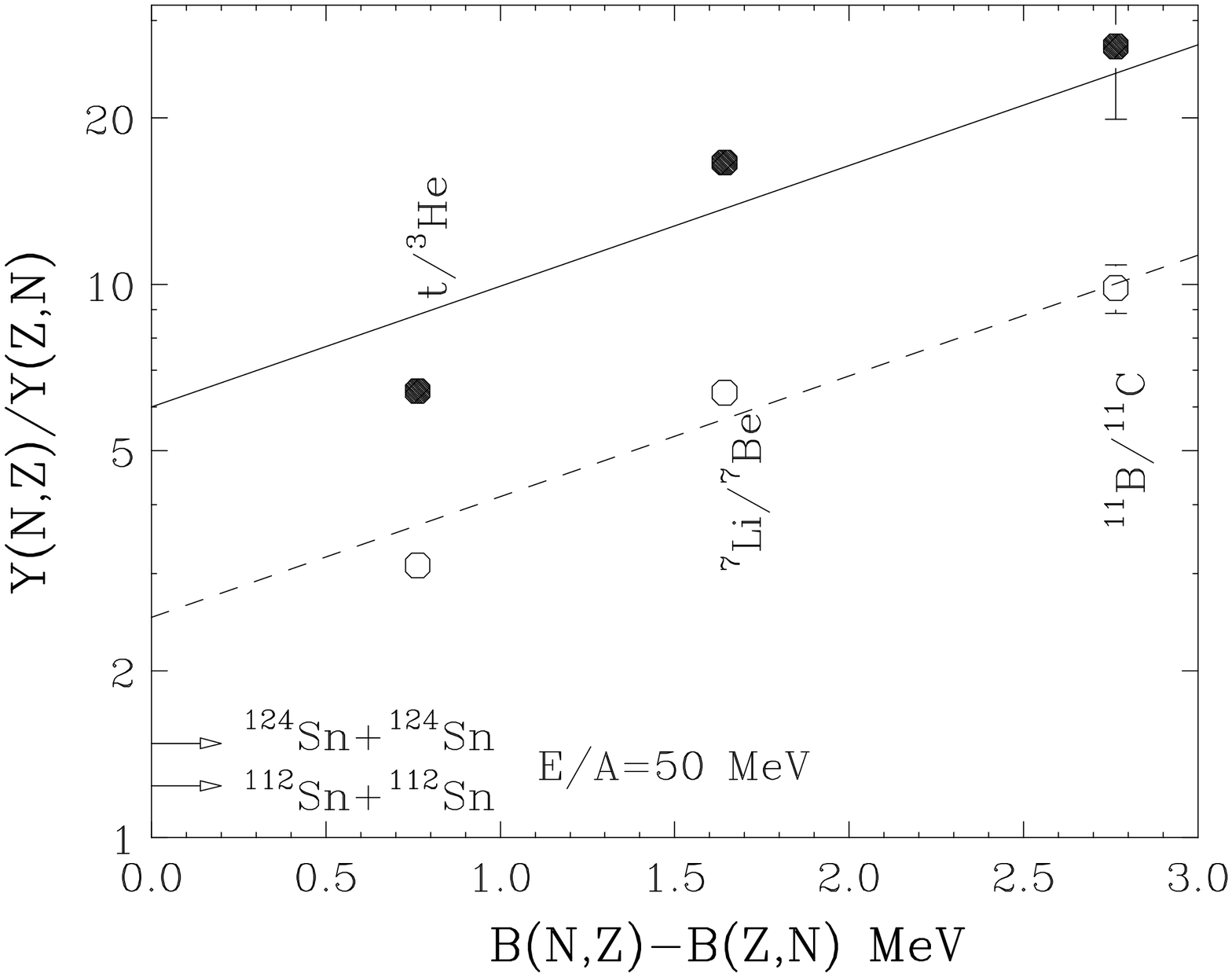}}}
\vskip 0.2in

Fig. 14: Isobar ratios for three mirror nuclei obtained from the 
$^{112}$Sn+$^{112}$Sn
(open circles) and $^{124}$Sn+$^{124}$Sn (solid points) 
reactions \cite{Xu00}. The lines
are drawn to guide the eye.

\vskip 0.2in

Fig.14 shows the isobaric yield ratios of 3 pairs of
mirror nuclei as a function of the binding energy
difference, $\Delta$B for two reactions, $^{112}$Sn+$^{112}$Sn (open
points) and $^{124}$Sn+$^{124}$Sn (solid points) at E/A=50 MeV. If
the sequential decay and the Coulomb effects are
small, the dependence of the isobaric yield ratios on
the binding energy difference should be exponential,
i.e. of the form $(\rho_n/\rho_p)\exp(\Delta B/T)$  where 
$\rho_n$ and $\rho_p$ are the
neutron and proton densities. The experimental data
fluctuate around such a relationship. Extrapolation to
$\Delta B=0$ using best fit lines (dashed and solid lines)
allows one to obtain values for $\rho_n/\rho_p$  of 2.5 for the
$^{112}$Sn+$^{112}$Sn system (top line) and 5.5 for the 
$^{124}$Sn+$^{124}$Sn
system (bottom line).   Both of these numbers are larger
than the initial N/Z values of the two system, 1.24 and
1.48 for $^{112}$Sn+$^{112}$Sn and $^{124}$Sn+$^{124}$Sn, respectively.
The change in the N/Z values of the two
systems is about 20\%. However the changes between
any of the mirror nuclei ratios are of the order of 200\%,
much larger than what one expects if the extracted
neutrons introduced into the neutron rich systems are
homogenously distributed. This observation suggests
that the free neutron density needed to determine the
light particle yields emitted from multifragmentation
is much enhanced in the neutron rich system.

To bypass the sequential decay problems
completely and to avoid using only selected ratios, an
observable employing ratios of all measured isotope
yields is used. This method relies on extracting the
relative neutron and proton density from two similar
reactions, which differ mainly in isospin. Adopting the
approximation of a dilute gas in the Grand-Canonical
Ensemble limit with thermal and chemical equilibrium,
the production of isotopes with neutron number N and
proton number Z are governed by the nucleon
densities, $\rho_n,\rho_p$, the temperature T plus the
individual binding energies of the various isotopes
B(N,Z).
\begin{eqnarray}
Y(N,Z)=F(N,Z,T)\rho_n.\rho_p.\exp(B(N,Z)/T
\end{eqnarray}
The factor $F(N,Z,T)$ includes information
about the secondary decay from both particle stable
and particle unstable states to the final ground state
yields as well as the volume terms. (Some readers may
notice the similarity of Eq.(9.1) to the Saha equation
used to describe the nucleation of a neutron and proton
gas in astrophysics. In that case the prefactor F(N,Z,T)
is dominated by the entropy term.)
If one constructs the ratio of $Y(N,Z)$ from two
different related reactions, the observable called the
relative isotope ratio, $R_{21}(N,Z)$ has a simple
dependence on the relative neutron density and proton
density of the free nucleon gas.
\begin{equation}
R_{12}(N,Z)=\frac{Y_2(N,Z)}{Y_1(N,Z)}\approx (\frac{\rho_{n,2}}{\rho_{n,1}})^N
(\frac{\rho_{p,2}}{\rho_{p,1}})^Z=\hat{\rho_n}^N\hat{\rho_p}^Z   
\end{equation}
In the study of the central collisions of four Sn
systems at incident energy of 50 MeV per nucleon
\cite{Xu00}, the relative neutron and proton densities have
been measured for the $^{112}$Sn+$^{124}$Sn (N/Z=1.36),
$^{124}$Sn+$^{112}$Sn (N/Z=1.36), $^{124}$Sn+$^{124}$Sn (N/Z=1.48)with
respect to the $^{112}$Sn+$^{112}$Sn (N/Z=1.24) system. More than
20 isotope ratios are measured and they follow the
dependence of Eq.(9.2) very well \cite{Xu00,Tsang00}.

The extracted $\hat{\rho_n}$ and $\hat{\rho_p}$ ratios are
shown in Fig.15; $\hat{\rho_n}$  increases while
$\hat{\rho_p}$ decreases with the N/Z ratio of the total
system. The increase of $\hat{\rho_n}$ is consistent with
neutron enrichment in the gas phase while the
decrease of $\hat{\rho_p}$ suggests proton depletion, a
consequence of n-enrichment in the nucleon gas. The
experimental trend (data points with the solid line
drawn to guide the eye) is much stronger than the
trend expected if neutrons and protons were
homogeneously mixed (dashed lines) in the breakup
configuration.  Adopting an equilibrium breakup
model, results of Figs. 14 and 15 are consistent with
isospin fractionation, a signal predicted in the liquid-gas
phase transition. However, as with other
signatures for phase transition observed so far, since
the isospin fractionation is governed by the symmetry
energy of the neutron and proton, ``isospin
fractionation" is a more general characteristic of heavy-ion
reaction than the liquid-gas phase phenomenon. In
fact, dynamical models also give predictions of isospin
amplication, in qualitative agreement with the data
\cite{Baran98}.

\vskip 0.2in
\epsfxsize=4in
\epsfysize=4in
\centerline{\rotatebox{90}{\epsffile{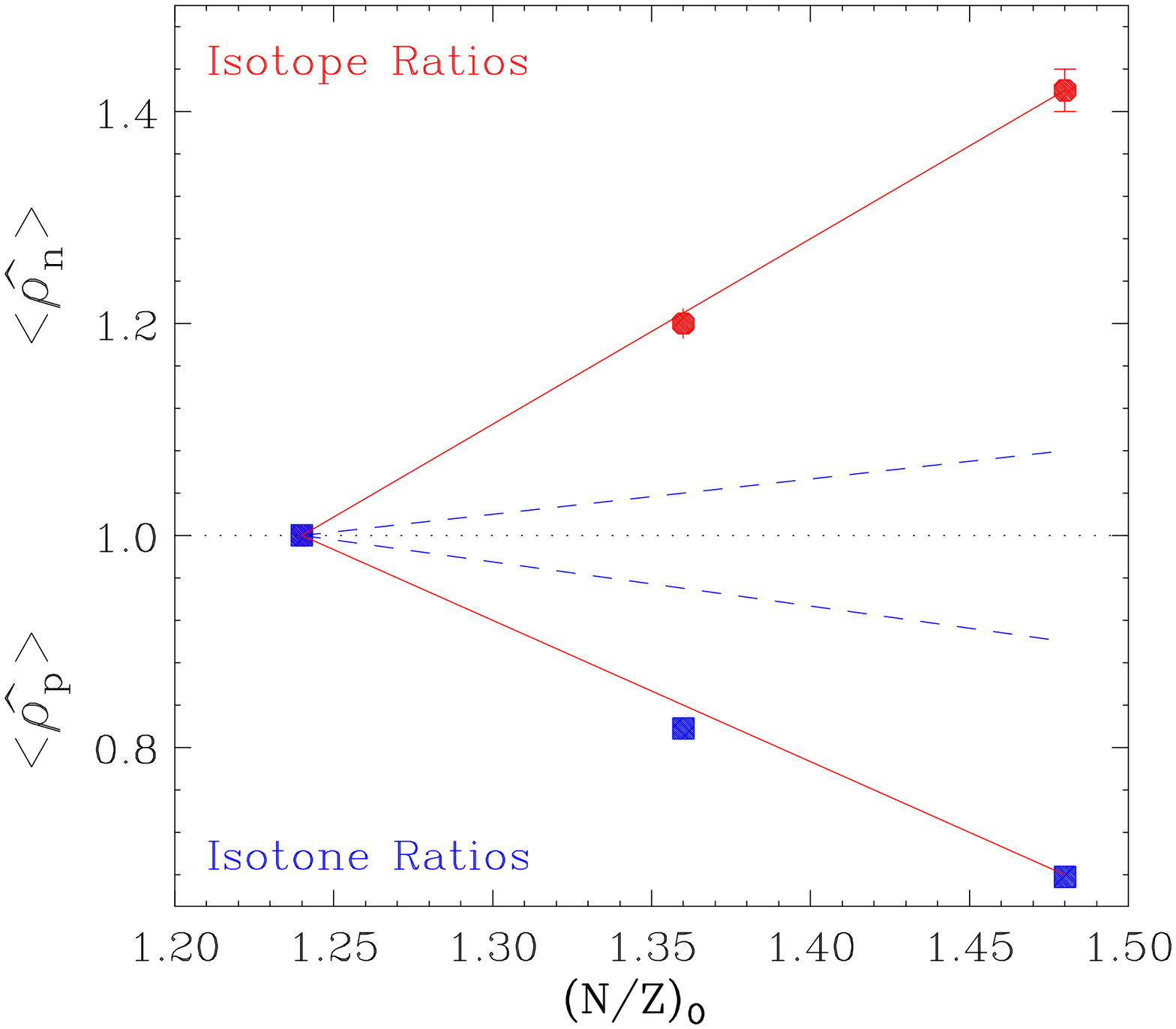}}}
\vskip 0.2in

Fig. 15: The mean relative free neutron and free proton density as
a function of $(N/Z)_O$. The dashed lines are the expected n-enrichment and
p-depletion with the increase of isospin of the initial systems. The solid
lines are drawn to guide the eye \cite{Xu00}.

\pagebreak
\section{A Class of Statistical Models}
A class of statistical models has been very successful in explaining
multifragmentation processes in heavy-ion collisions.  These models
assume the following scenario.  One defines a freeze-out volume.
At this volume an equilibrium statistical machanics calculation is 
done.  However, these statistical calculations do not 
start from a fundamental two-body interaction or even a
simplified schematic two-body interaction.  
Instead the inputs are the properties of the composites
(which appear as bound objects because of the fundamental two-body
interaction); their binding energies and the excited states.  Their
populations are solely dictated by phase-space.  This is very
similar to chemical equilibrium between perfect gases as, for
example, discussed in \cite{Reif}.  The only interaction between
composites is that they can not overlap with each other in the
configuration space.  Coulomb
interaction between composites can be taken into account in different
stages of approximation.  These models have the virtue that they can
be used to calculate data for many experiments whether these
experiments relate to phase transition or not.  The Copenhagen model,
a statistical multifragmentation model abbreviated SMM 
(also referred to as SMFM), has become, {\it de facto}, the ``shell-model''
code for intermediate energy heavy ion data.  An excellent review
of this model exists \cite {Bondorf95}.  The Berlin Model, a microcanonical
multifragmentation model, usually abbreviated MMMC, has also been used
to fit experimental data \cite{Gross97}.  Some other references for
microcanonical simulation of similar physics are \cite{Randrup87,Raduta97}.
While there have been tremendous improvements in techniques and details,
the roots of such models for heavy-ion collisions go back more than 20
years \cite {Mekjian77}.

With some simplifications, the model of composites within the freeze-out
volume at a given temperature can be exactly solved.  In order to 
distinguish this model from SMM and MMMC (which are much harder to
implement) we  will coin a name.  We will call this the thermodynamic model.
As phase transition aspects are easily studied in the model, we treat
this in detail.

\section{ A Thermodynamic Model}
 
If there are $A$ identical particles of only one kind in an enclosure, 
the partition function of the system can be written as
\begin{eqnarray}
Q_A=\frac{1}{A!}(\omega)^A
\end{eqnarray}
Here $\omega$ is a one-particle partition function of the particle.
For a spinless particle this is $\omega =\frac{V}{h^3}(2\pi mT)^{3/2}$;
$m$ is the mass of the particle; $V$ is the available volume within
which each particle moves;
$A!$ corrects for Gibb's paradox.  One might argue that this
is not a rigorous way of treating symmetry or antisymmetry of particles
but a recent paper \cite{Jennings}
demonstrates  that for nuclear disassembly this
correction is very adequate.  If there is only one species, eq.(11.1)
is trivially calculated.

If there are many species, the generalisation is
\begin{eqnarray}
Q_A=\sum \Pi_i \frac{(\omega_i)^{n_i}}{n_i!}
\end{eqnarray}
Here $\omega_i$ is the partition function of a composite which has $i$
nucleons.  We are at the moment ignoring the distinction between a 
neutron and a proton and thus our composites are bound states of $i$ nucleons.
For a dimer, $i=2$, for a trimer, $i=3$ etc.  In a more realistic version
we will introduce the distinction between neutrons and protons but this 
model of one type of nucleon is highly illustrative, so we will
continue with this for a while.

Eq.(11.2) is no longer trivial to calculate.  The trouble is with the
sum in the right hand side of eq.(11.2).  The sum is restrictive.
We need to consider only those
partitions of the number $A$ which satisfy $A=\sum in_i$ 
This restriction is hard to implement in an actual calculation
and already for $A$ of
the order of 100, the number of partitions which satisfies the sum
is enormous.  We can call a given allowed partition a channel.
The probablity of the occurrence of a given channel $(n_1,n_2,n_3.....)$
is 
\begin{displaymath}
P(\vec n)\equiv P(n_1,n_2,n_3........)
\end{displaymath}
\begin{eqnarray}
P(\vec n)=\frac{1}{Q_A}\Pi \frac{(\omega_i)^{n_i}}{n_i!}
\end{eqnarray}
The average number of composite of $i$ nucleons is easily seen from 
Eq.(11.3) to be
\begin{eqnarray}
<n_i>=\omega_i\frac{Q_{A-i}}{Q_A}
\end{eqnarray}
Since $\sum in_i=A$ one readily arrives at a recursion relation \cite {Chase}
\begin{eqnarray}
Q_A=\frac{1}{A}\sum_{k=1}^{k=A}k\omega_kQ_{A-k}
\end{eqnarray}
For one kind of particle, $Q_A$ above is easily evaluated on a computer
for $A$ as large as 3000 in matter of seconds.  Thus in this model we
can explore the thermodynamic limits.  It is this recursion relation
that makes computation so easy in the model.  In the realistic model
with two kinds of particles which we will introduce later, systems
as large as 400 particles are easily done.  It is important to 
realise that millions of channels possible in the partitions (Eq.(11.3))
are exactly taken into account, although numerically.  No Monte-Carlo
sampling of the channels is required.

We now need an expression for $\omega_k$ which can mimic the nuclear
physics situation.  We take
\begin{eqnarray}
\omega_k=\frac{V}{h^3}(2\pi mT)^{3/2}k^{3/2}\times q_k
\end{eqnarray}
where the first part arises from the centre of mass motion of the 
composite which has $k$ nucleons and
$q_k$ is the internal partition function.  For $k=1, q_k=1$ and for $k\ge 2$
it is taken to be
\begin{eqnarray}
q_k=\exp [(W_0k-\sigma(T)k^{2/3}+T^2k/\epsilon_0)/T]
\end{eqnarray}
Here, as in \cite {Bondorf95}, $W_0$=16 MeV is the volume energy term,
$\sigma(T)$ is a temperature dependent surface tension term and the last
term arises from summing over excited states of the composite in the
Fermi-gas model.  For high temperatures, this will overestimate the
contribution of the excited states and a modified expression is sometimes
used to correct for this \cite {Randrup87}.
The explicit expression for $\sigma(T)$ used here
is $\sigma(T)=\sigma_0[(T_c^2-T^2)/(T_c^2+T^2)]^{5/4}$ with $\sigma_0$=
18 MeV and $T_c$=18 MeV.  The value of $\epsilon_0$ is taken to be 16 MeV.
The energy carried by one composite is given by 
$E_k=T^2\partial ln\omega_k/
\partial T=\frac {3}{2}T+k(-W_0+T^2/\epsilon_0)+\sigma(T)k^{2/3}-
T[\partial \sigma(T)/\partial T]k^{2/3}$.  Of these, the first
term comes from cm motion and the rest from $q_k$.  In \cite {DM98},
the term $T[\partial \sigma(T)/\partial T]k^{2/3}$ was neglected.  It
is included here but makes little difference.  In the nuclear case
one might be tempted to interpret the $V$ of Eq.(11.6) as simply the
freeze-out volume but it is clearly less than that; $V$ is the volume
available to the particles for the centre of mass motion.  In the Van der Waals
spirit we take $V=V_{freeze}-V_{ex}$ where $V_{ex}$ is taken here
to be constant and equal to $V_0=A/\rho_0$ \cite {Hahn}.  The
precise value of $V_{ex}$ is inconsequential so long it is
taken to be constant.  Calculations employ $V$; the value of $V_{ex}$
enters only if results are plotted against $\rho/\rho_0=V_0/(V+V_{ex})$
where $\rho$ is the freeze-out density.
The energy of the system
is $E=\sum <n_k>E_k$; the
pressure is $p=T\partial lnQ_A/\partial V=T\frac{1}{V}\sum <n_i>$.  These can
be deduced from Eqs.(11.2) and (11.4).

The surface tension term is crucial for phase transition in this model.
At a given temperature the free energy $F=E-TS$ has to be minimised.
It costs in the energy term $E$ to break up a system.  A nucleus of
$A$ nucleons has less surface than the total surface of 
two nuclei each of $A/2$ nucleons (the volume energy term has no preference
between the two alternatives).
Therefore at low temperature one will see a large chunk.  The $-TS$ 
term favours break up into smaller objects.  The competition between
these two effects leads to the following features seen in experiments.
At low temperature (low beam energy) each event will have one large
composite (the fission channel is not included in these models) and 
few small composites.
This leads to the inclusive cross-section being U-shaped (Fig.16).
(For illustration, starting from this figure and and upto and including Fig.20,
all calulations employ $V$ of Eq.(11.6) =$1.5V_0$).  As the temperature
increases, the peak on the large $a$ side begins to decrease, finally
disappearing entirely.  When this happens, one has crossed from the 
coexistence zone to the gas phase.  A graph of the yield $Y(a)$ against $a$
is shown in Fig.16.  At temperature 6.2 MeV one sees both a large 
residue and smaller clusters, at 6.7 MeV the large cluster just disappears
and at 7.2 MeV one has only the gas phase.

\pagebreak

\epsfxsize=4.0in
\epsfysize=4.0in
\centerline{\epsffile{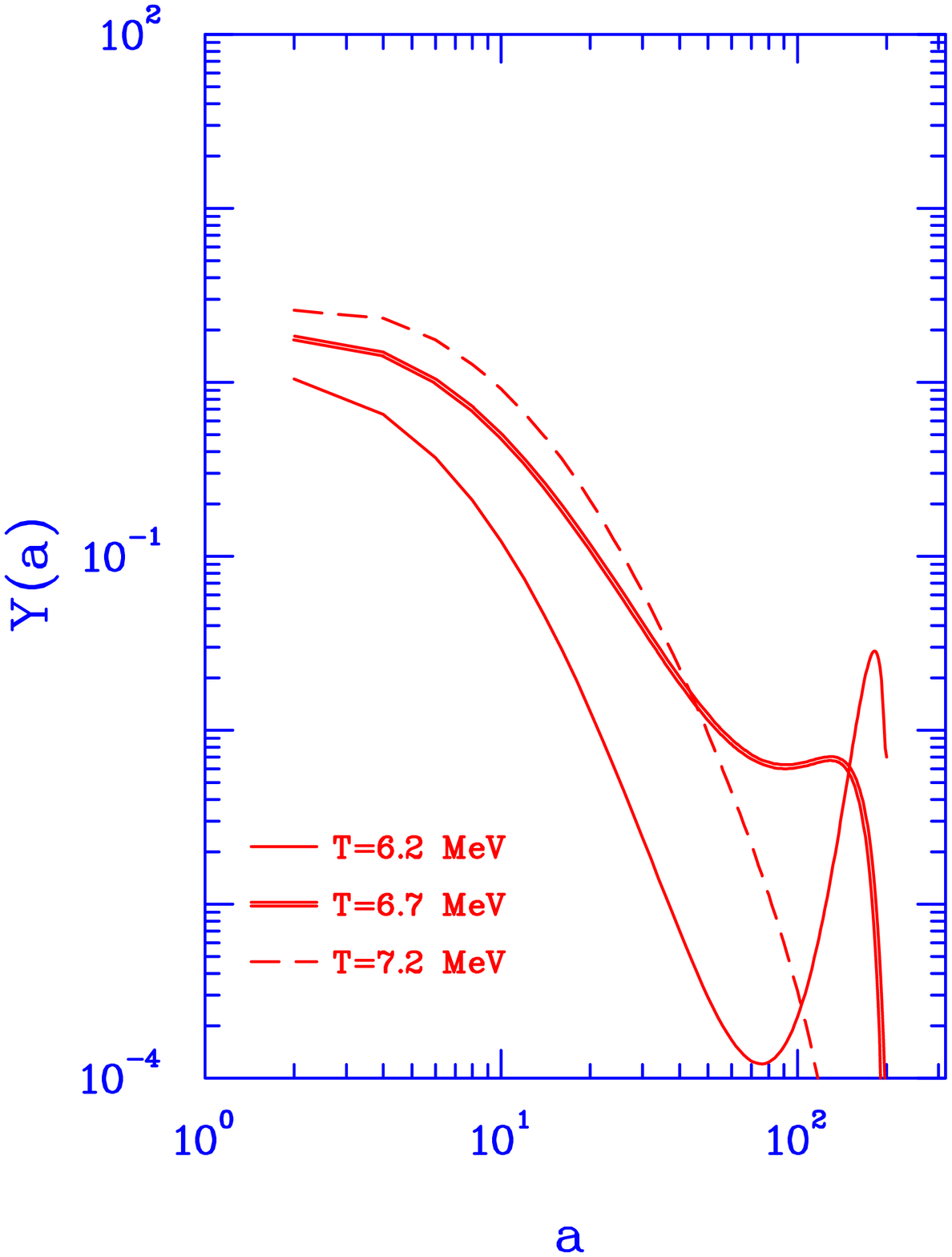}}
\vskip 0.2in

Fig.16.  In a model of one kind of nucleon, the yield $Y(a)$ against $a$
where the disintegrating system has 200 nucleons.  At temperature 6.2 MeV
one has coexistence: on the average, there is a large blob and also
lighter mass composites.  At 6.7 MeV, the maximum in the yield $Y(a)$
on the high mass side just disappears..  This is also the temperature
at which the specific heat $C_v$ per particle maximises (see Fig.17).
We call this the boiling temperature.  At higher temperature, the yield
$Y(a)$ falls monotonically.

\vskip 0.2in

It is instructive
to plot for the same nucleus the specific heat per nucleon
labelled by $C_v$, the total multiplicity
and $N_{IMF}$, the number of intermediate mass fragments as a function
of temperature (Fig.17).  One sees the specific heat maximising at the same
temperature as the one at which, in Fig.16, the peak in the high $a$ side
of the yield function $Y(a)$ just disappears.  One should remember that
fragments produced in this model appear at non-zero temperature.
They will further decay by sequential emissions.  Thus the total
multiplicity plotted here is lower than actual value to be expected
finally.  After the sequential decays, the yields of very light elements
such as monomers, dimers etc. will increase substantially  as the heavy
composites decay by emitting these.  In $N_{IMF}$
in Fig.17 we have included $a$=6 to 40.  With sequential decays included
$N_{IMF}$ will go down from the values shown in Fig.17.  

\vskip 0.2in
\epsfxsize=4.0in
\epsfysize=4.0in
\centerline{\epsffile{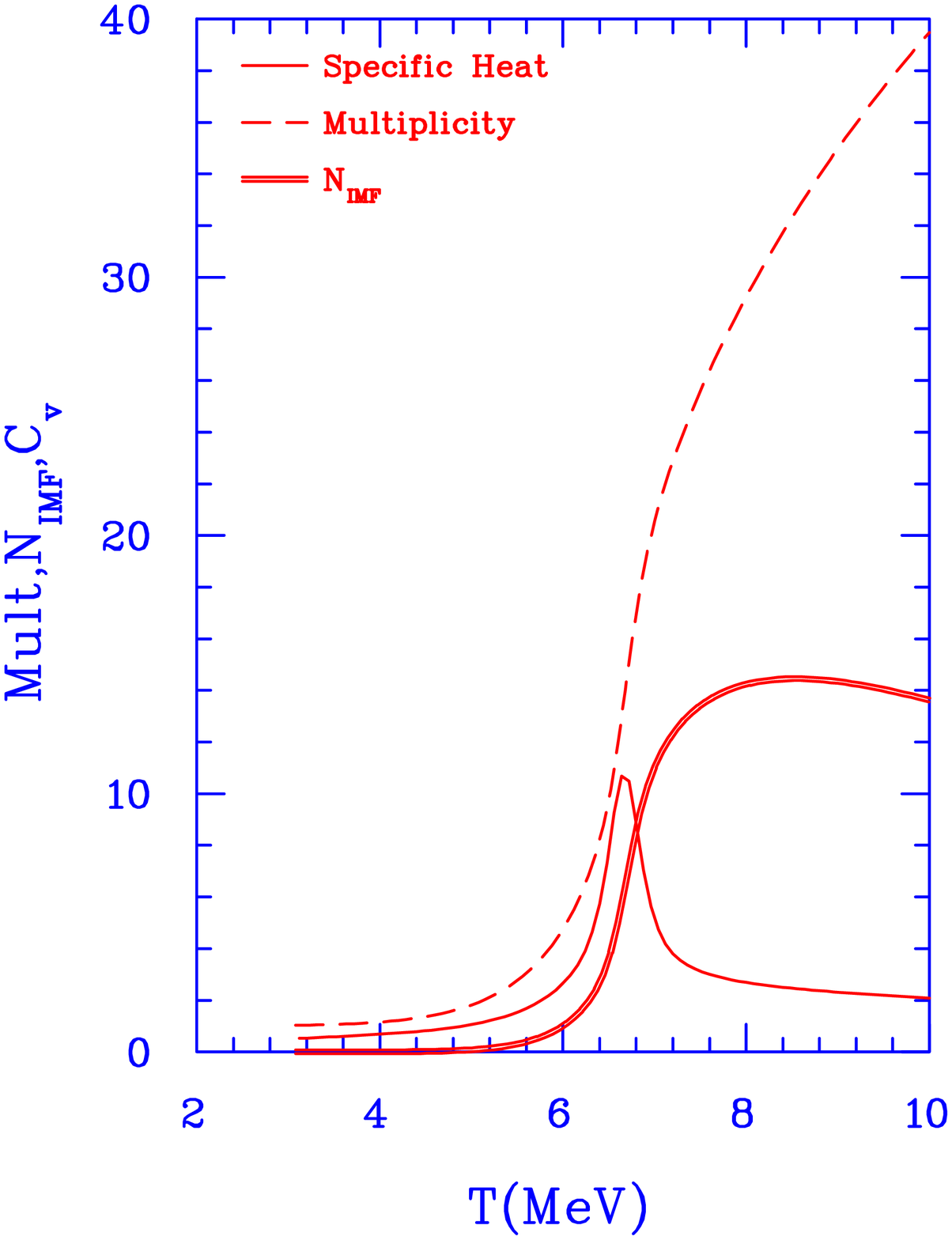}}
\vskip 0.2in

Fig.17. For the same model of 200 nucleons of one kind, the total
multiplicity, $N_{IMF}$ (the number of intermediate mass fragments defined
here as having $a$ between 6 and 40) and the specific heat $C_v$ per
particle as a function of temperature.  Note that the maximum of the 
specific heat coincides with the quick rise in $N_{IMF}$ and the disappearence
of the maximum in the yield $Y(a)$ on the high $a$ side.

\vskip 0.2in
Keeping these
reservations in mind we see in Fig.17 that 
the sudden increase of the multiplicity and 
$N_{IMF}$ imply coexistence.  At higher temperature, the system is in the
gas phase.  The cross-section for a large residue is very small (Fig.16).
In order to understand the nature of the phase transition we now go
to much larger systems so that one can feel confident about extrapolation
to the thermodynamic limit.  With this in mind we have plotted in Fig.18.,
for a system of 1400 and 2800 particles,
the free energy per particle as a function of temperature (The free 
energy is simply -$T lnQ_A$).  A brake appears to develop in
the first derivative of $F/A$ which signifies first order phase transition.

\vskip 0.2in
\epsfxsize=4.0in
\epsfysize=4.0in
\centerline{\epsffile{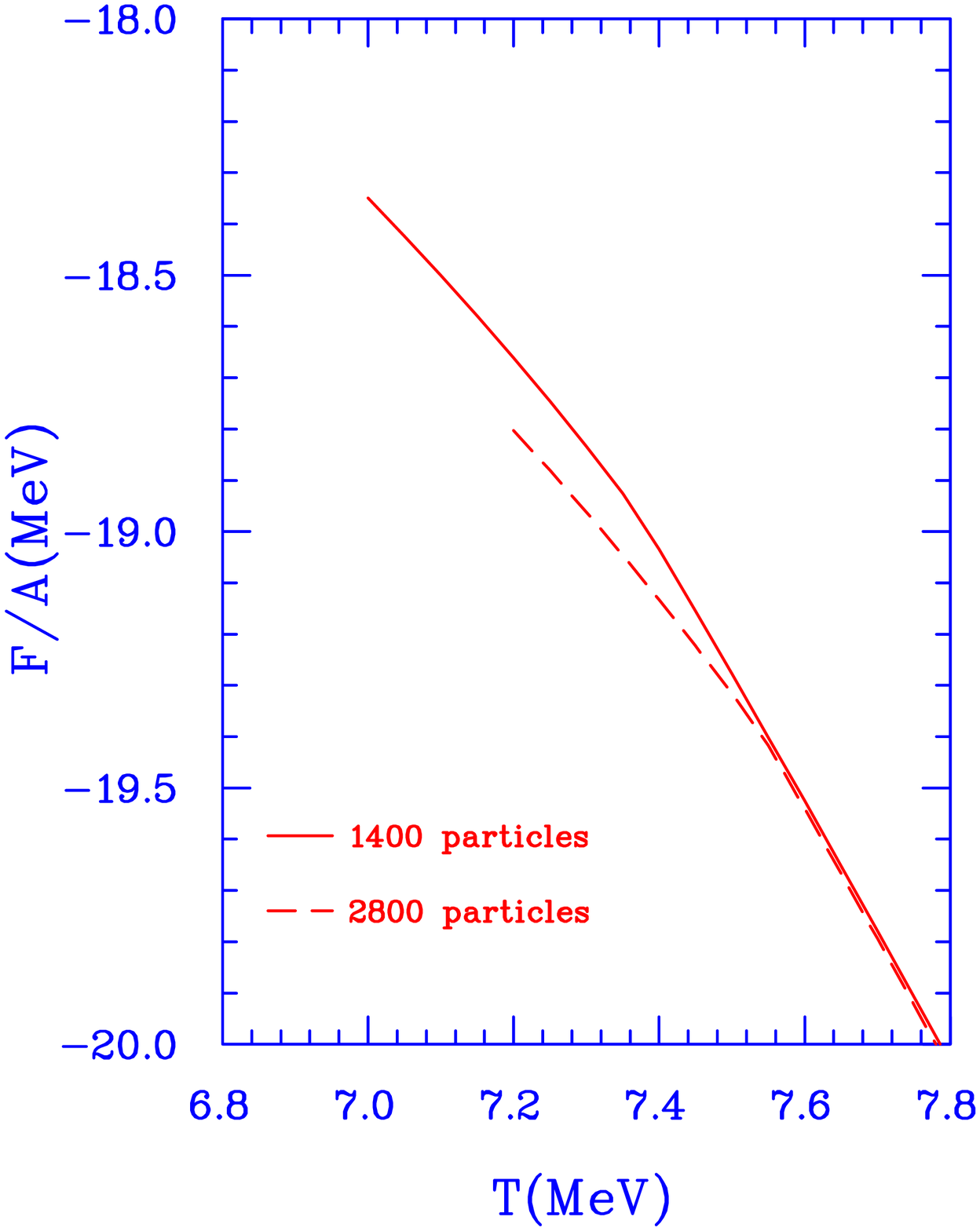}}
\vskip 0.2in

Fig.18.  The first derivative of the free energy per particle ($-TlnQ_A/A$)
will show a break as a function of temperature for a large system.  Shown
here are cases of 1400 and 2800 particles.

\vskip 0.2in 
We follow this up in Fig.19 by calculations of specific heat $C_v$ per
particle for 200, 1400 and 2800 particles.  As the number of particles
increases, the maximum in the $C_v$ per particle becomes sharper and the
height increases.  
In Fig.20 we have tried to understand the origin of this
singularity in greater detail.  Let us denote by $<a_{max}>$ the ensemble
average of the mass number of the heaviest composite (the technique
for this calculation is given in \cite{DM98}).  This should scale like
$A$ where $A$ is the number of particles in the disintegrating system.
In Fig.20 we have plotted $<a_{max}>/A$ as a function of $T/T_b$ where
$T_b$ is the temperature at which the specific heat maximises.  As $A$
becomes large, the drop in the value of $<a_{max}>/A$ at 
$T=T_b$ becomes sharp.  The sudden disappearence of this large blob of
size $\approx A/2$ causes this behaviour of $C_v$.

\vskip 0.2in
\epsfxsize=3.5in
\epsfysize=3.5in
\centerline{\epsffile{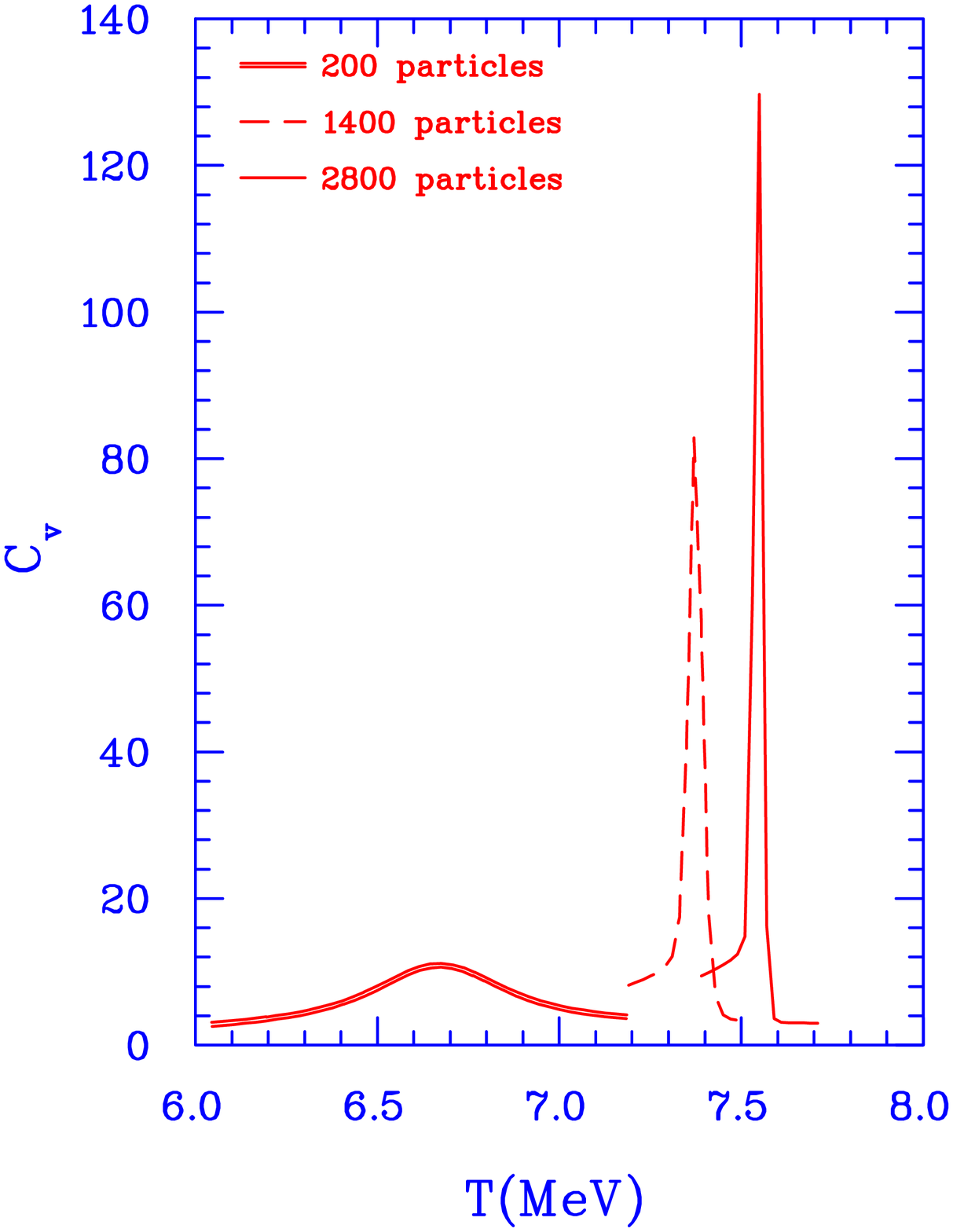}}

Fig.19.  As the number of particles increase, the maximum in $C_v$
per particle becomes very narrow and very high.

\vskip 0.1 in
\epsfxsize=3.5in
\epsfysize=3.5in
\centerline{\epsffile{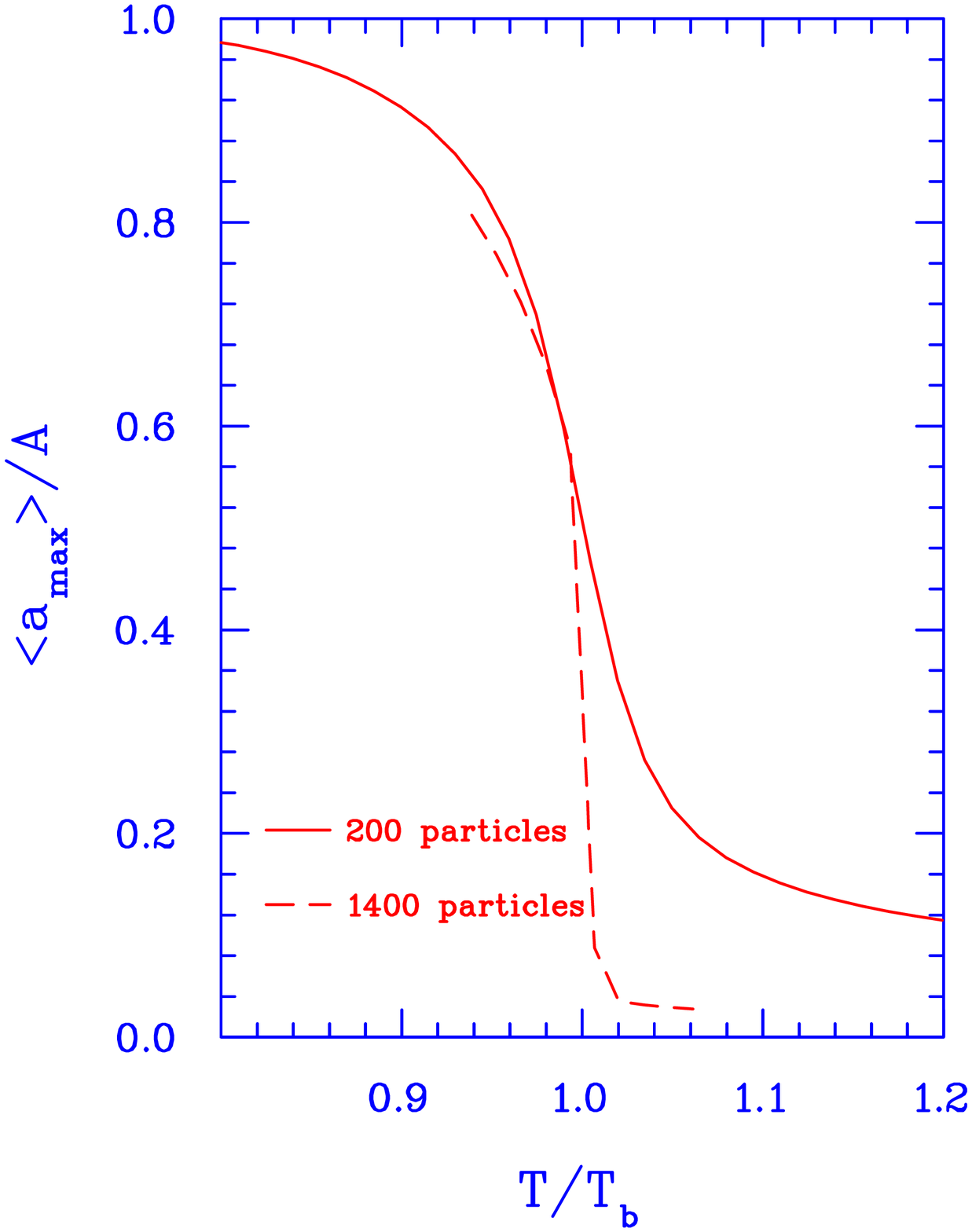}}
\vskip 0.1in

Fig.20.  For large systems a large blob of matter suddenly disappears
at $T_b$.

\vskip 0.2in
In Fig.21 we have drawn a $p-\rho$ diagram for a system of 200 particles
at various temperatures.  We have also drawn a line that is labelled
co-existence which passes through the points where the specific heat attains
the highest values.  For plotting this graph we have used 
$V_{ex}=200/\rho_0$.  We
stop below $\rho/\rho_0$=0.50.  At higher density the approximation
of non-interacting clusters (even after including
Van der Waals type correction
for finite volumes of the composites) would be very questionable.

\vskip 0.2in
\epsfxsize=4.0in
\epsfysize=4.0in
\centerline{\epsffile{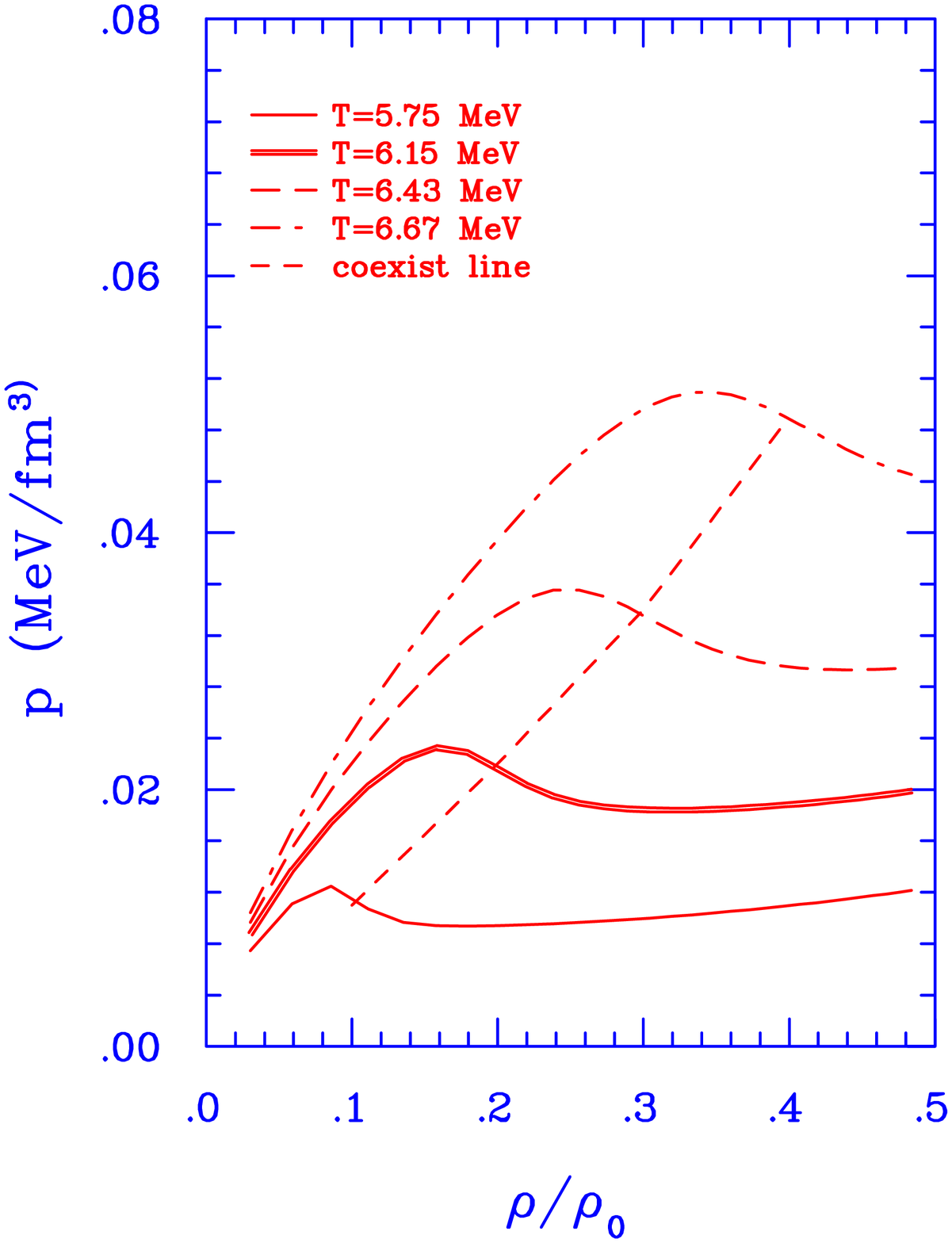}}
\vskip 0.1in

Fig.21. Isotherms for different temperatures for a system of 200 nucleons.
Here it is assumed that $V_{ex}=200/\rho_0$.  Line labelled coexistence
goes through points of highest $C_v$.

\vskip 0.2in
One approximation in the above calculation is the assumption
of constant excluded volume.  The excluded volume (as can be verified
in Monte-Carlo simulation) is a function of the total multiplicity .  It
is also a function of the freeze-out volume inside which the particles
are constrained to move.  For 200 particles, the effect of this variability
of the excluded volume on the $p-V$
diagram was investigated in \cite{Majumder99}.  The difference  is not
large.  However, this has not been studied in the thermodynamic limit.
It will be very interesting to investigate what effect it will have on the
nature of the phase transition in the very large $A$ limit.

The thermodynamic properties of this model have been further studied in
\cite {Jbe,Bug}.
\section{Generalisation to a more realistic model}

For comparisons with actual data the model must be made more realistic.
Towards that goal, a composite is now  
labelled by two indices: $\omega\rightarrow \omega_{i,j}$ where the
the first index in the subscript refers to the number of protons 
and the second, to the number of neutrons in the composite.  The
partition function for a system with $Z$ protons and $N$ neutrons
is given by
\begin{eqnarray}
Q_{Z,N}=\sum\Pi_{i,j}\frac{\omega_{i,j}^{n_{i,j}}}{n_{i,j}!}
\end{eqnarray}
There are two constraints: $Z=\sum i\times n_{i,j}$ and 
$N=\sum j\times n_{i,j}$.  These lead to two recursion relations any one
of which can be used.  For example,
\begin{eqnarray}
Q_{Z,N}=\frac{1}{z}\sum_{i,j}i\omega_{i,j}Q_{Z-i,N-j}
\end{eqnarray}
where
\begin{eqnarray}
\omega_{i,j}=\frac{V}{h^3}(2\pi mT)^{3/2}(i+j)^{3/2}\times q_{i,j}
\end{eqnarray}
Here $q_{i,j}$ is the internal partition function.  These could
be taken from experimental binding energies, excited states and 
some model for the continuum or from the liquid drop model or a 
combination of both.  The versatility of the model lies in being able
to accommodate any choices for $q_{i,j}$.  A choice of $q_{i,j}$
from a combination of the liquid-drop model for binding energies and 
the Fermi-gas model for excited states that has been used is
\begin{eqnarray}
q_{i,j}=\exp [(W_0(i+j)-\sigma(i+j)^{2/3}-\kappa\frac{i^2}{(i+j)^{1/3}}
-s\frac{(i-j)^2}{i+j}+T^2(i+j)/\epsilon_0)/T]
\end{eqnarray}
where $W_0$=15.8 MeV, $\sigma$=18.0 MeV, $\kappa$=0.72 MeV and $\epsilon_0$=
16 MeV.  One can recognise in the parametrisation above the volume
term, the surface tension term, the Coulomb energy term, the symmetry
energy term and contributions from excited states. 

The coulomb interaction is long range; some effects of the Coulomb 
interaction between different composites can be included
in an approximation called the Wigner-Seitz approximation.  We assume,
as usual, that
the break up into different composites occurs at a radius $R_c$
which is greater than normal radius $R_0$.  Considering this as a
process in which a uniform dilute charge distribution within 
radius $R_c$ collapses 
successively into denser blobs of proper radius $R_{i,j}$ we write
the Coulomb energy \cite{Bondorf85} as

\begin{eqnarray}
E_C=\frac{3}{5}\frac{Z^2e^2}{R_c}+
\sum_{i,j}\frac{3}{5}\frac{i^2e^2}{R_{i,j}}(1-R_0/R_c)
\end{eqnarray}
It is seen that the expression is exact in two extreme limits: very
large freeze-out volume ($R_c\rightarrow\infty$) or if the freeze-out volume
is the normal nuclear volume so that one has just one nucleus with the
proper radius. 

For the thermodynamic model that we have been pursuing, the constant
term $\frac{3}{5}\frac{Z^2e^2}{R_c}$ in the above 
equation is of no significance since the freeze-out volume is assumed 
to be constant.  In a mean-field sense then one would just replace
the Coulomb term in Eq.(12.4) by
$\kappa\frac{i^2}{(i+j)^{1/3}}(1.0-(\rho/\rho_0)^{1/3})$

Calculations with the thermodynamic model with two kinds of particles
and realistic $q_{i,j}$ were done in \cite {Bhatta1,Bhatta2}.  A
plateau in the caloric curve is found around 5 MeV which is in accordance
with experimental finding.  An interesting point in the calculation
is the following observation.  Without the Coulomb, the height in the
peak of the specific heat increases with $A$ (see previous section).
With Coulomb the height is reduced and the dependence on $A$
nearly disappears.  The growth in size is compensated by the growth
in Coulomb repulsion.  This means the caloric curve is approximately
universal,i.e., does not depend strongly on the specific system which
is disintegrating.  We show the caloric curves computed for three
disintegrating systems in Fig.22.  This is taken from \cite {Bhatta1}.

\vskip 0.2in
\epsfxsize=4.0in
\epsfysize=4.0in
\centerline{\rotatebox{-90}{\epsffile{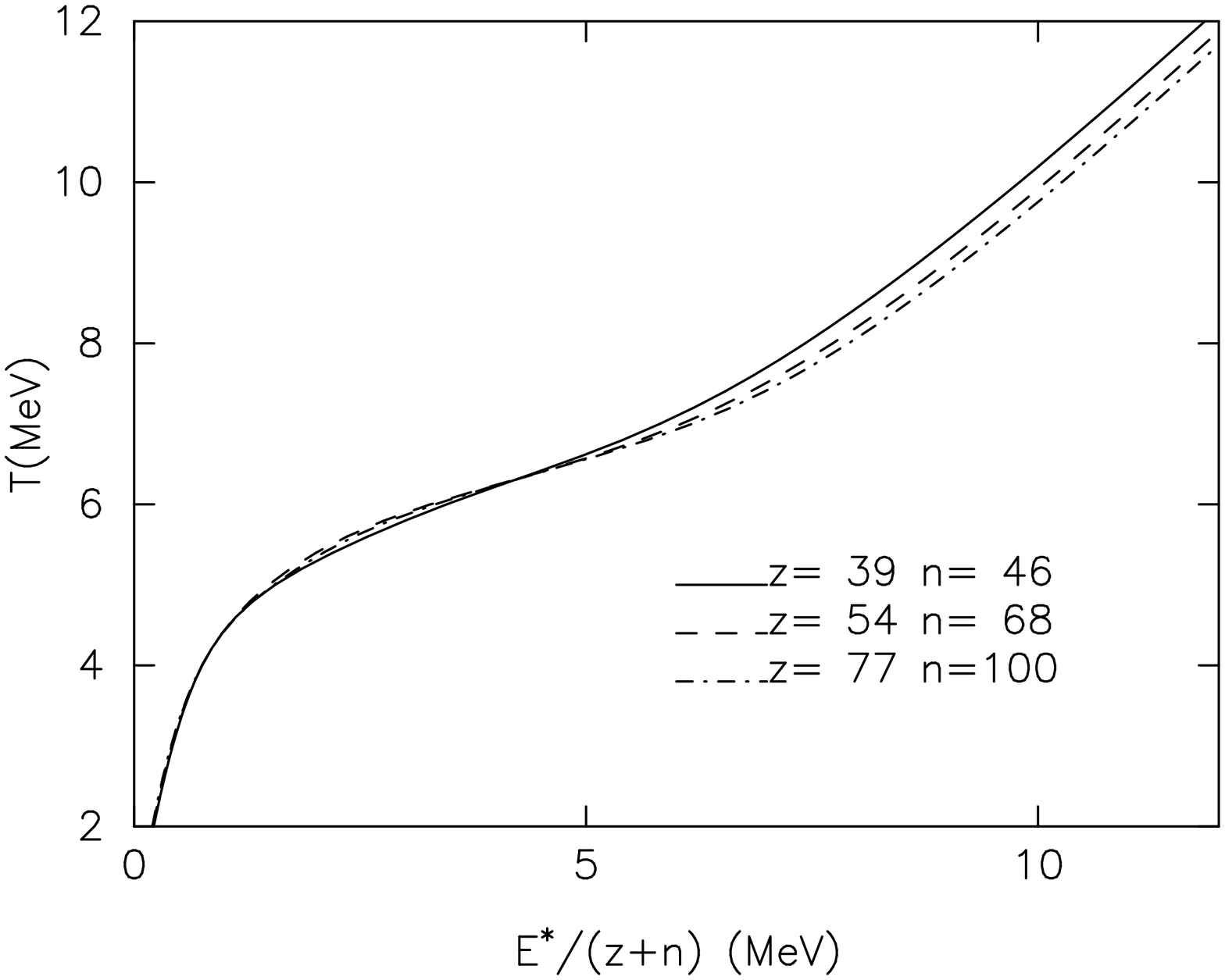}}}
\vskip 0.4in

Fig.22. The caloric curve for three different nuclei.  The Coulomb 
interaction is included in the Wigner-Seitz approximation.

\vskip 0.2in
Of course, the model can also be used to calculate other observables,
not necessarily related with any phase transition aspect.  However,
for many purposes an ``afterburner'' calculation is required.  The
composites obtained in the calculations are ``hot''.  They will 
subsequently decay by particle emissions.  In \cite{Majumder00} these
subsequent decays were included in an approximate manner
so that one can compare with 
experimental yields of boron, carbon and nitrogen isotopes.  This comparison
is shown in Fig.23.

\epsfxsize=4.0in
\epsfysize=6.0in
\centerline{\epsffile{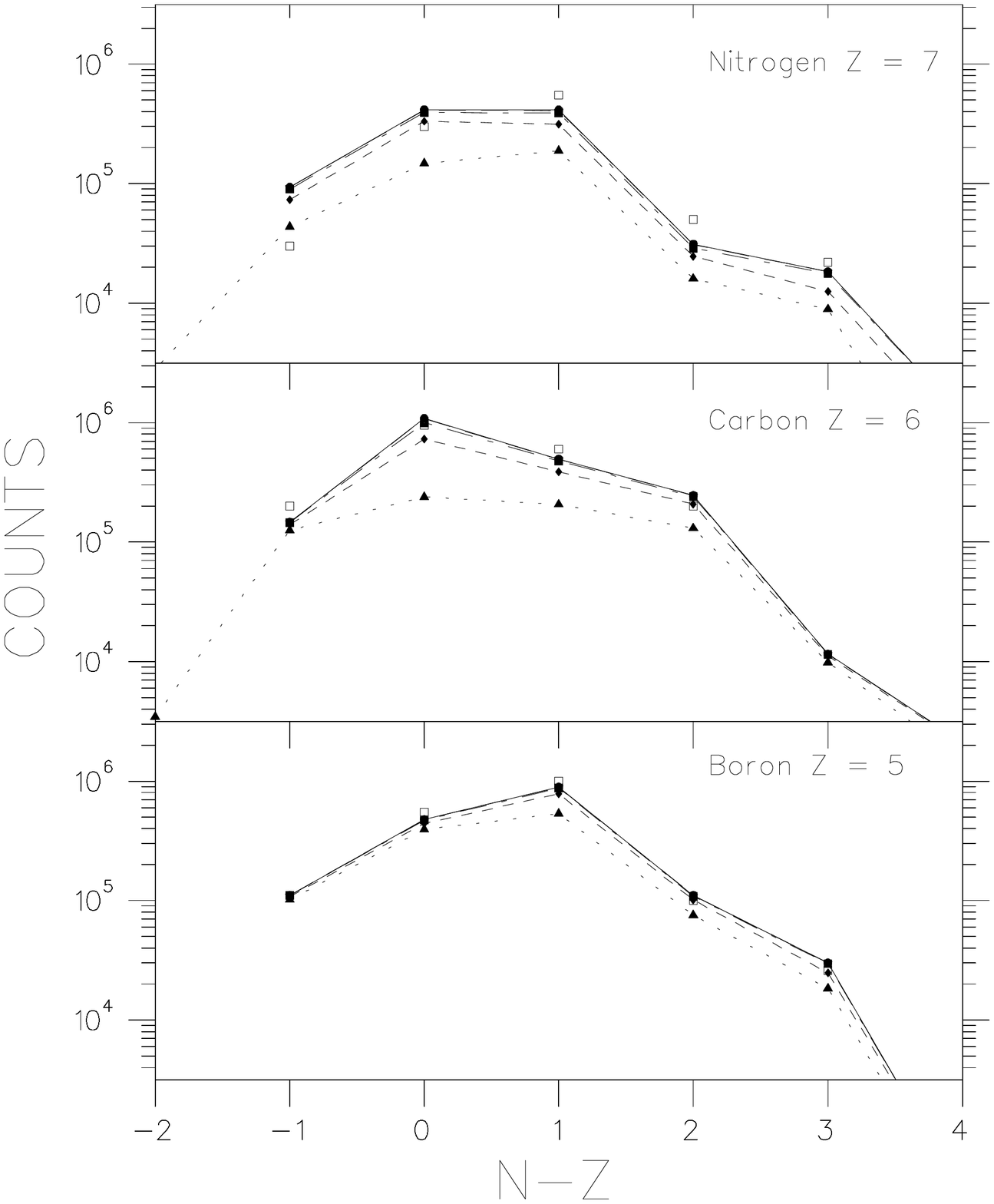}}
\vskip 0.2in

Fig.23. Neutron number (N), proton number (Z) vs. counts for the three
cases of boron, carbon and nitrogen.  The experimental data are from
\cite {Xu} S+Ag at 22.3$A$ MeV.  The open squares are the experimental
data.  The dotted line with the triangle plotting symbol is the 
thermodynamic model calculation as described in the text.  A temperature
of 5 MeV is used.  The other
curves are obtained by allowing sequential decays of hot nuclei.  The
calculation stops after four possible decays (the solid line) as virtually 
no changes are seen between triple and quadruple decays. 

\section{A brief review of the SMM model}
A very comprehensive review of this model exists \cite {Bondorf95} so
we only give a brief resum\'e here.  It is relevant to mention 
that a peak in the specific heat at about 5 MeV temperature was 
predicted in the model well before experiments were done \cite {Bondorf2}.
 
In the SMM each break-up channel is treated separately.  A given channel
is specified by the set of $n_{i,j}$'s, V and temperature $T$.  The volume
$V$ which is the free volume available for the motion of the cms of the 
composites is taken to be $V=\chi V_0=\chi A/\rho_0$ where $\chi$
is multiplicity dependent ($M=\sum n_{i,j})$.
\begin{eqnarray}
\chi=(1+\frac{d}{r_0A^{1/3}}(M^{1/3}-1))^3-1
\end{eqnarray}
In the above $d$ is taken to be 1.4 fm; $r_0A^{1/3}$ is the normal
radius of a nucleus of $A$ nucleons.
The concept of temperature is used but its primary use is to make the
energy in each channel the same value and to correspond to the
experimental situation.  It is therefore a microcanonical
calculation in spirit.  The energy in a given channel $E_{tot}$ is given by
\begin{eqnarray}
E_{tot}(T,V)=E^{tr}(T)+\sum E_{i,j}(T)n_{i,j}+E_{cou}(V)
\end{eqnarray}
where $E^{tr}(T)=\frac{3}{2}(M-1)T$, $E_{i,j}(T)$ gives the intrinsic
energies at temperature $T$ (includes binding energies, contribution
from excited states etc.; see the discussion in section XII)
and $E_{cou}(V)$ completes the Wigner-Seitz
estimation of Coulomb energy.  The crucial thing here is the choice of
channels.  It is impossible to include all channels.  For $A$=200
the number of possible multifragment partitions is $3.9\times 10^{12}$
so a Monte-Carlo
sampling which is geared to include the most important channels at 
a given excitation energy is needed.  This is a very elaborate story in 
itself and all we can do here is to provide references.  An important
paper which elaborates on the procedure is \cite {Sneppen87}.
The review article on the 
model \cite{Bondorf95} gives a more complete list of references.
As in all models of this type, sequential decays
of the hot fragments need to be included to compare with most 
experimental data. The fit with data on IMF multiplicity, mean energy
etc. is normally quite good.

\section{The Microcanonical Approach}
Pioneers of this approach were the Berlin group \cite {Gross87} and
Randrup and collaborators \cite {Randrup87}.  In the Berlin approach,
the clusters, which have finite sizes, are all totally inside the freeze-out 
volume.  This freeze-out volume is the same
for all channels.  Randrup et al. demand that centres of all the clusters
should be within  the freeze-out volume.  There are other practical 
differences between the two formulations.  Let us sketch the general procedures
that any microcanonical calculation will have to accomplish.  
Suppose that we are interested only in calculating the average number
of clusters of a composite which has $k$ protons and $l$ neutrons, i.e.,
$<n_{k,l}>$, when the total energy is $E$ and the
disintegrating system has $Z$ protons and $N$ neutrons.  How do we proceed?
For simplicity
we will say that the only interaction between the clusters is that
they can not overlap.  We will also assume that the composites have only
ground state.

There are many possible break-up channels.  All divisions
(that satisfy $\sum in_{i,j}=Z$ and $\sum jn_{i,j}=N$ where $n_{i,j}$
is the number of cluster which has $i$ protons and $j$ neutrons) are
allowed and equally likely to occur.  The phase space available to each
channel, however, will be, in general, different and will strongly affect 
the final result.  The phase space associated with each break-up channel
is
\begin{eqnarray}
\Omega[n_{i,j}]=\frac{1}{h^{3M}}\Pi_{p,q}\frac{1}{n_{p,q}!}
\int \Pi d^3r_{i,j}d^3p_{i,j}\delta[E+\sum B_{i,j}-\sum p_{i,j}^2/
2m_{i,j}-\sum v(i,j:m,n)]
\end{eqnarray} 
Here $M$ is the multiplicity in the channel $[n_{i,j}]$, $B_{i,j}$ the
binding energy of the cluster $(i,j)$ and $v(i,j:m,n)$ is the potential
energy between the clusters $(i,j)$ and $(m,n)$.  In our case it is 
either 0 (when they are separated) or $\infty$ (when they overlap).
The momentum integral is analytic:
\begin{eqnarray}
\int d^3p_1....d^3p_f\delta [K-\sum_1^f\frac{p_i^2}{2m_i}]=\frac{2\pi}
{\Gamma(3f/2)}(m_1....m_f)^{3/2}(2\pi K)^{3f/2-1}
\end{eqnarray}
Once the momentum integral is done we still need to do the configuration
space integral.  This is by no means trivial but one can estimate it
in a Monte-Carlo procedure.  The first particle is placed at a random
position inside the freeze-out volume.  Having placed the first one we 
try to place the second one, again at a random position in the freeze-out
volume.  We may succeed but we may fail also if, by chance, the second 
chosen position was such that the new particle overlaps with the
particle already in.  A successful run occurs if we are able to put
$M$ particles in without failing once.  
An unsuccessful run happens if anytime in the chain we failed to put
a particle.  Then the volume integral is $(V_{freeze-out})^M\times 
\frac{N_s}{N_{un}+N_s}$ where $N_s$ is the number of successful runs
and $N_{un}$ is the number of unsuccessful runs.  The quantity just
calculated is $V^M$ where $V$ is the volume of the thermodynamic model.

Provided all this is done for each channel and we have calculated the
phase space $\Omega[n_{i,j}]$ for each channel, the average number of
particles of a composite with $k$ protons and $l$ neutrons will be 
given by
$\sum n_{k,l}\Omega[n_{i,j}]/\sum \Omega[n_{i,j}]$ where $n_{k,l}$
is the number of composites with proton number $k$, neutron number $l$
in the channel labelled by $[n_{i,j}]$ and $\Omega[n_{i,j}]$ is the
phase space integral associated with this channel.  The sum is over 
all $[n_{i,j}]$.

In practice, this procedure is impossible to carry out as the number
of channels is inordinately large.  An ``importance sampling'' \cite{Allen}
of the phase space is necessary.  This is usually done with the 
Metropolis method \cite{Allen,Metropolis}.  We will have occasion to use
this technique later also.  One attempts elementary moves by which one
migrates from one channel (here with multiplicity $M$) to a neighbouring
channel (multiplicity $M+1$ or $M-1$).  These moves are ``fission''
(take a composite and arbitrarily break it into two pieces) and ``fusion''
(join two composites).  Let us call $\Omega$ the phase space integral
before the move.  One also calculates the phase space integral after
the attempted move.  Let us call this $\Omega'$.  If $\Omega'>\Omega$
the move is accepted.  If $\Omega'<\Omega$, the probability of switching
is given by the ratio $\Omega'/\Omega$ (see \cite{Allen} why these 
give the correct transition probabilities).  An event is accepted every
$N$ attempted moves (some successful and some not) and averages 
will be calculated with many such events; $N$ should be sufficiently
large to avoid event-to-event correlation and of course one should
have sufficient number of events to reduce statistical errors.

There are a great many details which need to be worked out before
such a program can be instituted for practical calculations.
The interested reader should consult the original literature.  The
techniques for microcanonical calculations were developed in the
mid eighties spanning several papers, each one an improvement over
the previous one.  What we have outlined here are the general principles. 

\section{ The Percolation Model}
This model is strikingly different from the models described above.
The model has been extensively studied in condensed matter physics.
A delightful monograph exists \cite {Stauffer92} which has all the
material needed to follow the application in nuclear physics.  The
applications in nuclear physics were made by Campi and collaborators
\cite {Campi84,Campi88} and by Bauer and collaborators 
\cite {Bauer84,Bauer88}.

There are two types of percolation models: site percolation and bond
percolation.  For applications to nuclear physics, bond percolation
was used.  In bond percolation 
there are $N$ lattice sites.  One uses a three-dimensional cubic   
lattice, thus $N=5^3,6^3$ etc.   The number of nucleons is also
$N$.  Each lattice site contains one nucleon.  We do not distinguish
between neutrons and protons.  The crucial parameter
is the bonding probability $p$ whose value can vary between 0 and 1.
The probability that two nearest neighbour nucleons will be part 
of a cluster is given by the value of $p$.  If $p$=0 (high excitation
energy) all $N^3$ nucleons will emerge as singles.  If $p$=1,
the nucleons stay together as one nucleus (low excitation energy,
not enough to break up the nucleus).  For the values of $p$ between
the two extremes Monte-Carlo sampling is needed to generate events
and determine in each event the occurrence of clusters of different
sizes.  There is a phase transition in this model.  One can define
a percolating cluster; this is a cluster, which, if it exists,
spans the walls, i.e., connects opposite walls through an unbroken
cluster.  For $N$ very large, this appears at the value of $p=0.2488$.
This value of $p$ will be labelled by $p_c$.  The order parameter
in this model is the probability that an arbitrary site (equivalently
an arbitrary nucleon) is part of this percolating cluster.
Below $p_c$, this is zero since there is no percolating cluster.  It
starts from zero at $p_c$ and continuously moves towards the value 1
as the value of $p$ is increased.  The phase transition in this model
is continuous and not a first-order phase transition.  This aspect had a very
important and strong influence in the history of search for phase 
transition in heavy-ion collisions.  
Near critical points, One can define critical
exponents and try to evaluate them from experiment.  We will see later
that even though we now regard the phase transition in nuclear
heavy ion collisions to be first order, it is meaningful to try to
measure certain exponents.  In the lattice gas model (to be described
below) these retain significance even when one is far from a critical
point and is in the vicinity of a first-order phase transition.

We give the values of the more common exponents in the thermodynamic
limit (i.e., $N\rightarrow \infty$).
One of these exponents we have already encountered many times.  Near the
critical point the yield of mass $a$ is given by
\begin{eqnarray}
Y(a,p)=a^{-\tau}f[(p-p_c)a^{\sigma}]
\end{eqnarray}
At the critical point the yield is a power law.
The value of $\tau$ in the percolation model is 2.18.  The value of $\sigma$
in the above equation is 0.45.  Let us denote the
mass of the largest cluster by $a_{max}$.  The second moment is defined by
\begin{eqnarray}
S_2=\sum ~ ' a^2Y(a)/A
\end{eqnarray}
Here the summation excludes the $a_{max}$ and $A=N^3$ is the number of
nucleons.  The second moment $S_2$
diverges: $S_2\propto |p-p_c|^{-\gamma}$ where the value of $\gamma$
is 1.80.  In finite systems $S_2$ will not become infinite but will
go through a maximum as $p_c$ is traversed.  Above the percolation point
the order parameter is given by $a_{max}/A\propto |p-p_c|^{\beta}$ 
where $\beta=0.18$.

In spite of its simplicity, the percolation model was an aid in
understanding various phenomena.  It has now been replaced by a lattice
gas model which is more realistic, more versatile and indeed contains
the percolation model as a special case.

\section {The lattice Gas Model (LGM)}
The advantage of the percolation model is that clusters are easily
obtained.  This break-up can be compared with experiment.  But
there is no equation of state in the usual sense.
The equation of state requires two
variables: $p$ and $V$, then $T$ is automatically known from the 
EOS: or $p$ and $T$ then the EOS gives $V$ etc.  There is only one parameter
in the percolation model, the bond probability.  There is no obvious
way this model can be linked to finite temperature Hartree-Fock theory
or the thermodynamic model or SMM or the microcanonical model.  There
is no Hamiltonian.
 
In \cite {Pan195,Pan295} LGM was introduced so that one has an EOS 
as in Hartree-Fock theory but also has the capability of predicting
clusters as in the percolation model.  The EOS of LGM in mean-field
theory in a grand canonical ensemble is done in 
textbooks \cite{Huang}.  To obtain clusters in the model
an extension of the wellknown model
is necessary.  Although LGM today is more complete with
the inclusion of isospin dependence and Coulomb interaction, we introduce
first the simplest version.  This will be very easily generalised later.

As in percolation, we have $N_s$ lattice sites but now, in general,
$N_s$ is greater than $A$, the number of nucleons that need to be 
put in these sites.  When $N_s=A$ the nucleus has normal density.  We
are not allowed to put more than one nucleon on a site.  Thus the 
model is limited to normal volume or larger.  Because cluster formation
presumably takes place in a volume significantly larger than normal
volume, this restriction is not debilitating.  The nearest neighbours
have a bond $\epsilon$ which is negative.  Only nearest neighbours
interact.  The exclusion of the possibility of two nucleons occupying
the same site mimics a short range repulsion.  The attractive nearest
neighbour interaction simulates the attractive interaction which
is also short range (but longer than the short range repulsive interaction).

Let $N_{nn}$ be the number of $nn$ bonds in a specific lattice configuration.
The energy carried by these bonds is $\epsilon N_{nn}$.  Thus the
partition function is 
\begin{eqnarray}
Q=\sum_{N_{nn}}g(N_s,A,N_{nn})e^{-\beta\epsilon N_{nn}}
\end{eqnarray}
Here $g(N_s,A,N_{nn})$ is the number of configurations which have $N_{nn}$
nearest neighbour bonds and which can be formed from $A$ nucleons in
$N_s$ lattice sites.  This is not analytically solvable.  Hence calculation
of observables where configurations have the above weighting requires
Monte-Carlo simulations.

The simulations are usually done in the Metropolis algorithm.  Starting
from an initial configuration chosen suitably \cite {Sam}, one attempts
a switch between an unoccupied site and an occupied site.  If the resulting
change of energy $\Delta E$ is negative, the switch is accepted.  If 
$\Delta E$ is positive, it is accepted with a probability $e^{-\Delta E/T}$.
After a large number $N$ of attempted switches (some successful and some
unsuccessful) an event is accepted.  $N$ should be large enough
to avoid event to event correlation.

The grand canonical ensemble of the LGM (sum over all possible $A$'s) can
be mapped onto a three dimensional Ising model \cite {Huang,Lee}.  The
latter has been extensively studied and indeed serves as a model
for liquid-gas phase transition.  Many of the known results of the Ising
model can be directly applied.  For example in the large $A$ limit
the critical temperature will be $1.1275\epsilon$ and the critical density
$\rho/\rho_0=A/N_s=0.5$.

We consider now an extension of the model so that clusters can be 
computed.  Suppose we have generated a configuration.
At finite temperature, the nucleons will not be frozen at
the lattice sites.  They will have momenta.  In this configuration
each of the $A$ nucleons can be given a momentum by Monte-Carlo
sampling of a Maxwell-Boltzmann distribution at the given temperature.
Thus in an event we have nucleons at definite lattice sites with
definite momenta.  There may be some isolated nucleons which have
no nearest neighbours.  These clearly are monomers.  The next case
is when there is a cluster of two nucleons which are nearest neighbours
of each other.  They will form a bound cluster of two
if the kinetic energy of relative motion is insufficient to overcome
the attraction between the two nucleons, i.e., $p_r^2(1,2)/2\mu
+\epsilon<0$.  Here $\vec p_r(1,2)=\frac{1}{2}(\vec p_1-\vec p_2)$
and $\mu=m/2(m$=mass of one nucleon).

It turns out that this prescription which is rigorously correct for a
cluster of two also works {\it statistically} for larger clusters. 
That is, we can formulate a rule that independent of other neighbours,
two nearest neighbours form part of the same cluster if the relative 
kinetic energy of the two is insufficient to overcome their attraction.
It is obvious that this recipe, introduced in \cite {Pan195,Pan295},
reduces the many body problem of recognising
a cluster of many nucleons into a sum of independent two body problems.
For brevity we refer to this as PD recipe.  To see why this works
{\it statistically} even if not individually let us specifically consider
a three body cluster \cite {Sam}.  

For three particle clusters, nearest neighbours are either linear or L
shaped.  In either case there is only one particle which has two bonds
(label this by particle 2) and two others (label them 1 and 3) which have 
one bond each.  According to the PD recipe this will form a three body
cluster if $p_r^2(1,2)/2\mu+\epsilon <0$ and $p_r^2(2,3)/2\mu+\epsilon <0$.
To check if particle 3 is part of a three body cluster (similar arguments
will be needed for particles 1 and 2) we should instead verify if
$p_r^2(12,3)/2\tilde \mu+\epsilon <0$.  Here $\vec p_r(12,3)$ is the
relative momentum between the centre of mass of (1+2) 
and 3; $\tilde \mu=(2/3)m$
is the reduced mass for this relative motion.  Thus there may be cases
where the PD recipe gives a three body cluster wheras in reality the
third one will separate.  But there will also be cases where the PD recipe
will deem that the third one will separate whereas in reality it stays
attached.  Statistically overestimation will cancel out underestimation
because for a Maxwell-Boltzmann distribution all relative motions are
also Maxwellian at the same temperature.  That is, in Monte-Carlo
simulation, $p_r^2(12,3)/2\tilde \mu$ will be as many times below the value
$-\epsilon$ as $p_r^2(2,3)/2\mu$ will be.

The same argument applies to particle 1.  For particle 2. it can be verified
that if $p_r^2(1,2)/2\mu+\epsilon<0$ and 
$p_r^2(2,3)/2\mu+\epsilon <0$ then
$p_r^2(13,2)/2\tilde\mu+2\epsilon <0$ is always satisfied.

Campi and Krivine have used a different approach and come to the
conclusion \cite{Campi97} that the PD recipe gives the correct number
of particle stable clusters.  

Recognition of clusters in a many body system is a complicated issue.
The PD recipe was tested in \cite{Subal96} in numerical simulations
and found reliable.  In the PD recipe once the configuration of
$A$ nucleons and their momenta are given, the cluster decomposition
is immediate.  One may however, starting from this initial condition,
switch to a different model.  One may
propagate particles using molecular dynamics.  At asymptotic times
clusters are easily identified as different clusters will separate
from each other.  Of course the result will depend upon the interaction
potential used for molecular dynamics propagation.
To test the PD recipe one must use an interaction
consistent with the assumptions of the lattice gas model.  Let $a$
be the length of each side of the elementary cubic lattice.  The interaction
between particles must become repulsive when the distance between
them gets to be less than $a$.  At distance $\sqrt{2}a$ it is deepest
at -5.33 MeV.  At distance beyond it must go to zero
rapidly.  Given the same initialisation and such interaction,
molecular dynamics produced very similar results as the PD recipe.

It follows that with this recipe of calculating composites, we
do not need to worry about subsequent evaporation as one needs to
in many other models; thermodynamic, SMM and microcanonical.  This is
a tremendous advantage.  Evaporation was already taken into account when
we applied the PD recipe.  One does not take the size of the cluster
to be given by just the number of nucleons which are connected to each
other through the nearest interaction \cite {Biro}.  Some of these
will fly away.  The rest that remain and are counted, are particle
stable.

With a prescription for obtaining clusters, the LGM will show many
features in common with the percolation model.  This leads to
interesting properties.

\section{Phase Transition in LGM}
Whether one later ascribes momenta to nucleons and calculates clusters
or not, there is phase transition in the traditional LGM.  The
thermodynamics of the system does not depend upon the definition of
clusters.  The 
coexistence curve can be drawn.  This diagram is simply transferable 
from studies in the three dimensional Ising model.  We show this in
Fig.24.  The thermal critical point is shown in the diagram as C.P.

\vskip 0.2in
\epsfxsize=3.0in
\epsfysize=3.0in
\centerline{\epsffile{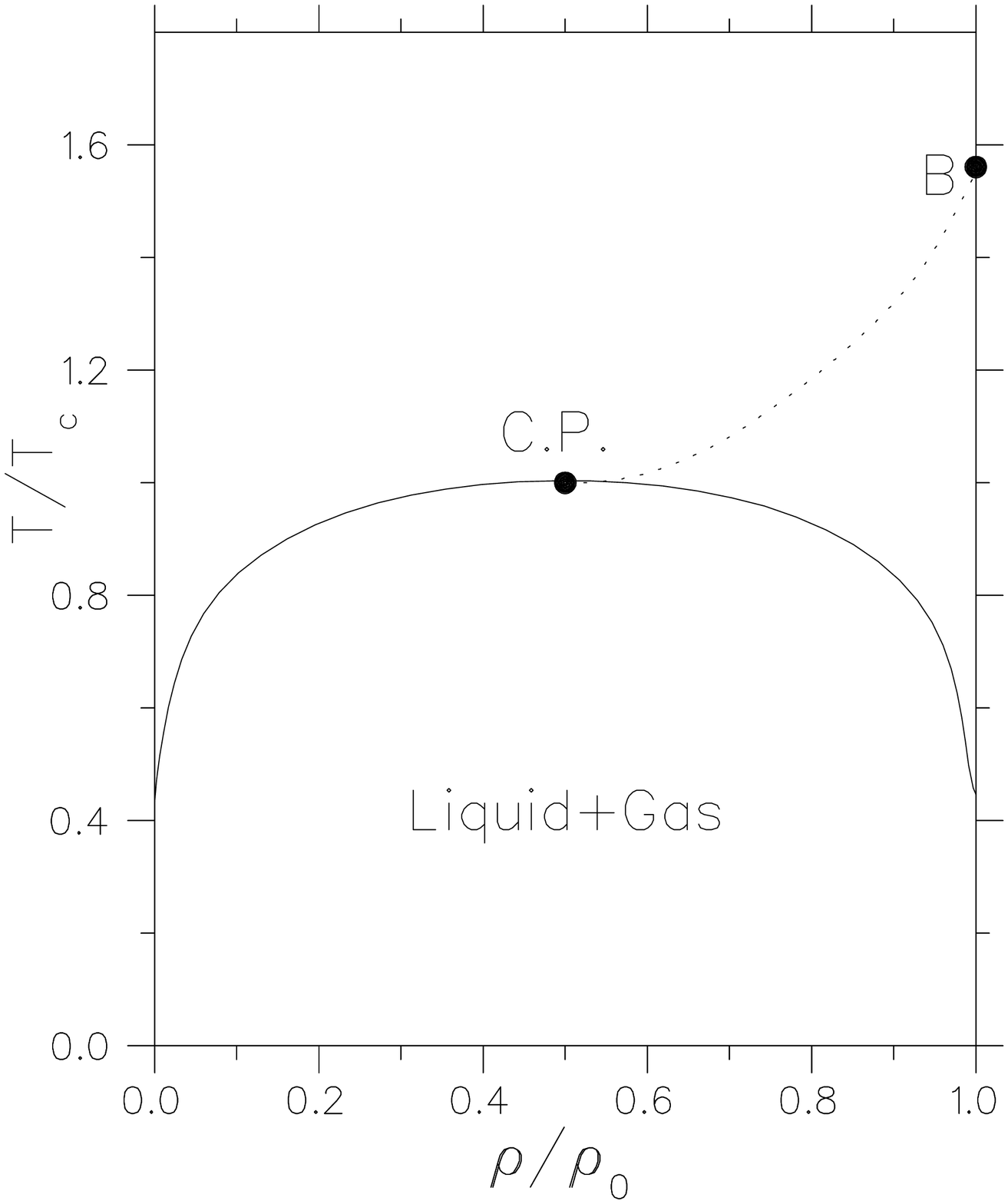}}
\vskip 0.1in

Fig.24. Phase diagram of the three dimensional lattice gas model.
The full line is the coexistence curve.  Percolation sets in along
the dotted line and continues along the coexistence line to the left 
of C.P.

With a rule for calculating clusters it will be very interesting if
the thermal critical point also coincides with the onset of a percolating
cluster.  This aspect was studied by Coniglio and Klein \cite{Coniglio}.
They propose that the probability that two nearest bonds have an active
bond between them be given by 
\begin{eqnarray}
p=1-\exp (-\beta |\epsilon|/2)
\end{eqnarray}
With this definition, these authors, using renormalisation group techniques,
proved that at $(\rho_c,T_c)$ percolation just sets in.  However, 
percolation sets in not just at the thermodynamic critical point but rather
along a continuous line in the $(\rho,T)$ plane (the dotted line in Fig.24).
This was studied in \cite {Kertesz} and the line is called
Kert\'esz line.  Thus the critical exponents $\tau,\beta$ and $\gamma$
are meaningful not only at the critical point but along an entire line.

It turns out the PD recipe which is a natural choice for calculation
of clusters in the case of nuclear disintegration is very close to
the Coniglio-Klein (CK) formulation.  The PD formula for $p$ (using
the fact that in a Maxwell-Boltzmann distribution the relative motion
is also Maxwell-Boltzmann) is
\begin{eqnarray}
p=1-\frac{4\pi}{(2\pi\mu T)^{3/2}}\int_{\sqrt{2\mu |\epsilon|}}^{\infty}
\exp (-p_r^2/2\mu T)p_r^2dp_r \nonumber
=1-\frac{2}{\sqrt{\pi}}\int_{|\epsilon|/T}^{\infty}e^{-q}q^{1/2}dq
\end{eqnarray}
A comparison of $p$ according to the above formula to the Coniglio-Klein
formula is shown in Fig.25.  They are very close.  As far as we
know, all cluster calculations in nuclear physics use the PD recipe.

With the aid of Fig.24 we can now discuss phase transition in
nuclear disintegration according to LGM.  The freeze-out volume that fits
the data best \cite{Pan295} is bigger than twice the normal nuclear
volume.  In that case as the temperature of the disintegrating system
is raised from a low value to a high value (either by changing the
beam energy or by gating on appropriate impact parameter) the system
will cross the coexistence curve on the low density side of the 
critical point (to the left of C.P. in Fig.24).  Thus we will have
first-order phase transition \cite {Pan98}.  As the line is crossed
one will see a discontinuity in specific heat, a peak in $S_2$ and
other features.

\vskip 0.2in
\epsfxsize=3.0in
\epsfysize=3.0in
\centerline{\epsffile{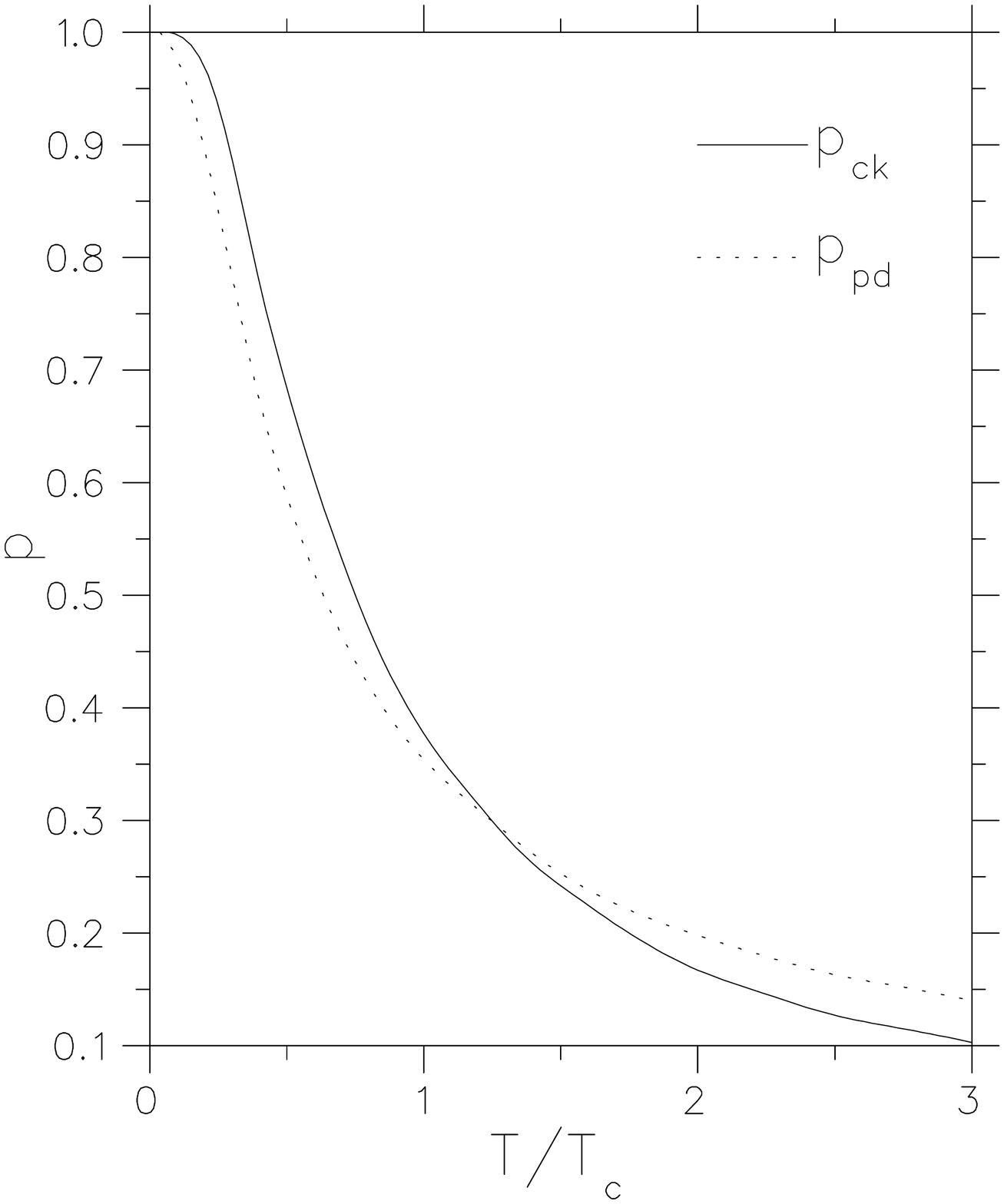}}
\vskip 0.2in
Fig.25. The bond probability p is plotted as a function of temperature.
The solid line ($p_{ck}$) is the Coniglio-Klein formula, the dotted line
($p_{pd}$) is the PD recipe.

\section{Isospin dependent LGM including coulomb interaction}
For many practical applications, it becomes necessary to distinguish
between like particle interactions (bond between proton and proton
or/and between neutron and neutron) and unlike particle interaction
(bond between neutron and proton); $\epsilon$ between like particles
must be repulsive or zero otherwise we can obtain dineutron or
diproton bound states.  The bond between unlike particles will be
attractive.  At zero temperature in nuclear matter, energy considerations
imply that sites will be alternately populated by neutrons and protons.
Thus the only nearest neighbour interactions will be those between unlike
particles.  Nuclear matter binding energy then dictates that 
$\epsilon_{np}=-5.33$MeV.  This however does not fix 
the value of $\epsilon_{pp}$ or $\epsilon_{nn}$.

It is clear that the Monte-Carlo technique of generating events
for finite nuclear systems can also be used when the interactions
between like and unlike particles are different.  The Coulomb
interaction between protons can also be included.  When this is done
at zero temperature we obtain the ground state energies.
This was done in \cite{Sam} for a range
of nuclei for $\epsilon_{pp}=\epsilon_{nn}=0$.  The binding
energies of these nuclei thus computed were then fitted
to a simple liquid-drop mass formula:
\begin{eqnarray}
E/A=-a_v(1-\kappa I^2)+a_s(1-\kappa I^2)A^{-1/3}+a_c\frac{Z^2}{A^{4/3}}
\end{eqnarray}
Here $I=(N-Z)/A$ is the neutron-proton asymmetry of the nucleus.  
The fit of LGM binding energies 
to this four parameter formula is quite good.  We compare
the four parameters deduced from LGM to liquid-drop model values 
\cite{Myers} in the table.  Considering the simplicity of the model,
the agreement is gratifying.  We also notice that the asymmetry
parameter $\kappa$ is larger than the liquid-drop value.  
Since $\epsilon_{np}$
is fixed from nuclear matter binding energy, the only way we can bring
down $\kappa$ is to make $\epsilon_{pp}$ and $\epsilon_{nn}$ go negative.  
As explained already this is not permissible.  We are therefore led
to the conclusion that $\epsilon_{np}=-5.33$MeV and $\epsilon_{pp}=
\epsilon_{nn}=0$ are the best choices for isospin dependent LGM.
\vskip .5in
Table I. Lattice gas and phenomenological liquid-drop model parameters.

-------------------------------------------------------------------------------

~~~~~Model~~~~~~~~~~~~~~~~$a_v$~~~~~~~~~~~~~$a_s$~~~~~~~~~~~~$\kappa$~~~~~~~~~~~~$a_c$

~~~~~~LGM~~~~~~~~~~~~~~16.0~~~~~~~~~~16.03~~~~~~~~~~2.14~~~~~~~~0.746

Phenomenological   ~~15.68~~~~~~~~~18.56~~~~~~~~~~1.79~~~~~~~~0.717

-------------------------------------------------------------------------------

\section{Calculations with isospin dependent LGM}
We mentioned earlier that the isotopic content of the gas phase
can be different from that of the liquid phase.  This comes out 
nicely in LGM.  Fig.26, taken from \cite{Sam} demonstrates this.
Here one considers breakup of $^{197}$Au at different temperatures.
The average value of the charge of the largest residue at each
temperature is denoted by $<Z_{max}>$.  This will drop in value as the
temperature is increased.  Also calculated is the average value of
neutron content $<N_{max}>$ of the largest cluster.  We may regard the
largest cluster as the liquid phase and the rest of the nucleons as
primarily belonging to the gas phase.  The disintegrating system has
$N/Z=1.49$.  With an isospin dependent LGM, the $<N_{max}>/<Z_{max}>$
is much closer to 1.  This means the gas phase has a higher $N/Z$
ratio, higher than that of the parent system.  The reason for this
behaviour is that the symmetry energy drives the $N/Z$ ratio of the
largest cluster towards the value unity.  The Coulomb effect will
offset this as shown in the figure.  The simplest version
of the LGM which had no isospin dependence and no Coulomb term will
keep the $<N_{max}>/<Z_{max}>$ at the value pertaining to the 
disintegrating system.  This is also shown in the figure.  This 
contradicts experiment.

\vskip 0.2in
\epsfxsize=3.0in
\epsfysize=3.0in
\centerline{\epsffile{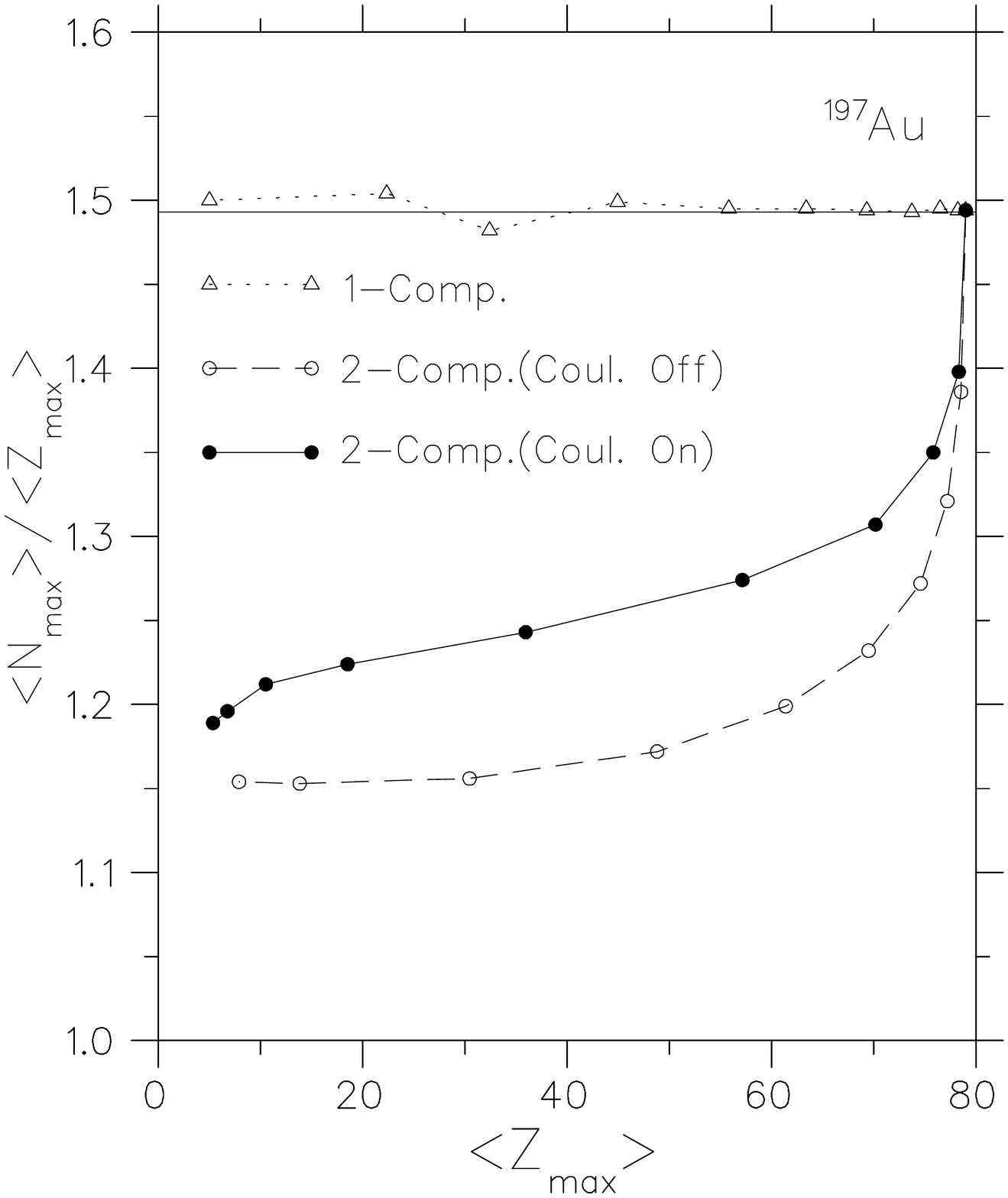}}
\vskip 0.2in

Fig.26 The ratio $<N_{max}>/<Z_{max}>$ as $<Z_{max}>$ changes (because
temperature changes).  The temperature increases by 0.5 MeV between
two successive symbols as we move from right to left starting with
2.5 MeV.

\vskip 0.2 in
For $^{197}$Au we show several quantities as a function of temperature.
In experiments one extracts the quantity $\tau$ where the yield $Y(Z)$
as a function of $Z$ is fitted to $Y(Z)\propto Z^{-\tau}$.  The
power law comes out quite well in LGM.  We notice that the maximum in
$C_v$, the maximum in $S_2$ and the minimum in $\tau$ all bunch 
around T=4.2 MeV which we then associate with the crossing of the 
coexistence curve.  The maximum in $N_{IMF}$ is at a slightly higher
temperature.

\vskip 0.2in
\epsfxsize=3.0in
\epsfysize=3.0in
\centerline{\epsffile{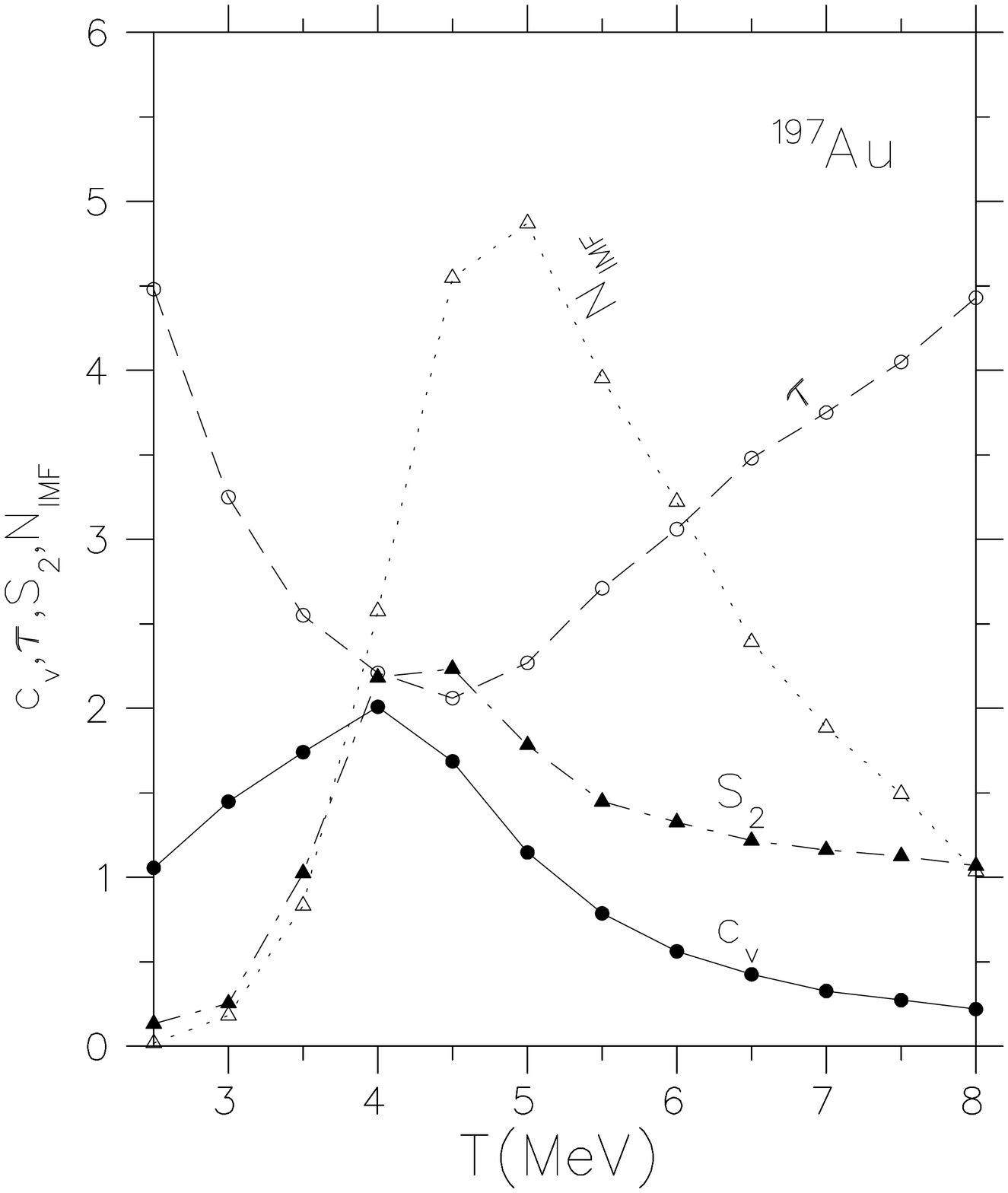}}
\vskip 0.2in

Fig.27 The specific heat $C_v$, the exponent $\tau$ for the power law
fit to the yields of isotopes, the second moment $S_2$ and the IMF yield are
shown as a function of temperature for $^{197}$Au.  The calculations are
done in an $8^3$ lattice.

The EOS for isospin dependent LGM in mean-field theory using
the Bragg-Williams and the Bethe-Peierls approximation can be found
in \cite {Pan298}.  Phase transition aspects, much more than what
we have covered here, can be found
in \cite {Ch99,Gu99,Borg,Ma199}.  The model has been used successfully
to obtain several experimental results, for example,
$t/^3He$ ratios as the isotopic content of the parent
system changes \cite {Ch99,Sam}.  Some applications were made in 
\cite {Ma99}.  The shortcomings of the model for detailed fittings
to experimental data are obvious.  The model has cubical symmetry
rather than spherical symmetry.  The shell effects are missing
although the smooth part of binding energy is approximately
reproduced.  The excitation spectrum of the composites is incorrect.  But
it has many nice features not present in other models.  Here composites are
formed directly out of fluctuations.  One starts with interactions
between two nucleons.  The inclusion of Coulomb effects is
precise, though numerical.  Best of all, it includes interactions
between composites. 

\section{Fragment yields from a model of nucleation}

A phenomenological droplet model based on homogeneous nucleation theory
has also been used to describe mass yields provided in heavy-ion
collisions \cite{Good}.  The nucleation model is an extension
of the Fisher droplet model which was originally used to describe
such yields.  The extension allows for the possibility
that supersaturated systems are produced during the brief encounter
of two colliding nuclei.  Homogeneous nucleation occurs in
supersaturated systems when chance collisions of particles in the
gas phase yield to local density inhomogeneties.  These inhomogenities
are droplets of particles of the liquid phase that will grow in size
if the systems lived for a long time.  Specifically, in the
supersaturated phase, a critical size droplet exists which is determined
by the surface tension and the difference of liquid and gas chemical
potentials.  Droplets larger than the critical size will grow by 
accumulating nucleons in order to lower the free energy of the system while
droplets of smaller size will evaporate nucleons also lowering the
free energy of the system.  This behaviour in growth and evaporation
reflects itself in a yield distribution which is U-shaped with an
initial decrease with $A$ until the critical size $A_c$ is reached
and then an increase in yield with $A$ above $A_c$.  In this
nucleation description, the probability of formation of droplets
is determined by calculating the change in Gibbs free energy with
and without a drop at constant temperature and pressure.  For
example, if a droplet of size $A$ is surrounded by $B$ droplets
of the gas phase, then 
$G_{with drop}=\mu _lA+\mu _gB+4\pi R^2\sigma(T)+T\tau lnA$
and $G_{no drop}=\mu _g(A+B)$.  Here $\mu_l$ and $\mu_g$ are the
liquid and gas chemical potentials, $R$ is the radius of the drop
(with $R=r_0A^{1/3}$) and $\sigma(T)$ is the surface free energy
such that $4\pi r_0^2\sigma(T)=18$ MeV at $T$=0.  The $T\tau lnA$ term
is a term introduced by Fisher to account for the
power law fall-off of yield distributions at a critical point with
$\tau$ a critical exponent.  The probability of forming a drop of
$A$ nucleons is $P\propto \exp(-\Delta G/T)$ where $\Delta G=
G_{with drop}-G_{no drop}$.  This gives a cluster distribution
$N(A)$, to a constant $C$,
\begin{eqnarray}
N(A)=\frac{C}{A^{\tau}}\exp[\frac{\mu_g-\mu_l}{T}A-
\frac{4\pi R^2\sigma}{T}A^{2/3}]
\end{eqnarray}
On the coexistence curve of a liquid-gas phase transition 
$\mu_g=\mu_l$ and $N(A)$ is a monotonically decreasing function of $A$.
For a supersaturated system $\mu_g>\mu_l$ and $N(A)$ has a minimum
value at a critical size droplet $A_c$.  Neglecting the $\tau lnA$
term, $A_c$ is given by $A_c^{1/3}=\frac{2}{3}\frac{4\pi r_0^2\sigma}
{\mu_g-\mu_l}$.

The model fits yields of many heavy-ion collisions.

\section{Isospin fractionation in mean-field theory}

Early studies of the liquid-gas phase transition were carried out using
a Skyrme interaction and focussed mostly on a one component system made of
nucleons even though expressions were developed for two component system 
of protons and neutrons \cite{jaquaman2}.  The one 
component aspects are given in
section II of this review.  The two component aspects will now be discussed
in a mean-field approach with this section based mostly on the work of
Muller and Serot \cite{Muller95}.  Extensions of the results of 
\cite{jaquaman2} are now being carried
out in \cite{Lee00} using a Skyrme interaction while the Muller-Serot analysis
is based on a relativistic mean-field model.  Initial results in \cite{Lee00}
are qualitatively similar to those of \cite{Muller95}.  
Both approaches allow a
complete calculation various thermodynamic properties, such as the
pressure and proton and neutron chemical potentials.  In one component 
systems, the Skyrme interaction and the relativistic mean-field model
lead to an S-shaped behaviour of pressure versus volume or density at 
fixed temperature with stable, unstable, supersaturated and supercooled
regions.  The standard Maxwell construction describes the liquid to gas
phase transition with both phases having the same pressure $p_L=p_G$,
and the chemical potential $\mu_L=\mu_G$.  The equality of chemical
potentials of the liquid and gas phases in phase equilibrium is
equivalent to equal areas of the regions above and below the Maxwell or
vapour pressure line in the S-shaped loop in $p$ vs. $V$.  For two 
component systems, phase equilibrium becomes more complicated since the
proton to neutron ratio can be different in the two phases because of 
the symmetry energy which favours $N=Z$.  Since the symmetry energy will
be large in the denser liquid phase, the proton-neutron asymmetry will be
bigger in the gas phase than in the liquid phase.  In two component
systems, the phase equilibrium conditions consist of setting, at fixed
pressure, the proton chemical potentials in the two phases equal to
each other and similarly, the neutron chemical potentials are equal
in the liquid and gas phases.  The properties of the phase separation
boundaries (binodals) and instability boundaries (spinodals) are studied
as a function of a quantity labelled $y=\rho_p/\rho$ which is the proton
fraction, with the neutron fraction given as $\rho_n/\rho=1-y$, and
$\rho=\rho_n+\rho_p$.  For two component systems, the binodals are now
a surface in plots in ($p,T,y$) space.  By contrast, for one
component systems, the binodal is a line for the vapour or Maxwell pressure
vs. $T$ which terminates at the critical temperature $T=T_c$.  The line
is at $y=0.5$ in the ($p,T,y$) space of two component systems.  The
two dimensional binodal surface of the phase coexistence boundary now
contains a line of critical points for different values of $y$ and a line
of maximal asymmetry.  In a $(p,T,y)$ plot of the binodal surface, the
intersection of a fixed $T$ plane with the surface gives a loop of $p$
vs. $y$.  The maximal asymmetry point is at $(dy/dp)_T=0$ and physically
corresponds to the smallest proton ratio or the largest neutron ratio
on the binodal surface at each $T$.  The critical
point is at $(dp/dy)_T=0$ on this loop.  The loop degenerates to a point
at the critical temperature of a symmetric system $y=0.5$ (see fig. 7 in
\cite {Muller95}).  If $T$ and $p$ are both fixed, the binodal 
surface has two values
of $y$: $y_1(T,p)$ and $y_2(T,p)$ corresponding to different values of the
proton fraction in the liquid and gas phase.  These different values 
arise because of the difference in symmetry energy in the liquid and the gas
phase.  One of the interesting conclusions of the mean-field two component
model of the liquid-gas phase transition is that the first-order transition
of a one component system becomes a second-order transition.  Because of
the greater dimensionality in the physical situation, the phase transition
is continuous.  The role of dimensionality due to the number of components
of the system on the order of the phase transition was also pointed out
by Glendenning \cite{Glendenning}.

\section{Dynamical Models for Fragmentation}
The one common characteristic of all the theoretical models considered so
far is that they are static, i.e., they all assume that equilibrium
is achieved and hence laws of eqiulibrium statistical mechanics apply.  A
more fundamental calculation would use a transport equation.  Here
two nuclei approach each other in their ground state.  By solving a time
dependent equation we see what the final outcomes are.

The BUU model does start with two nuclei, in their ground states, boosted
towards each other.  One does not have to introduce a temperature.
The model has a mean-field as well as hard collisions and has indeed 
proven to be highly successful in predicting sideward flow, squeeze-out
etc. \cite{Bertsch88}.  While BUU is good for predicting 
expectation values of one-body operators,
it does not have fluctuations.  Thus it will not produce clusters.  A great
deal of effort went into introducing fluctuations in BUU type formalism.  
A very short list of
references are \cite {Gsdg,Bbsdg,Ayik1,Ayik2,Burgio}.  Unfortunately, practical
calculations are extremely time consuming.  In rare cases calculations have
been done to compare with experimental data\cite {gallego} but the method 
has not been pushed to see, for example, the rise and fall of IMF production,
an accurate estimation of the $\tau$ parameter etc.  Maybe, in future
such calculations will be done.

While exact calculations for heavy-ion collisions based on quantum mechanics
are impossible to carry out, classical calculations for ion-on-ion
collisions for $\approx$400 particles are entirely feasible with present day
computers.  This requires solving for each particle 
$i: d\vec r_i/dt=\vec p_i/m$
and $d\vec p_i/dt=-\sum_{j\ne i}\vec \nabla_{r_i} V(\vec r_i,\vec r_j)$.  
Here $V(\vec r_i,\vec r_j)$ is the two body interaction.
One starts at time $t=0$ with initial values of
$\vec r_i, \vec p_i$ and numerically integrates out time.  In a set
of three papers \cite{Sch,Lenk,Vicentini} Pandharipande and coworkers
studied disassembly of hot classical drops as well as collisions between
cold charged argon balls containing $A_1$ and $A_2$ particles.  The chosen
values of $(A_1,A_2)$ were (108,108), (200,16) and (65,65) \cite{Sch}. 
The interaction
between atoms was a truncated (12,6) potential and a Coulomb interaction
was added.  Although this is thirteen years later, the reader will
find reference \cite {Sch} still very relevant and revealing.  As a
function of time, the evolution of temperature and density of
the central region is plotted so that one can see under what
initial conditions the central region reaches spinodal instability
and what the final products are in such cases.  They point out
that the mass yields calculated in the disassembly of hot equilibrated drops
having 216 particles and density somewhat less than normal density is 
very similar to mass yield of 108+108 collisions.  Thus 
the assumption of statistical equilibrium is valid. 
The authors stress that the classical argon balls used in the 
study are not intended to be mock nuclei, but instead to provide simple
systems whose time evolution can be studied exactly by solving Newton's
equations of motion.  Direct comparisons with nuclear data are difficult.
Nonetheless, there are many similarities.  A power law for fragment
yields followed from these calculations, the minimum value being 
about 1.7.  Another remarkable feature is that 
the apparent temperature deduced from the fragment kinetic energies is
much larger than that of the system.  The large kinetic energies of
fragments in the simulation appear to come from collective motion of
expansion and Coulomb repulsion.

In a later paper, the Illinois group devised a simple nucleon-nucleon
potential for classical calculations of nuclear heavy ion collisions
\cite{Vijay}.  They did not use this potential to study liquid-gas phase
transition but instead used this at Bevalac energies to investigate 
flow angles and transverse momenta.  Later, disassembly of a collection
of nucleons which start with initial temperature and density of
interest in this article, 
and interact via this potential was considered by other groups
\cite {Latora,Belkacem,Finocchiaro,Dorso}.  Pratt
et al. \cite {Pratt95} used a truncated (8,4) potential to study 
similar dissociation.
It has not been demonstrated that such simulations apply to nuclear cases 
very well since actual nuclear data for specific cases have not been
compared.

The Frankfurt group proposed a simulation which they called quantum
molecular dynamics(QMD).  This was used for Bevalac energies initially.
Relativistic versions exist and are in frequent use.  This has also
been used for energy region of interest here.  Detailed exposition
of the model exists \cite {Aichelin}.  Here each nucleon is represented
by a Gaussian in coordinate and momentum space with widths consistent
with uncertainty principle.  The centroids of the Gaussians move but the
widths are kept fixed.  The centroids move under the influence of 
mean-field except when the centroids move very close to each other 
($b<\sqrt{\sigma/\pi}$).  Then they scatter as in two-body scattering. Pauli 
blocking is taken into
account for scattering.  Because each particle is represented by a centroid
in phase space with fixed widths in momentum and coorinate space, there
is fluctuation built into the system and in the final stage one can recognise
clusters.  In calculations reported in reference \cite {Tsang93} fragment 
multiplicities were underpredicted in the energy region of interest here.
The Copenhagen group has done simulations using a prescription
which they dubbed nuclear molecular dynamics \cite {cop}.
This is quite similar to QMD.

All such simulations which provide clusters at the end are quite
computer intensive..  The clusters are easily recognised if each
event is run until ``asymptotic'' times so that different clusters
are well separated from each other.  But that requires
considerable computer time.  Quite sophisticated
algorithms have been introduced for early recognition of clusters 
\cite {Dorso97}.  This is of practical importance.

Many other models are on the market which will simulate ion-on-ion collisions.
They were not necessarily introduced to study liquid-gas phase transition.
Some references are \cite {Boal,Feld,Horiuchi}.  Phase transition aspects
were discussed in reference \cite{Schnack}.

A very attractive model, expanding emitting source  (EES) model was
proposed by Friedman \cite {Fried90}.  This model assumes that initially the
hot system evaporates as well as expands.  For low initial temperature
the system will cease expanding and will revert towards normal density.
But beyond a certain temperature at the end of this slow expansion the system
will explode.  The relationship of this model to liquid-gas phase transition
will be interesting to explore.

\section{outlook}

The possible links of many experimental observables to expected liquid-gas
phase transition in finite nuclear systems continue to be a fascinating
story.  Much has been learnt and much remains to be learnt.  The topic
has forced nuclear physicists to delve into realms that were not familiar
to them.  Necessity has prompted us to do interesting theoretical
work, for example, \cite {Gu99} in finite-size scaling and in general,
about phase transitions in ``small'' systems \cite{Gro}.  We are exploring 
ideas which are of relevance in other fields. There is a broadening 
of the horizon 
which is refreshing.  We foresee substantial effort in several
directions, for example, in dynamical models.

Just fifteen years ago, multifragmentation was barely mentioned in the
literature. However, in the past decade
tremendous progress has been made in understanding the
multifragmentation process and its relationship to the liquid-gas phase
transition in nuclear matter.  Even though we have not found one definitive
experimental signature pinpointing the phenomenon, 
we know that a mixed phase can be created in the heavy-ion reactions, that
multifragmentation occurs within 50-200 fm/c after the initial collisions
with a freeze-out density of less than 1/3 of the normal nuclear density,
and that the freeze-out temperature is probably in the range of 4-6 MeV.
In the near future, experiments will be designed to measure
excitation energy, reaction time and freeze-out densities and other
observables more precisely. Nonetheless, the availability
of comprehensive experimental data has stimulated intense interest
on the theoretical front leading to better understanding of the
statistical and dynamical nature of nuclear collisions. 
More exciting developments will be awaiting in the
exploration of the isospin degree of freedom in the liquid-gas phase
transition with the availability
of high to moderate intensity radioactive beams.

In writing this article we have also realised that a vast amount 
of work has been done in this field
by groups widely spread geographically.  The literature on this subject
is colossal.  Our reporting had to be necessarily
selective, influenced largely by our own involvement and experience.
We are particularly aware that in writing an article about a subject whose
scope is this large we must have left out a significant amount
of interesting work.  For example, several review articles can be written
on the subject matter of the last section alone.  We finish therefore by 
apologising for all the omissions that occurred. 

\section{Acknowledgements}

Subal Das Gupta thanks J. Lopez for an advanced copy of a 
monograph on heavy-ion collision reactions  written by 
Lopez and Dorso\cite {Lopez}.  Research support was
provided by Natural Sciences and
Engineering Research Council, {\it le Fonds pour la Formation de
Chercheurs et l'aide \`a la Recherche du Qu\'ebec}, 
the National Science Foundation
Grant No. PHY-95-28844 and the U.S. Department
of Energy Grant No. DE FG02-96ER40987.


\end{document}